\documentclass[%
 reprint,
superscriptaddress,
nofootinbib,
nobibnotes,
 amsmath,amssymb,
 aps, prd
]{revtex4-2}

\usepackage{graphicx}
\usepackage{dcolumn}
\usepackage{bm}
\usepackage{hyperref}
\usepackage{aas_macros}
\usepackage{amsmath} 
\usepackage{physics}
\usepackage{amssymb}
\usepackage{multirow}
\usepackage{graphicx}
\usepackage[table,xcdraw]{xcolor}

\usepackage{orcidlink}

\newcommand{\LANL}{\affiliation{Center for Nonlinear Studies, Los Alamos National Laboratory, Los Alamos, NM 87545, USA}}
\newcommand{\UNH}{\affiliation{Department of Physics \& Astronomy, University of New Hampshire, 9 Library Way, Durham, NH 03824, USA}}
\newcommand{\UCB}{\affiliation{Department of Physics, University of California, Berkeley, Berkeley, CA 94720, USA}}
\newcommand{\KU}{\affiliation{Department of Physics and Astronomy, KU Leuven, Celestijnenlaan 200D, B-3001 Leuven, Belgium}}
\newcommand{\WSU}{\affiliation{Department of Physics \& Astronomy, Washington State University, Pullman, Washington 99164, USA}}

\graphicspath{{./}{figures/}}

\begin{document}

\preprint{APS/123-QED}

\title{Influence of neutrino-electron scattering and neutrino-pair annihilation on hypermassive neutron star}

\author{Patrick Chi-Kit \surname{Cheong}~\orcidlink{0000-0003-1449-3363}}
\email{patrick.cheong@berkeley.edu}
\LANL
\UNH
\UCB

\author{Francois \surname{Foucart}~\orcidlink{0000-0003-4617-4738}} \UNH

\author{Harry Ho-Yin \surname{Ng}~\orcidlink{0000-0003-3453-7394}}
\affiliation{Institut f\"ur Theoretische Physik, Goethe Universit\"at, Max-von-Laue-Str. 1, 60438 Frankfurt am Main, Germany}

\author{Arthur \surname{Offermans}~\orcidlink{0000-0002-8313-5976}} \KU

\author{Matthew D. Duez~\orcidlink{0000-0002-0050-1783}} \WSU

\author{Nishad Muhammed~\orcidlink{0000-0001-8574-0523}} \WSU

\author{Pavan Chawhan~\orcidlink{0000-0002-3694-7138}} \WSU

\date{\today}

\begin{abstract}
We investigate the influence of inelastic neutrino microphysics in general-relativistic magnetohydrodynamics simulations of a hypermassive neutron star. 
In particular, we include species/energy groups coupled neutrino-matter interactions, such as inelastic neutrino-electron scattering and electron-positron annihilation kernels, into simulations up to 50~ms.
Neutrino-electron inelastic scattering is known to have effective neutrino-matter energy exchange.
We show that, with neutrino-electron inelastic scattering, simulations predict 75\% higher disc mass with slightly different mass-averaged compositions, and 18\% more ejected mass with similar distributions.
The enhancement of the mass of the disc and the ejecta results in stronger baryon pollution, leading to less favourable jet launching environments. 
Furthermore, neutrino luminosities are about 50, 40, and 30\% higher for electron neutrino, electron anti-neutrino, and heavy-lepton neutrinos. 
In contrast, we do not see any significant impacts due to electron-positron annihilation.
\end{abstract}

\maketitle


\section{\label{sec:intro}Introduction}
Neutron star mergers that power kilonovae via the radioactive decay of the outflow of the system~\citep{1998ApJ...507L..59L, 2010MNRAS.406.2650M, 2016AdAst2016E...8T, 2019LRR....23....1M} are one of the source of $r$-process elements production.
Our current understanding of neutron star mergers~\citep{2017arXiv171005931M, 2018ASSL..457.....R} has been confirmed via the groundbreaking multimessenger detections of a binary neutron star merger on 17 August 2017~\citep{2017PhRvL.119p1101A, 2017ApJ...848L..13A, 2017ApJ...848L..12A}. 
However, details of the kilonova transients are still largely unclear.

Neutrino transport and neutrino-matter interactions play import roles in the mass ejection process of neutron star mergers, and hence the observables of kilonovae.
Specifically, neutrino microphysics could significantly affect the composition, mass, and velocities of the merger outflows.
Consequently, the ejecta properties dominate the brightness, duration, and colours of the kilonovae~\citep{2013ApJ...775...18B}.
Comprehensive neutrino microphysics inputs are necessary to better model the matter outflows and hence observational signatures of kilonovae.

Energy-dependent neutrino transport together with species/energy groups coupled neutrino-matter interactions is essential for accurate kilonovae modellings.
Most of the neutron star merger simulations adopted energy-integrated neutrino transport schemes with neutrino interactions that have no species and energy coupling (e.g. \cite{2014ApJ...789L..39W, 2015PhRvD..91f4059S, 2015PhRvD..91l4021F, 2016PhRvD..94l3016F, 2018MNRAS.475.4186F, 2022MNRAS.512.1499R, 2022PhRvD.105j4028S, 2024PhRvD.109d4012S, 2024PhRvD.109d3044I, 2024MNRAS.528.5952M}).
Despite its success in neutron star merger simulations, these simulations have two major limitations:
(i) spectral information of neutrinos cannot be captured, and
(ii) neutrino-matter interactions that require detailed spectral information, such as neutrino-lepton inelastic scatterings and pair processes, cannot be included naturally.

The importance of resolving the spectral information of neutrinos has been shown recently in simulations with Monte-Carlo \cite{2020ApJ...902L..27F, 2024arXiv240715989F} and two-moment \cite{2024arXiv240716017C} neutrino transports.
In these works, the authors have shown that estimating the neutrino distributions based on \cite{2016PhRvD..94l3016F} in energy-integrated transports could underestimate the neutrino average energies and luminosities, as well as the ejected mass.
Therefore, to accurately model the potential observational signature of neutron star merger, multi-group transport is highly desired.

So far, neutrino-matter interactions coupling different species/energy groups have not been directly included in general-relativistic neutron star merger simulations, except for approximate treatment of neutrino-antineutrino annihilation~\citep{2015ApJ...813...38R, 2017JPhG...44h4007P, 2017ApJ...846..114F, 2016ApJ...816L..30J} or in the Newtonian framework~\citep{2023ApJ...951L..12J} (but see a study with advanced neutrino treatment in the context of collapsars \cite{2020ApJ...902...66M}).
Although the effective opacity of interactions beyond beta processes and elastic scatterings (e.g. neutrino-electron inelastic scattering) are expected to be secondary in the merger context \cite{2023LRCA....9....1F, 2024ApJS..272....9N}, their impacts on hypermassive neutron star in dynamical simulations remain unknown. 
Some of these interactions involve neutrino-matter energy exchange, which can be important for disc and ejecta modellings.

Inelastic neutrino-electron scattering is known to be very important during the deleptonisation of the iron core in the context of core-collapse supernovae \cite{1993ApJ...410..740M}.
This interaction is very effective in thermally equilibrating neutrino and matter.
In the context of core-collapse supernovae, this interaction is shown to be important when the density reaches $5\times 10^{11}~\rm{g~cm^{-3}}$, which enhances the deleptonisation process and neutrino luminosities during the infall phase.
Despite the importance, it has not been applied in the context of hypermassive neutron star simulations.

Electron neutrino-pair production/annihilation has been proposed to power relativistic jet.
As a result, studies have attempted to include this interaction and see if it could result in relativistic outflows and thus explain the short gamma-ray bursts from neutron star mergers.
However, this mechanism alone appears unable to overcome the baryon pollution issue \cite{2016ApJ...816L..30J}, and the corresponding heating rates have been shown to be too low in various studies \cite{2015ApJ...813...38R, 2017JPhG...44h4007P, 2017ApJ...846..114F} (but see a recent counter-example~\cite{2024arXiv241002380K}).

The main goal of this work is to better understand the importance of the inelastic neutrino interactions in the evolution of post-merger hypermassive neutron stars.
To this end, we perform energy-dependent two-moment neutrino transport general-relativistic magnetohydrodynamics simulations of a post-merger-like hypermassive neutron star.
To investigate the influence of neutrino microphysics, we consider 3 sets of neutrino microphysics.
Specifically, we focus on the impact of the neutrino-electron inelastic scattering and thermal neutrino-pair production/annihilations.

The paper is organised as follows.
In section~\ref{sec:methods} we outline the methods we used in this work.
The results are presented in section~\ref{sec:results}.
This paper ends with a discussion in section~\ref{sec:discussion}.

\section{\label{sec:methods}Methods}
The initial profile and simulation setups are identical to our previous work~\cite{2024arXiv240716017C}, except for the input neutrino microphysics.
Here, we highlight the essential ingredients.

A post-merger-like quasi-equilibrium hypermassive neutron star with central energy density $\epsilon_{c}/c^2 = 1.2604\times10^{15}~\rm{g \cdot cm^{-3}}$ and angular momentum $J = 5~G M_{\odot}^2 / c$, constructed in \cite{2024PhRvD.110d3015C}, is set as the initial profile. 
The gravitational mass of this star is $2.77~{\rm M_{\odot}}$.
This equilibrium model is constructed with the ``DD2'' equation-of-state~\citep{2010NuPhA.837..210H} with a constant entropy-per-baryon $s = 1 ~ k_{\rm{B}} / \text{baryon}$ and in neutrinoless $\beta$-equilibrium.
The star is differentially rotating, following the 4-parameter rotation law of \cite{2019PhRvD.100l3019U} with angular velocity ratios $\left\{{\Omega_{\max}}/{\Omega_{\rm c}}=1.6,{\Omega_{\rm eq}}/{\Omega_{\rm c}}=1\right\}$, where $\Omega_{\max}$, $\Omega_{c}$, and $\Omega_{\rm eq}$ are the maximum, central, and equatorial angular velocities of the neutron star, respectively.
Magnetic fields are superimposed via the vector potential in orthonormal form in spherical coordinates: 
\begin{equation}
	\left(A^{\hat{r}}, A^{\hat{\theta}}, A^{\hat{\phi}}\right) = \frac{r_0^3 }{2\left(r^3+r_0^3\right)}\left(0, 0, B_{\rm pol} r \sin\theta \right),
\end{equation}
where we set $r_0 = 10~{\rm km}$ and $B_{\rm pol} = 10^{15}~{\rm G}$.

The general-relativistic neutrino magnetohydrodynamics code \texttt{Gmunu}~\citep{2020CQGra..37n5015C, 2021MNRAS.508.2279C, 2022ApJS..261...22C, 2023ApJS..267...38C, 2024ApJS..272....9N} is used to perform our simulations.
The code solves the general-relativistic magnetohydrodynamics equations via staggered-meshed constrained transport~\citep{1988ApJ...332..659E}.  
The Einstein field equations are solved in the conformally flat approximation.
Axisymmetric simulations are performed in cylindrical coordinates $(R, z)$, where the computational domain covers $0 \leq R \leq 2000~{\rm km}$ and $0 \leq z \leq 2000~{\rm km}$, with the resolution {$n_R \times n_z = 128 \times 128$} and allowing 6 adaptive-mesh-refinement levels on top of that.
The finest grid size at the centre of the star is $\Delta R = \Delta z \approx 488 ~\rm{m}$, while the grid size elsewhere is adaptively changed based on the gradient of the logarithmic rest mass density.
The finite temperature equation-of-state ``DD2''~\citep{2010NuPhA.837..210H} is used for evolutions.
The Harten, Lax and van Leer (HLL) approximated Riemann solver~\citep{harten1983upstream}, the piecewise parabolic method (PPM)~\citep{1984JCoPh..54..174C}, and the implicit-explicit time integrator IMEXCB3a~\citep{2015JCoPh.286..172C} are adopted in the simulations. 

We use a two-moment \emph{multi-group} neutrino transport scheme~\citep{2023ApJS..267...38C} and the maximum-entropy closure to set the second moment of the neutrino distribution function (pressure tensor)~\citep{1978JQSRT..20..541M}.
We use a range of implicit treatments described in \cite{2023ApJS..267...38C}, depending on the input neutrino interactions.
Specifically, if the neutrino interactions set considered has no species or groups couplings, we use the single-specie single-group (SSSG) implicit solver to limit computational cost.
Similarly, if the interactions set has groups couplings, we use the single-specie multi-groups (SSMG) implicit treatment.
Finally, we use the multi-species multi-groups (MSMG) implicit solver when the considered set has all species and groups couplings. 
We refer readers to \cite{2023ApJS..267...38C, 2024ApJS..272....9N} for the details of the expression of radiation four-force for neutrinos, as well as the implementations of the neutrino transport.

We list the neutrino microphysics considered here in table~\ref{tab:nu_int}.
We consider three species of neutrinos: the electron type (anti)neutrinos $\nu_e$ and $\bar{\nu}_e$, and heavy-lepton (anti)neutrinos $\nu_x$, where the muon and tau neutrinos (i.e. $\nu_{\mu}, \bar{\nu}_{\mu}, \nu_{\tau}$, and $\bar{\nu}_{\tau}$) are grouped into $\nu_x$.
The tabulated interaction rates and kernels are generated via \texttt{NuLib}~\citep{2015ApJS..219...24O}, and are tabulated during the simulations.
The neutrino opacity and emissivity table is logarithmically spaced in rest-mass density $\rho$ (82 points in $[10^6,3.2\times 10^{15}]\,{\rm g/cm^3}$) and fluid temperature $T$ (65 points in $[0.05,150]\,{\rm MeV}$), and linearly spaced in the electron fraction $Y_e$ (51 points in $[0.01,0.6]$).  
In addition, the table of the kernels of neutrino-electron inelastic scattering and neutrino-pair production/annihilation is logarithmically spaced in fluid temperature $T$ (65 points in $[0.05,150]\,{\rm MeV}$), and electron degeneracy parameter $\eta_e \equiv \mu_e / T$ (61 points in $[0.1,100]$), where $\mu_e$ is the electron chemical potential.  
The neutrino energies $\varepsilon$ is logarithmically spaced in 16 groups up to $528\,{\rm MeV}$ in both tables.
\begin{table*}[]
\centering
\resizebox{\textwidth}{!}{%
\begin{tabular}{|ccccllc|}
\hline
\multicolumn{3}{|c|}{Implicit solvers}                                                                                                                                                                          & \multicolumn{1}{c|}{Interactions}                                     & \multicolumn{1}{c|}{Type}               & \multicolumn{1}{c|}{Reference}                                       & labels                \\ \hline
\multicolumn{1}{|c|}{{\color[HTML]{228833} }}                        & \multicolumn{1}{c|}{{\color[HTML]{CCBB44} }}                        & \multicolumn{1}{c|}{{\color[HTML]{4477AA} }}                       & $\nu_e+n \leftrightarrow p+e^{-}$                                     & Beta process                            & \cite{2002PhRvD..65d3001H, 2006NuPhA.777..356B}                      & (a)                   \\
\multicolumn{1}{|c|}{{\color[HTML]{228833} }}                        & \multicolumn{1}{c|}{{\color[HTML]{CCBB44} }}                        & \multicolumn{1}{c|}{{\color[HTML]{4477AA} }}                       & $\bar{\nu}_e+p \leftrightarrow n+e^{+}$                               & Beta process                            & \cite{2006NuPhA.777..356B}                                           & (b)                   \\
\multicolumn{1}{|c|}{{\color[HTML]{228833} }}                        & \multicolumn{1}{c|}{{\color[HTML]{CCBB44} }}                        & \multicolumn{1}{c|}{{\color[HTML]{4477AA} }}                       & $\nu_{e}+(A,Z-1) \leftrightarrow (A,Z)+e^-$                           & Beta process                            & \cite{1985ApJS...58..771B}                                           & (c)                   \\
\multicolumn{1}{|c|}{{\color[HTML]{228833} }}                        & \multicolumn{1}{c|}{{\color[HTML]{CCBB44} }}                        & \multicolumn{1}{c|}{{\color[HTML]{4477AA} }}                       & $\nu+N \leftrightarrow \nu+N$                                         & Elastic scattering                      & \cite{2002PhRvD..65d3001H, 2006NuPhA.777..356B}                      & (d)                   \\
\multicolumn{1}{|c|}{{\color[HTML]{228833} }}                        & \multicolumn{1}{c|}{{\color[HTML]{CCBB44} }}                        & \multicolumn{1}{c|}{{\color[HTML]{4477AA} }}                       & $\nu+(A,Z) \leftrightarrow \nu+(A,Z)$                                 & Elastic scattering                      & \cite{1997PhRvD..55.4577H, 1985ApJS...58..771B, 2006NuPhA.777..356B} & (e)                   \\
\multicolumn{1}{|c|}{{\color[HTML]{228833} }}                        & \multicolumn{1}{c|}{{\color[HTML]{CCBB44} }}                        & \multicolumn{1}{c|}{{\color[HTML]{4477AA} }}                       & $\nu+\alpha \leftrightarrow \nu+\alpha$                               & Elastic scattering                      & \cite{1985ApJS...58..771B, 2006NuPhA.777..356B}                      & (f)                   \\
\multicolumn{1}{|c|}{{\color[HTML]{228833} }}                        & \multicolumn{1}{c|}{{\color[HTML]{CCBB44} }}                        & \multicolumn{1}{c|}{{\color[HTML]{4477AA} }}                       & $N+N \leftrightarrow N+N+\nu_{x}+\bar{\nu}_{x}$                       & Neutrino-pair process                   & \cite{2006NuPhA.777..356B}                                           & (g)                   \\ \cline{4-7} 
\multicolumn{1}{|c|}{{\color[HTML]{228833} }}                        & \multicolumn{1}{c|}{{\color[HTML]{CCBB44} }}                        & \multicolumn{1}{c|}{\multirow{-8}{*}{{\color[HTML]{4477AA} SSSG}}} &                                                                       &                                         & SSSG: \cite{2006NuPhA.777..356B}                                     &                       \\ \cline{3-3}
\multicolumn{1}{|c|}{{\color[HTML]{228833} }}                        & \multicolumn{1}{c|}{{\color[HTML]{CCBB44} }}                        & \multicolumn{1}{c|}{}                                              & \multirow{-2}{*}{$e^{-}+e^{+} \leftrightarrow \nu_{x}+\bar{\nu}_{x}$} & \multirow{-2}{*}{Neutrino-pair process} & SSMG/MSMG: \cite{1985ApJS...58..771B}                                & \multirow{-2}{*}{(h)} \\ \cline{4-7} 
\multicolumn{1}{|c|}{{\color[HTML]{228833} }}                        & \multicolumn{1}{c|}{\multirow{-10}{*}{{\color[HTML]{CCBB44} SSMG}}} &                                                                    & $\nu+e^{-}  \leftrightarrow \nu+e^{-}$                                & Inelastic scattering                    & \cite{1985ApJS...58..771B}                                           & (i)                   \\ \cline{2-7} 
\multicolumn{1}{|c|}{\multirow{-11}{*}{{\color[HTML]{228833} MSMG}}} &                                                                     &                                                                    & $e^{-}+e^{+} \leftrightarrow \nu_e+\bar{\nu}_e$                       & Neutrino-pair process                   & \cite{1985ApJS...58..771B}                                           & (j)                   \\ \hline
\end{tabular}%
}
\caption{	
	Table of different implicit treatments and the corresponding included neutrino interactions.
	The single-specie single-group (SSSG) implicit solver is used when there is no species or groups couplings.
	Similarly, the single-specie multi-groups (SSMG) or the multi-species multi-groups (MSMG) implicit treatment is used when there are energy-groups or both species and energy-groups couplings, respectively.
	We denote the electron, anti-electron and heavy-lepton neutrino as $\nu_e$, $\bar{\nu}_e$ and $\nu_x$, respectively.
	$\nu$ represents all three species of neutrino.
	Interactions that involve a specific type of neutrino are expressed explicitly.
	$(A,Z)$ represents a heavy nucleus with a mass number of $A$ and a proton number of $Z$, without including $\alpha$ particle.
	In \texttt{NuLib}, the nucleon-nucleon Bremsstrahlung process (g) is treated approximately as effective emissivity/absorption opacity so the SSSG implicit solver is used.
	Since \texttt{NuLib} does not provide the kernels of process (g), detailed study of the impact of such approximation is left as a future work.
	On the other hand, the neutrino-pair process (h) is treated approximately as effective emissivity/absorption opacity when the SSSG implicit solver is used, while it is described as the full production/annihilation kernels when the SSMG/MSMG solver is used.
	A comparison of this approximation can be found at appendix~\ref{sec:epa_approx}.
	In this work, we consider 3 neutrino microphysics sets, which correspond to SSSG, SSMG, and MSMG sets.
	The tabulated interaction rates and kernels are generated via \texttt{NuLib}~\citep{2015ApJS..219...24O}, and are tabulated during the simulations.}
\label{tab:nu_int}
\end{table*}

\section{\label{sec:results}Results}
\begin{figure}
	\centering
	\includegraphics[width=\columnwidth, angle=0]{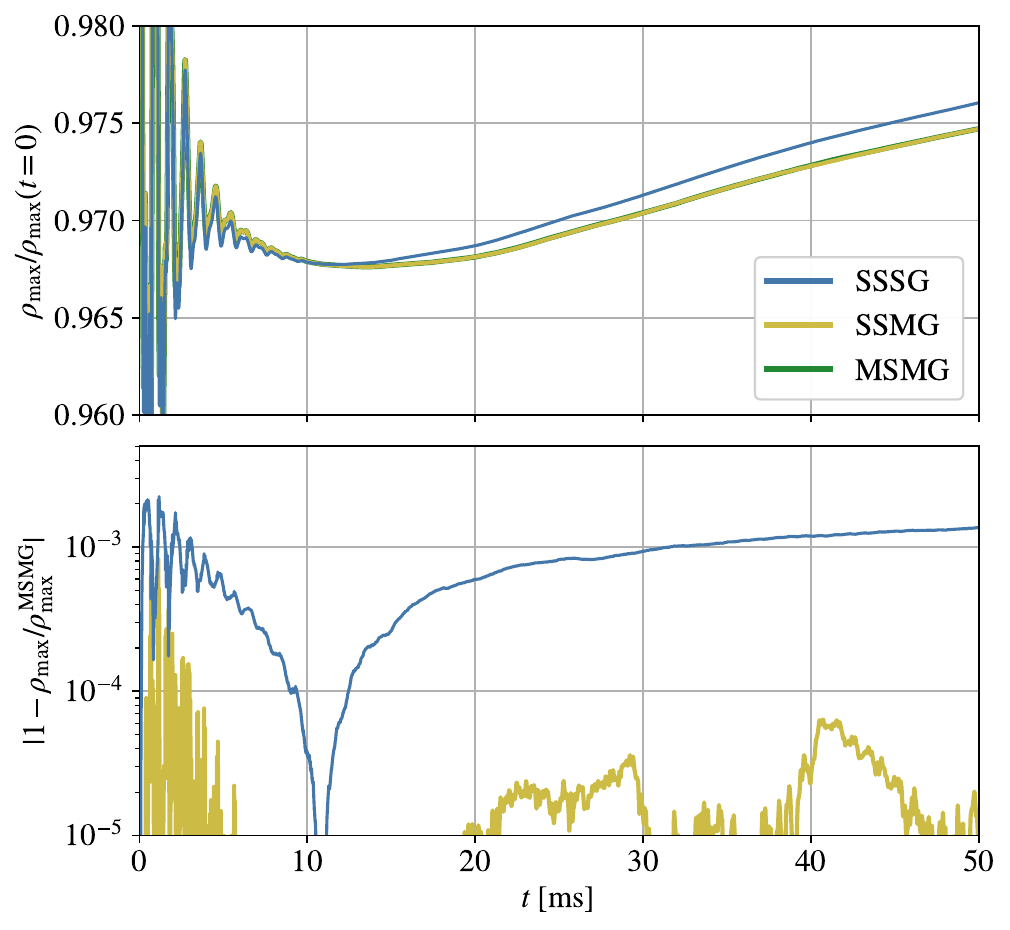}
	\caption{
		Evolutions of the maximum value of the rest mass density rescaled with its initial value $\rho_{\max}/\rho_{\max}\left(t=0\right)$ (\emph{top panel}) and their relative error with respect to the fully-coupled (multi-species multi-group) cases (\emph{bottom panel}).
		In the high density regions, the impact due to inelastic neutrino-electron scattering ($\nu+e^{-} \leftrightarrow \nu+e^{-}$) is stronger than the electron neutrino-pair production via electron-positron annihilation ($e^{-}+e^{+} \leftrightarrow \nu_e+\bar{\nu}_e$). 
		Overall, the maximum rest-mass density is affected $\lesssim \mathcal{O}\left( 0.1\% \right) $ via these interactions.
		}
	\label{fig:ns_rho_max}
\end{figure}

The magnetohydrodynamical evolutions of the regions where the rest-mass density is above neutrino trapping density (i.e. for $\rho \gtrsim \rho_{\rm trap} \approx 10^{12.5}~{\rm g/cm^3}$) are very similar in all cases.
Figure~\ref{fig:ns_rho_max} shows the time evolutions of the maximum value of the rest-mass density rescaled with its initial value $\rho_{\max}/\rho_{\max}\left(t=0\right)$ and their relative error with respect to the fully-coupled (multi-species multi-group) cases.
Inelastic neutrino-electron scattering ($\nu+e^{-} \leftrightarrow \nu+e^{-}$) causes modest changes to the maximum density evolutions: the maximum rest-mass density is affected at the 0.1\% when ignoring these interactions.
The impacts of electron neutrino-pair production via electron-positron annihilation ($e^{-}+e^{+} \leftrightarrow \nu_e+\bar{\nu}_e$) are minimal.

\begin{figure*}
	\centering
	\includegraphics[width=\textwidth, angle=0]{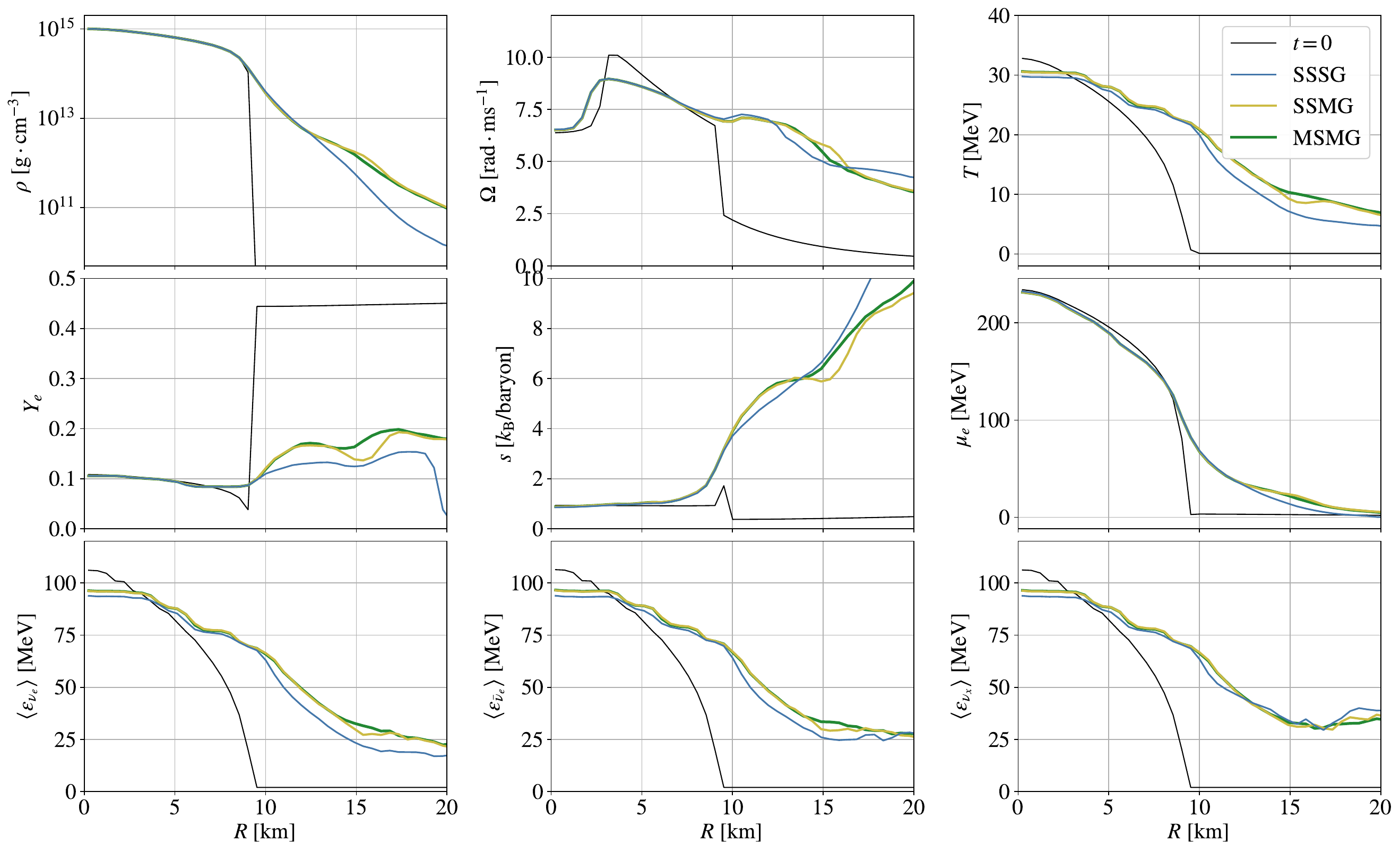}
	\caption{
		Comparison of the radial profiles of several quantities at $t=50~{\rm ms}$.
		For instance, we compare rest mass density $\rho$, angular velocity $\Omega$, matter temperature $T$, electron fraction $Y_e$, matter entropy per baryon $s$, electron chemical potential $\mu_e$, and neutrino averaged energies $\langle\epsilon_{\nu_i}\rangle$.
		Noticeable difference among different neutrino sets present when the rest mass density is below the neutrino trapping density (i.e. when $\rho \lesssim 10^{12.5}~{\rm g \cdot cm^{-3}}$).
		In the cases with inelastic neutrino-electron scattering, the lower density matter is more spread out, with higher temperature.
		The angular velocity, however, is not changed significantly.
		Inclusion of electron neutrino-pair production via electron-positron annihilation do not impact the overall evolution much in the shown regions.
		}
	\label{fig:ns_1d_profiles}
\end{figure*}

Despite all simulations behaving similarly in high density regions, differences can be seen when the rest mass density is lower than the neutrino trapping density.
Figure~\ref{fig:ns_1d_profiles} compares the radial profiles along the $R$-axis (i.e. $z=0$, at the equatorial plane) at the end of the simulations for all cases.
When inelastic neutrino-electron scattering is included, the materials that are below the neutrino trapping density are hotter and more spread out.
However, the angular velocity remains the same with or without this scattering.
Whether we include electron neutrino-pair production via electron-positron annihilation or not does not noticeably impact the evolution of the regions shown here.
This is because this interaction is expected to be important only in the low-density and low-temperature regions. 

\begin{figure*}
	\centering
	\includegraphics[width=\textwidth, angle=0]{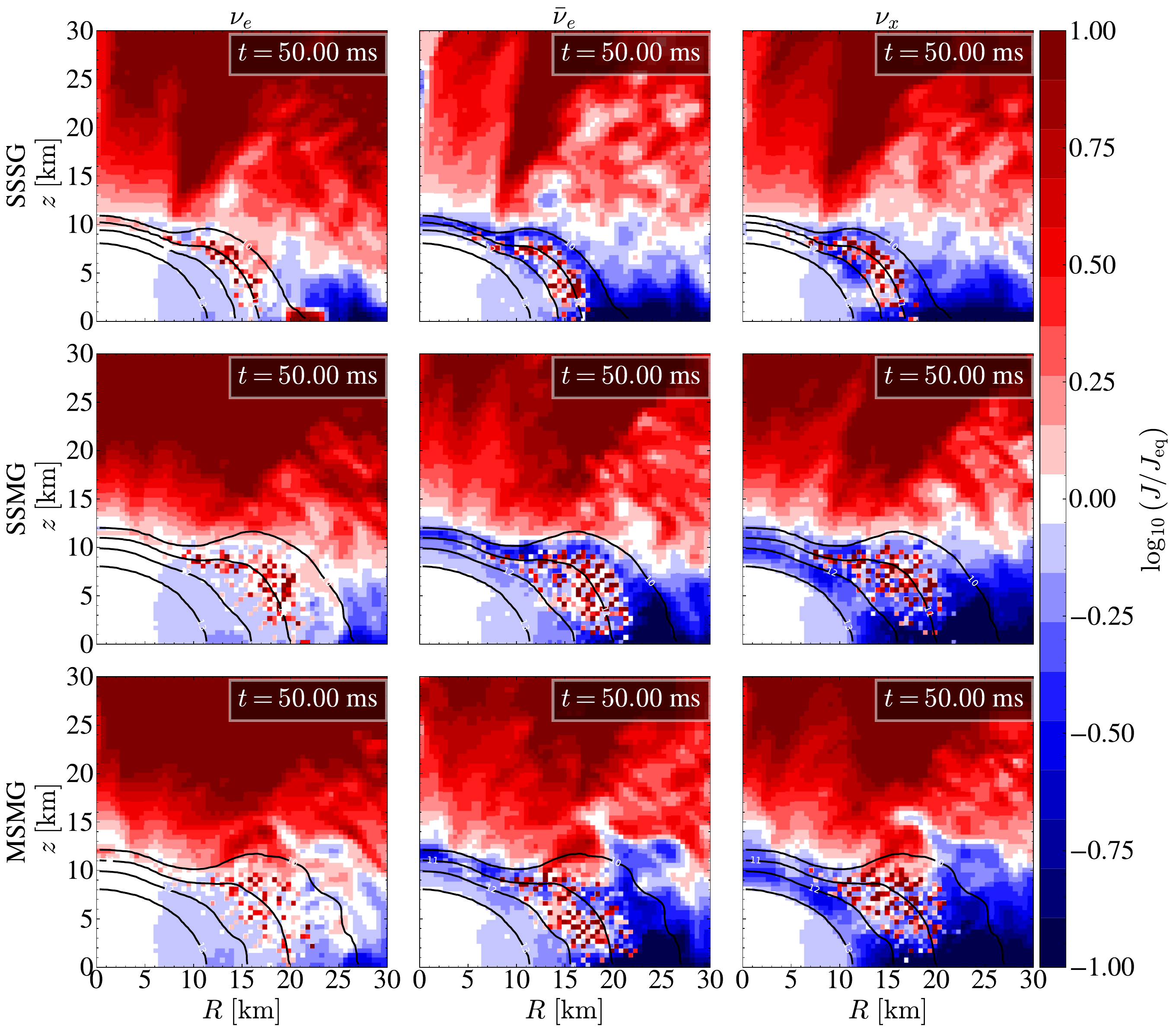}
	\caption{
		Ratios of the fluid frame neutrino energy density $J$ and the corresponding equilibrium energy $J_{\rm eq}$ at the end of the simulations.
		Black contours correspond to densities of $10^{13, 12, 11, 10}~{\rm g \cdot cm^{-3}}$.
		As references, the neutrino trapping density is $\gtrsim 10^{12.5}~{\rm g \cdot cm^{-3}}$ while the inelastic neutrino-electron scattering is strongest when $\rho \gtrsim 5\times{10^{11}}~{\rm g \cdot cm^{-3}}$.
		Irradiated regions ($J > J_{\rm eq}$) are shown in red, while the semi-transparent regions ($J < J_{\rm eq}$) are shown in blue.
		As shown at the \emph{middle} and \emph{right columns}, the $10^{13 - 11}~{\rm g \cdot cm^{-3}}$ regions turn into semi-transparent for $\bar{\nu}_e$ and $\nu_x$ in a more efficient way when the inelastic neutrino-electron scattering is included.
		However, no notable difference is found when the electron neutrino-pair production via electron-positron annihilation is included.
		}
	\label{fig:J_over_Jeq}
\end{figure*}

The influence due to inelastic neutrino-electron scattering can be visualised by comparing the neutrino energy densities in the fluid frame.
Figure~\ref{fig:J_over_Jeq} shows the ratios of the neutrino energy density $J$ (in fluid frame) and its value if the neutrinos were in equilibrium with the fluid $J_{\rm eq}$.
This plot enables us to identify where the fluid is being irradiated by neutrinos ($J > J_{\rm eq}$), where neutrinos are in equilibrium with the fluid ($J \approx J_{\rm eq}$), as well as semi-transparent ($J < J_{\rm eq}$) regions.
As shown in the figure, inelastic neutrino-electron scattering results in more transparent regions for $\bar{\nu}_e$ and $\nu_x$ in the regions where the rest mass density is between the neutrino trapping density and the inelastic neutrino-electron scattering starts to be important (i.e. when $10^{12.5}~{\rm g \cdot cm^{-3}} \lesssim \rho \lesssim 5 \times 10^{11}~{\rm g \cdot cm^{-3}}$).
However, $\nu_e$ is not affected as much as $\bar{\nu}_e$ and $\nu_x$.

\begin{figure}
	\centering
	\includegraphics[width=\columnwidth, angle=0]{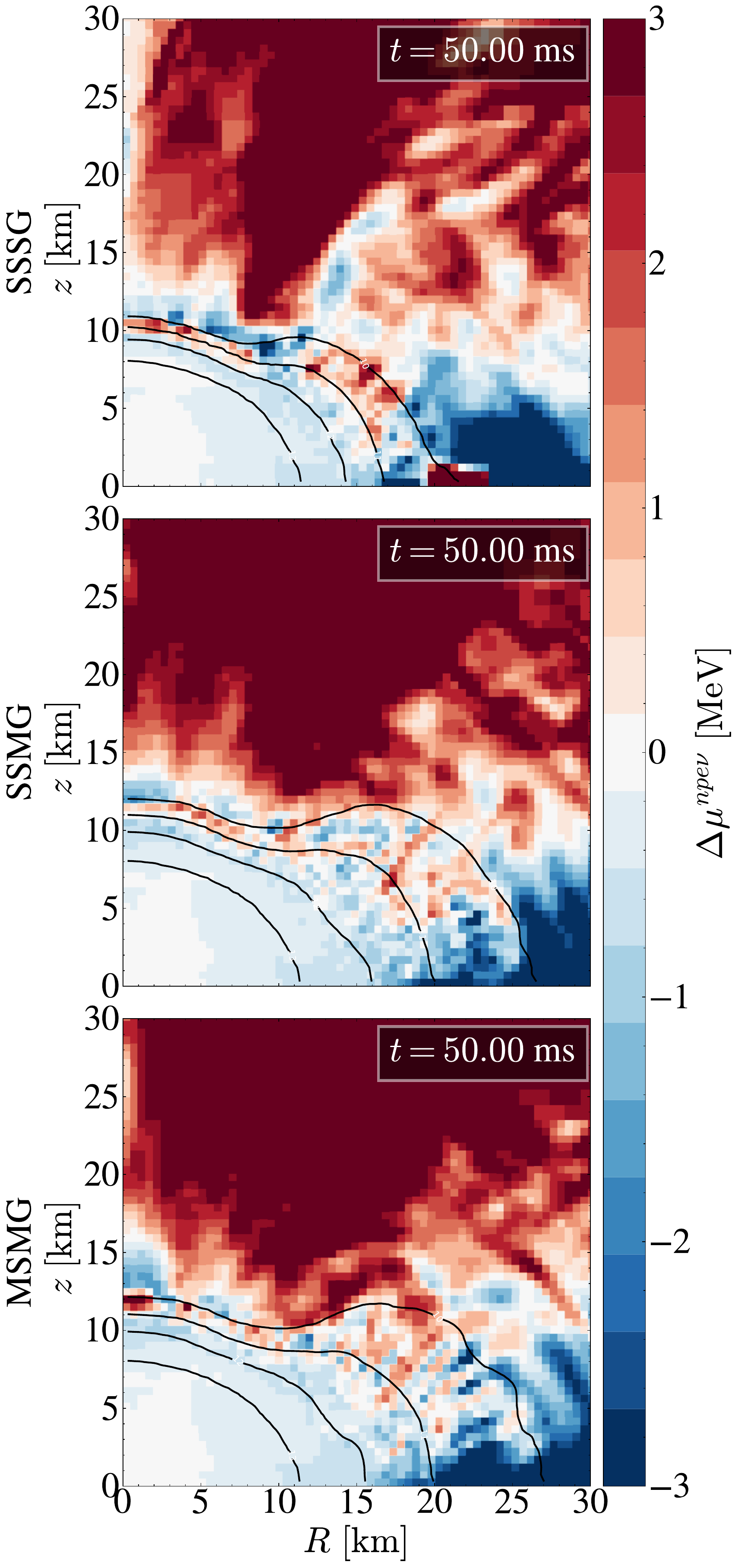}
	\caption{
		Out-of-equilibrium chemical potential $\Delta \mu^{npe\nu}$ at the end of the simulations.
		While inelastic neutrino-electron scattering alter the evolution mostly via energy exchange, it does not affect the weak-beta equilibrium.
		This is expected, as this is dominated mostly by the beta-processes, which are dominating when the density beyond neutrino trapping density.
		}
	\label{fig:delta_mu_nu}
\end{figure}

Although inelastic neutrino-electron scatterings noticeably alter the energy equilibration condition, the weak-beta equilibrium is still dominated by beta processes.
To better present the weak-beta equilibrium condition, we compute the chemical imbalance by following \cite{2021PhRvD.104j3006H, 2023MNRAS.525.6359L, 2024PhRvL.132u1001E}:
\begin{align}\label{eq:delta_mu}
	\Delta \mu^{npe\nu} &= \mu^{\rm eq}_{{\nu}_e} - \mu_{{\nu}_e},
\end{align}
where $\mu_{{\nu}_e} = \mu_e - \left( \mu_n - \mu_p \right) $ is the chemical potential of electron neutrino $\nu_e$, which can be obtained directly from the equation-of-state table. 
The chemical potential of $\nu_e$ in equilibrium (i.e. $\mu^{\rm eq}_{{\nu}_e}$) is obtained by numerically solving \cite{2019EPJA...55..124P}
\begin{align}\label{eq:Ynue}
	Y_{\nu_e} = \frac{4 \pi m_{\rm u}}{\rho \left(hc\right)^3}F_2\left(\frac{\mu^{\rm eq}_{\nu_e}}{k_{\rm B}T}\right) \left(k_{\rm B}T\right)^3 ,
\end{align} 
where $m_{\rm{u}}$ is the atomic mass unit, $F_2\left(\frac{\mu_\nu}{k_{\rm B}T}\right)$ is the $2$-th order complete Fermi-Dirac integral of the degeneracy parameter $\frac{\mu_\nu}{k_{\rm B}T}$, and
\begin{align}
	Y_{\nu_e} = \frac{m_{\rm u}}{W \rho} \int_0^\infty \frac{W \mathcal{J} + \alpha \mathcal{H}^t}{\varepsilon} \dd{V_\varepsilon}
\end{align} 
is the $\nu_e$ fraction extracted from simulations~\cite{2016ApJS..222...20K}.
Figure~\ref{fig:delta_mu_nu} shows the out-of-equilibrium chemical potential $\Delta \mu^{npe\nu}$ at the end of the simulations for all cases.
As shown in the figure, the weak-beta equilibrium is satisfied in the regions where $\rho \gtrsim 10^{12}~{\rm g \cdot cm^{-3}}$, where the beta processes are the dominant neutrino-matter interactions. 

It is important to note that equation~\eqref{eq:Ynue}, which is used to estimate the chemical potential in equilibrium $\mu^{\rm eq}_{{\nu}_e}$, assumes that neutrinos are in equilibrium with the fluid. 
However, this assumption may not hold in the outer regions (i.e., $\rho \lesssim 10^{11}~{\rm g \cdot cm^{-3}}$), where neutrinos are no longer in equilibrium. 
As a result, the out-of-equilibrium chemical potential $\Delta \mu^{npe\nu}$, as defined in equation~\eqref{eq:delta_mu}, may not serve as an accurate indicator in these decoupled regions.

\begin{figure}
	\centering
	\includegraphics[width=\columnwidth, angle=0]{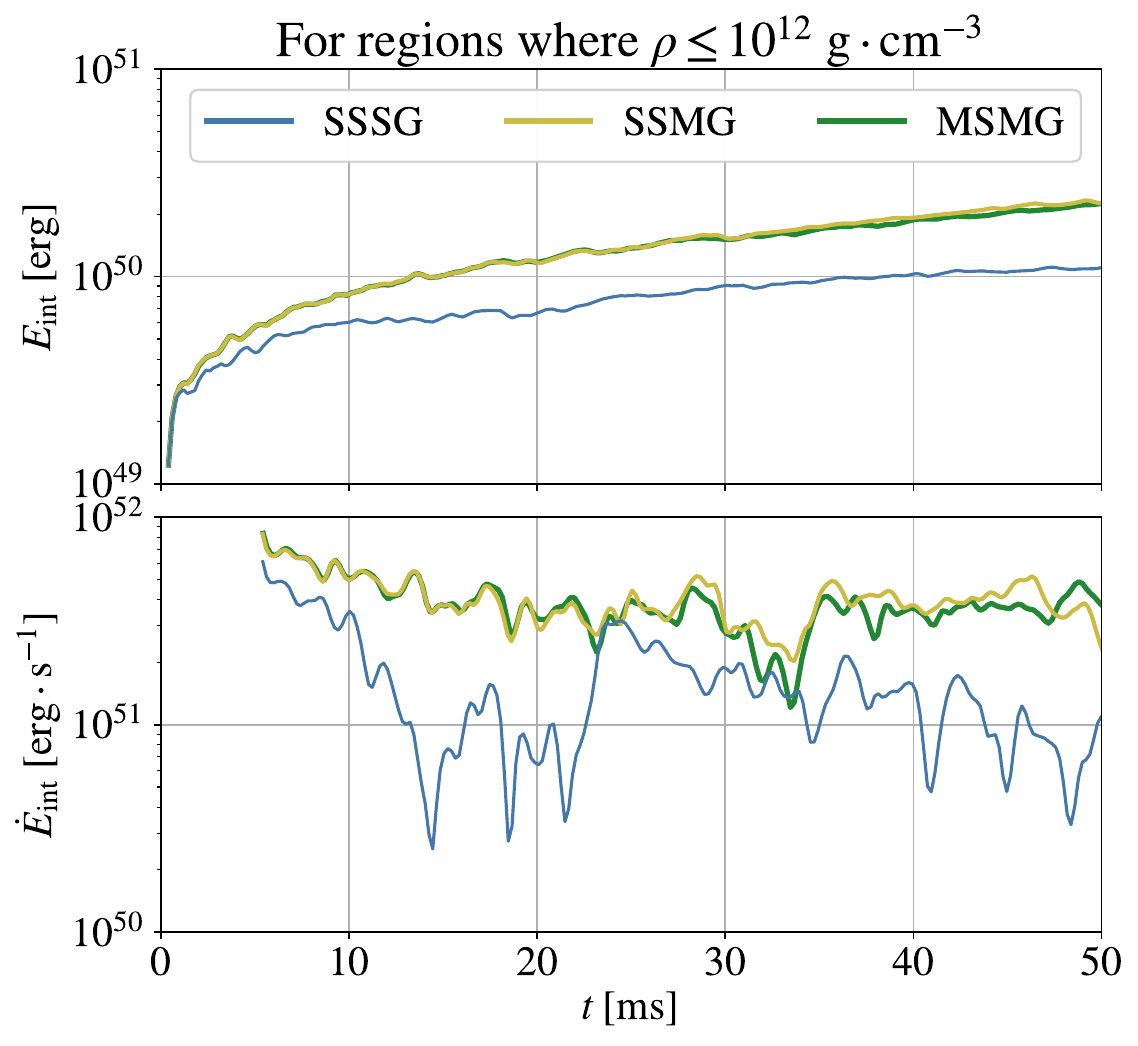}
	\caption{
		The time evolutions of internal energy (\emph{top panel}) and its rate of change (\emph{bottom panel}) for the matter where the rest-mass density is below $10^{12}~{\rm g \cdot cm^{-3}}$, using different neutrino interaction sets for a post-merger-like hypermassive neutron star. 
		High-density regions (i.e., for $\rho > 10^{12}~{\rm g \cdot cm^{-3}}$) are excluded for better visualization, as they are almost identical in all cases and contribute the majority of the internal energy.
		}
	\label{fig:matter_eint}
\end{figure}

Simulations that include inelastic scattering result in higher energy in the matter surrounding the neutron star. 
In Figure~\ref{fig:matter_eint}, we plot the time evolutions of the internal energy and the heating/cooling rates of the matter in regions where the rest-mass density is below $10^{12}~{\rm g \cdot cm^{-3}}$. 
Since high-density regions account for most of the internal energy of the system and behave similarly across all cases considered here, we exclude these regions from the plot for better clarity.
As shown in Figure~\ref{fig:matter_eint}, both the heating rate and the internal energy of the matter are higher throughout the simulation when inelastic neutrino-electron scattering is included. 
However, the impact of electron neutrino-pair production via electron-positron annihilation is secondary compared to that of inelastic neutrino-electron scattering.

\begin{figure*}
	\centering
	\includegraphics[width=\textwidth, angle=0]{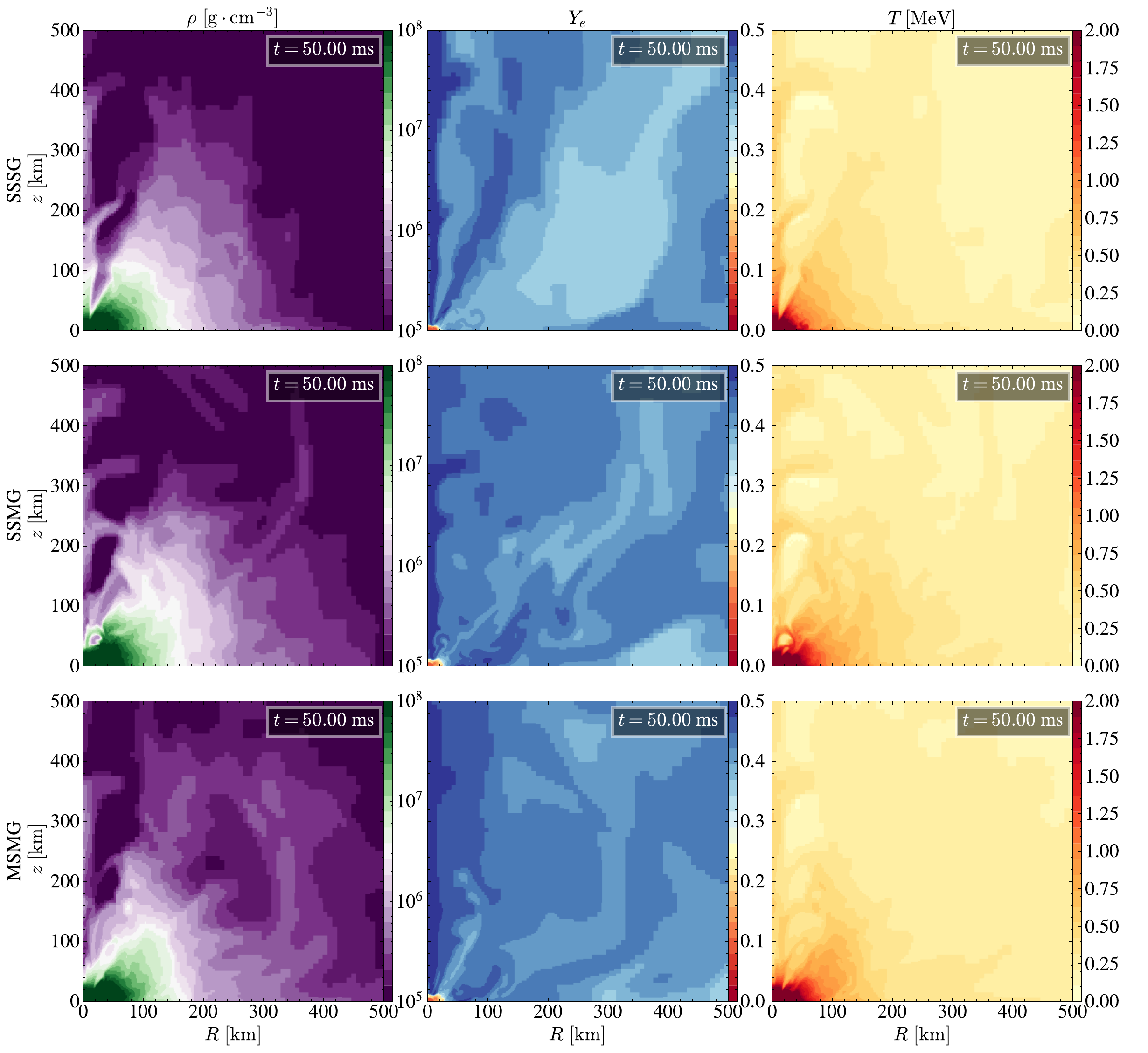}
	\caption{
		Profiles of rest-mass density $\rho$ (\emph{left column}), electron fraction $Y_e$ (\emph{middle column}) and temperature $T$ (\emph{right column}) with different neutrino microphysics sets (\emph{top to down rows}) at $t=50~{\rm ms}$ of a post-merger-like hypermassive neutron star.
		At later times, the magnetic winding is mostly saturated, and a quasi-equilibrium neutron star with disc system has formed.
		The density, temperature, and composition of the disc depends on the neutrino microphysics.
		The disc is slightly denser and more neutron-poor when the inelastic neutrino-electron scattering is included.
		}
	\label{fig:all_hydro_r500}
\end{figure*}

\begin{figure}
	\centering
	\includegraphics[width=\columnwidth, angle=0]{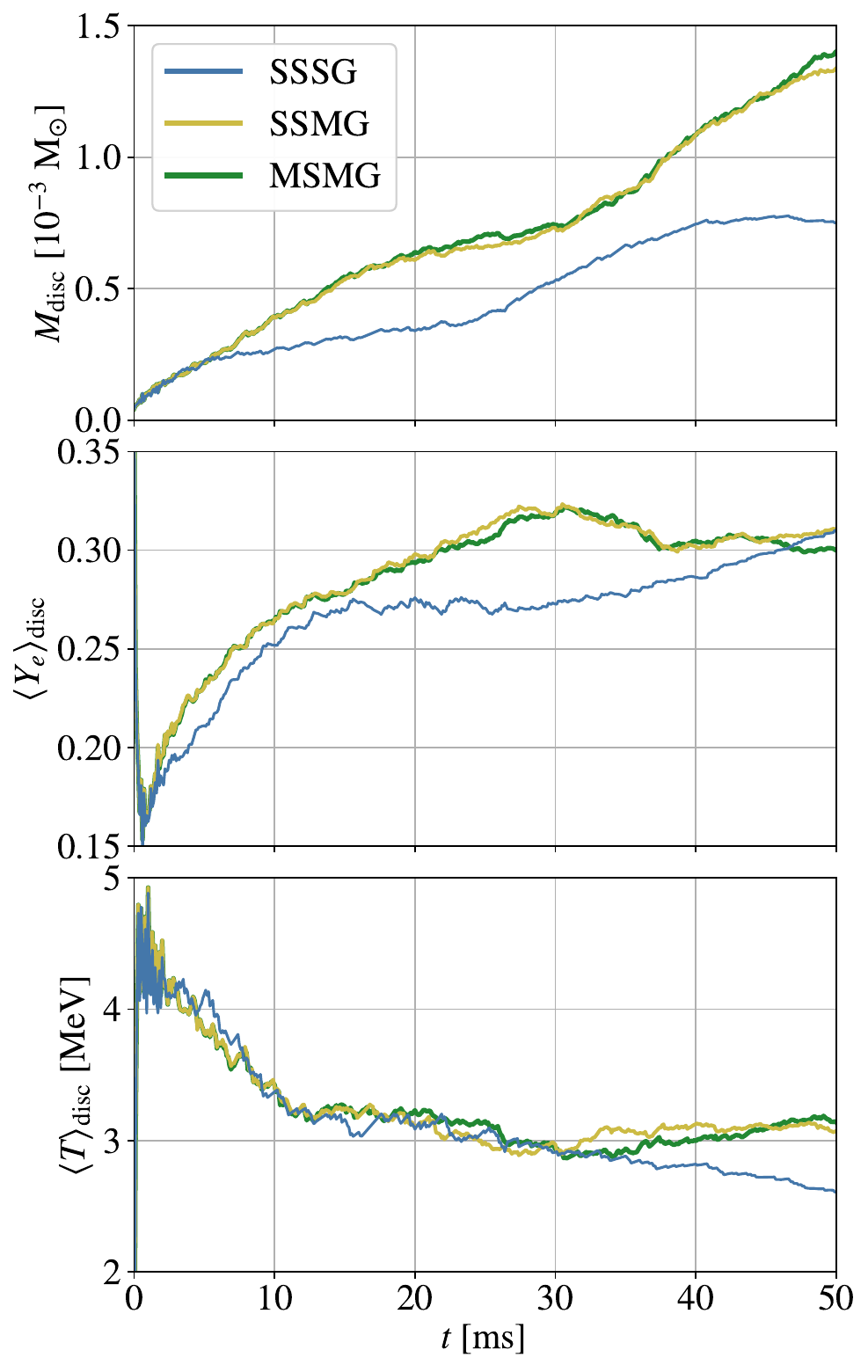}
	\caption{
		The time evolutions of rest-mass (\emph{top panel}), mass-averaged electron fraction (\emph{middle panel}) and temperature (\emph{bottom panel}) of the disc (where the rest-mass density below $10^{11}~{\rm g \cdot cm^{-3}}$) with different neutrino interaction sets of a post-merger-like hypermassive neutron star.
		The impact due to inelastic neutrino-electron scattering ($\nu+e^{-} \leftrightarrow \nu+e^{-}$) is significant, where the disc mass is $\sim 75\%$ larger with this process at the end of the simulations.
		Also, the disc is slightly more neutron-poor in these cases.
		The impact due to electron neutrino-pair production via electron-positron annihilation is secondary compared to inelastic neutrino-electron scatterings.
		Interestingly, the averaged temperature is not sensitive to all the neutrino interaction sets considered in this work.
		}
	\label{fig:ns_disc_avg}
\end{figure}

The discs surrounding the neutron star are found to be more massive in simulations with inelastic scattering, as shown in figure~\ref{fig:all_hydro_r500}.
To quantify the key properties of the disc, in figure~\ref{fig:ns_disc_avg} we plot the disc mass and the density-weighted averaged electron fraction and temperature.
When the inelastic neutrino-electron scattering ($\nu+e^{-} \leftrightarrow \nu+e^{-}$) is included, the disc can be 60-80\% more massive, and is slightly more neutron-poor, while the averaged temperature in all cases are qualitatively similar among the neutrino interaction sets considered in this work.
However, the impact due to electron neutrino-pair production via electron-positron annihilation ($e^{-}+e^{+} \leftrightarrow \nu_e+\bar{\nu}_e$) is minimal in all cases.

\begin{figure}
	\centering
	\includegraphics[width=\columnwidth, angle=0]{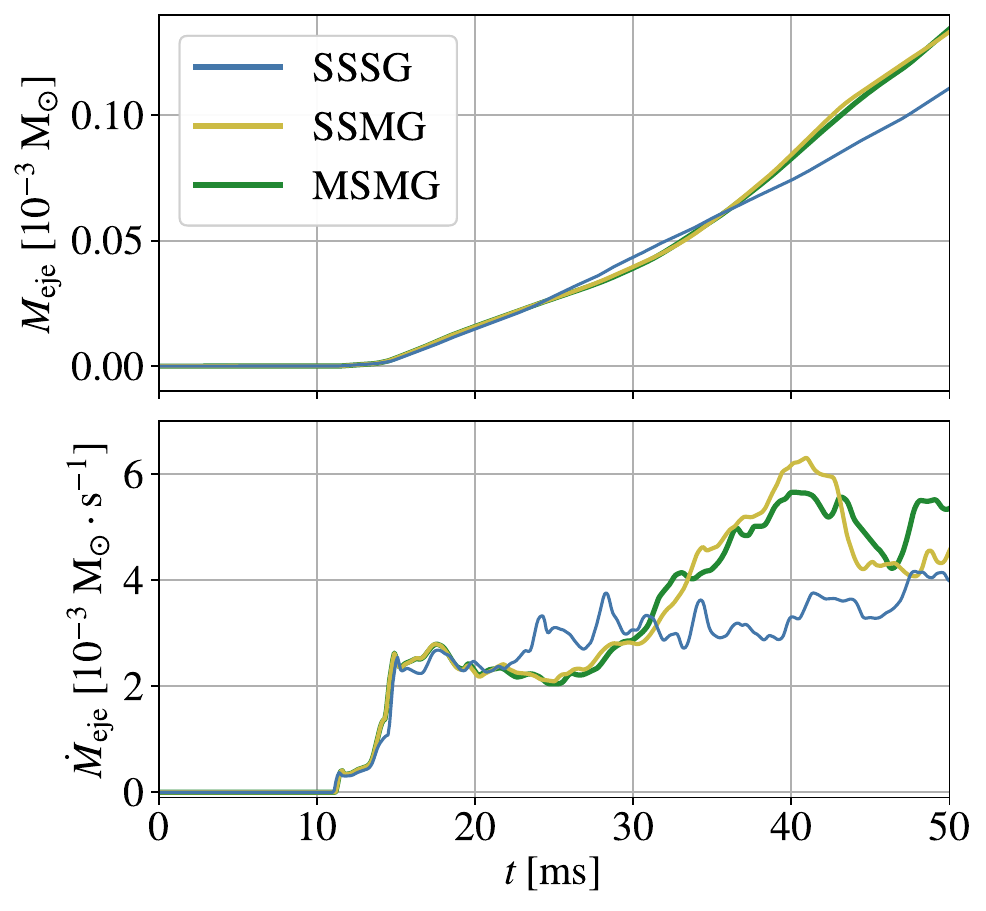}
	\caption{
		The total ejected rest-mass (\emph{upper panel}) and the mass ejection rate (\emph{lower panel}) evolutions of a post-merger-like hypermassive neutron star with different neutrino interaction sets.
		These quantities count the material that leaves the extraction cylinder ($R=z=1800~{\rm km}$).
		Similar to the disc mass (see figure~\ref{fig:ns_disc_avg}), cases with inelastic neutrino-electron scattering ($\nu+e^{-} \leftrightarrow \nu+e^{-}$) predicts higher mass ejection rate and hence more ejected mass than the cases without.
		At $t=50{\rm ms}$, the SSMG and MSMG cases result in $\sim 18\%$ more ejected mass. 
		The mass ejection rate is $\gtrsim 100\%$ higher at about 40~ms, but approaching to similar values at the end of the simulations.
		}
	\label{fig:ns_eje_mass}
\end{figure}

\begin{figure*}
	\centering
	\includegraphics[width=\textwidth, angle=0]{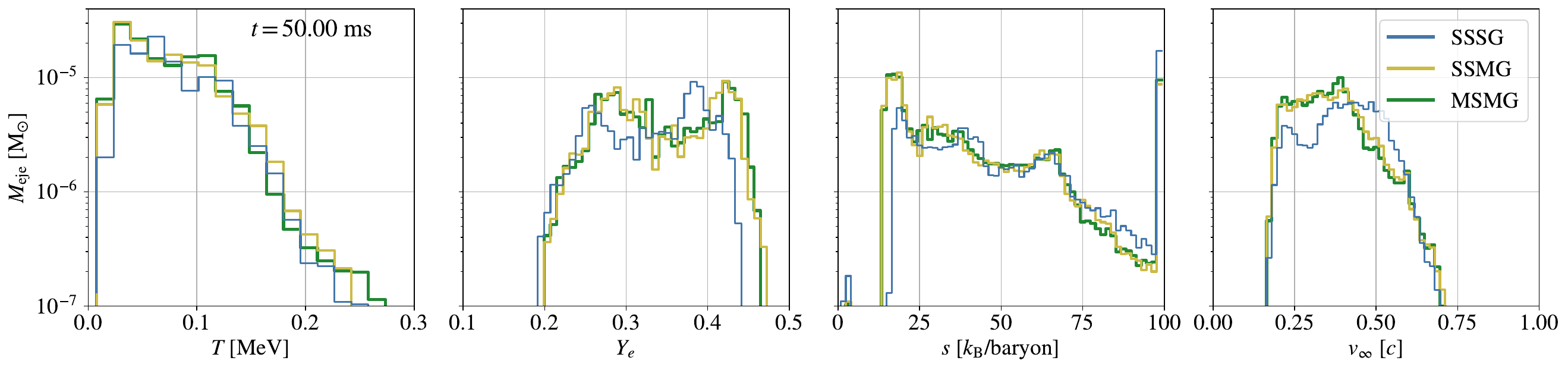}
	\caption{
		1-D histograms of the ejecta that leaves the extraction cylinder ($R=z=1800~{\rm km}$) of a post-merger-like hypermassive neutron star up to $t=50~{\rm ms}$.
		These plots compare the distributions of ejected mass as functions of temperature $T$, electron fraction $Y_e$, entropy per baryon $s$ and the asymptotic velocity $v_{\infty}$ with different neutrino interaction sets.
		Up to 50~ms, all the distributions are qualitatively the same.
		}
	\label{fig:eje_histogram}
\end{figure*}

Similar to the disc evolutions, the properties of the ejecta are affected by considering different neutrino interaction sets.
Figure~\ref{fig:ns_eje_mass} shows the time evolution of the total ejected mass and ejection rate.
The ejected mass follow similar trends as in the disc mass (e.g. see figure~\ref{fig:ns_disc_avg}).
In particular, mass ejection rate and therefore ejected mass are higher than the cases without including inelastic neutrino-electron scattering ($\nu+e^{-} \leftrightarrow \nu+e^{-}$).
Similarly, with or without electron neutrino-pair production via electron-positron annihilation are very minor in all cases.
Despite the difference in mass ejection, their distributions are qualitatively the same, as shown in figure~\ref{fig:eje_histogram}.

\begin{figure*}
	\centering
	\includegraphics[width=\textwidth, angle=0]{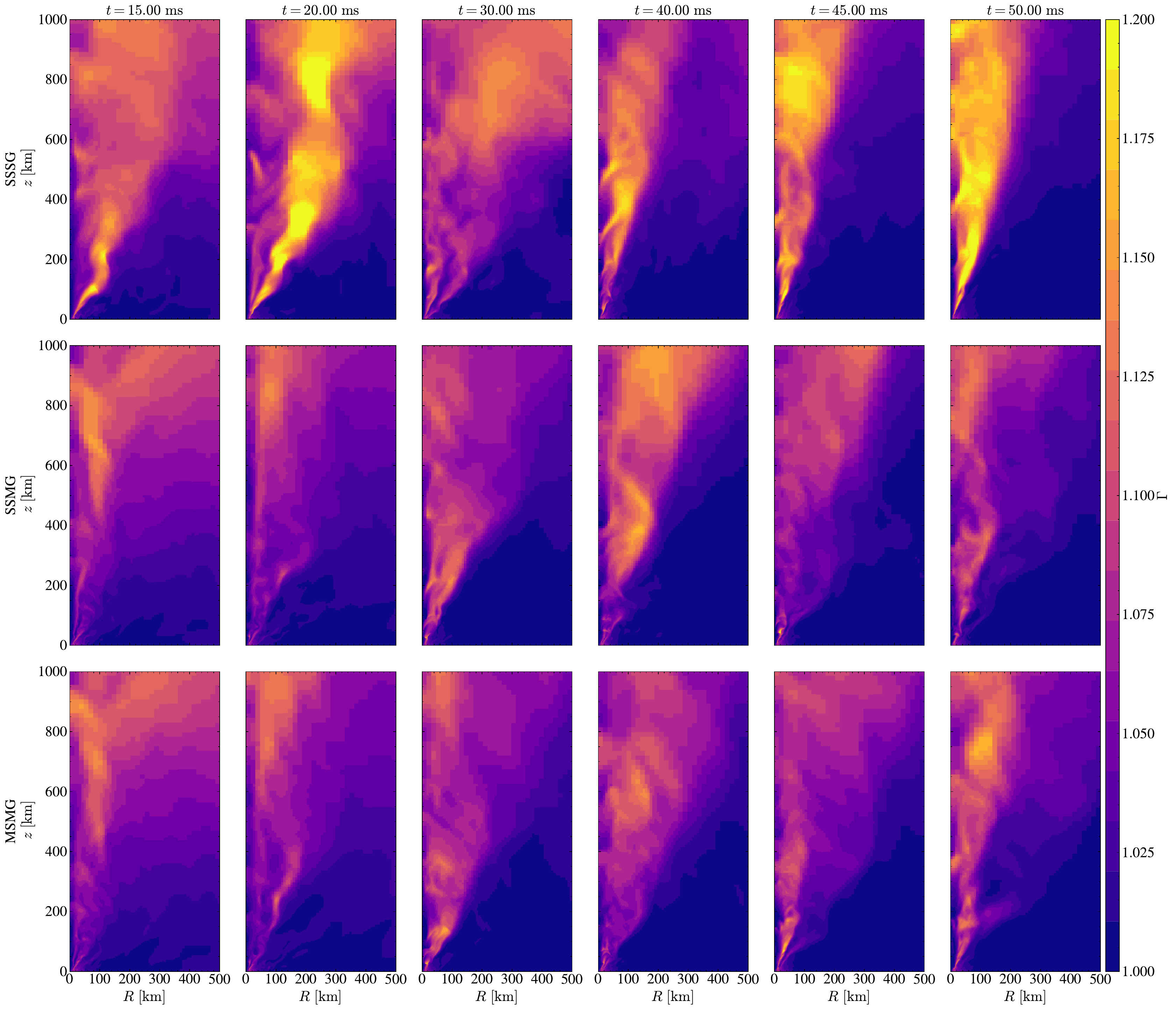}
	\caption{
		The time evolutions of the Lorentz factor $\Gamma$ with different neutrino interaction sets (\emph{top to bottom rows}) of a post-merger-like hypermassive neutron star.
		Different sets of neutrino interactions could alter the funnel structures.
		Since the existence of inelastic neutrino-electron scattering enhances the disc and quasi-isotropic ejecta mass, the baryon pollution in the polar region could also be enhanced.
		This makes jet launching more challenging.
		}
	\label{fig:gamma_time_series}
\end{figure*}

Enhancement of the disc and quasi-isotropic ejecta mass could result in less favourable environments for jet launching due to baryon pollution in the polar region.
As shown in figure~\ref{fig:gamma_time_series}, the Lorentz factor $\Gamma$ becomes weaker when the inelastic neutrino-electron scattering is included.
Including the electron neutrino-pair production via electron-positron annihilation leads to some structural changes in the funnel regions, but it does not noticeably enhance or suppress the accelerations of the polar outflow.

\begin{figure*}
	\centering
	\includegraphics[width=\textwidth, angle=0]{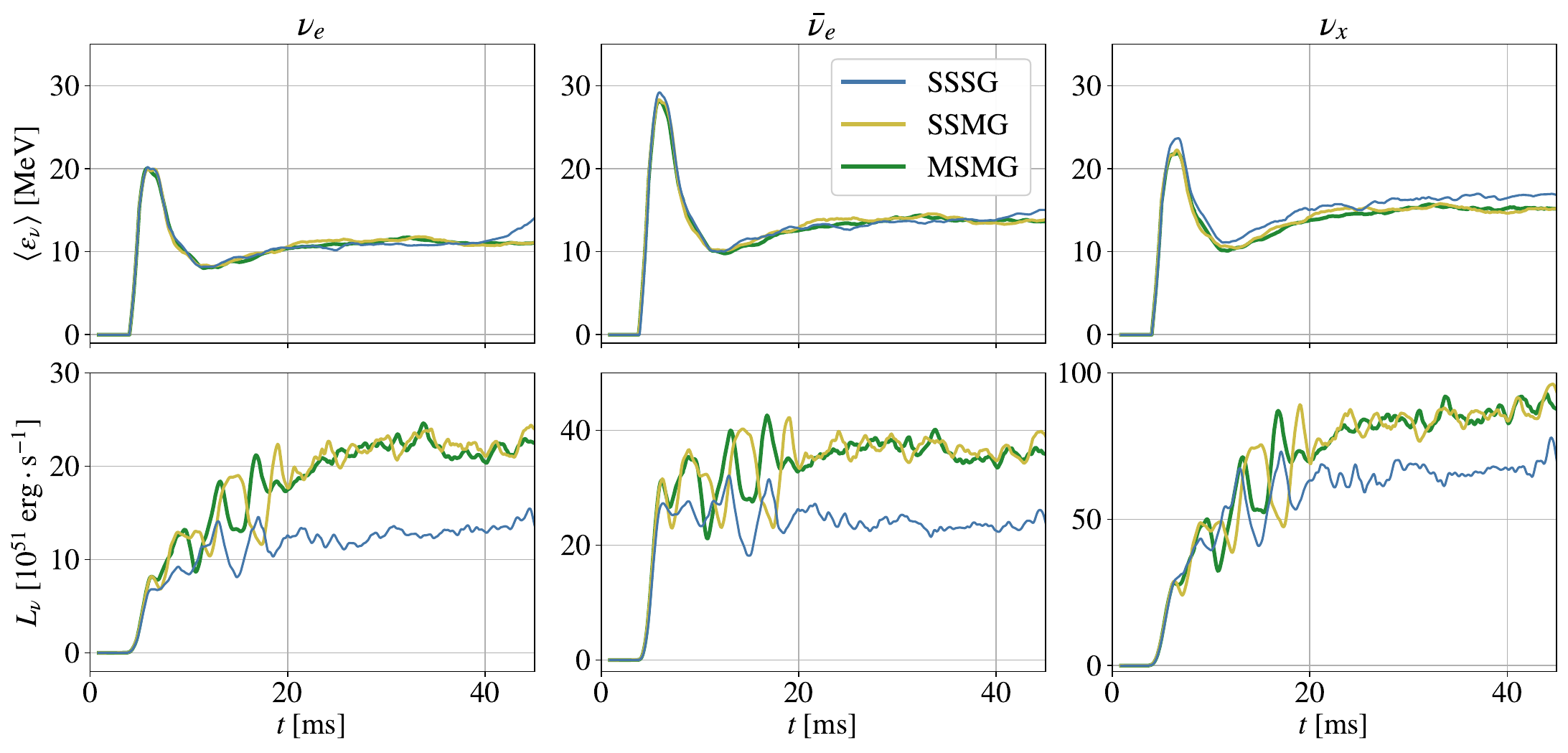}
	\caption{
	Time evolution of far-field averaged neutrino energies ($\langle {\epsilon_\nu} \rangle$, \emph{upper panels}) and luminosities ($L_\nu$, \emph{lower panels}) measured by an observer comoving with fluid at 1000 km of a post-merger-like hypermassive neutron star.
	The averaged neutrino energies are qualitatively the same among all the cases.
	The luminosities in the cases that include inelastic neutrino-electron scattering ($\nu+e^{-} \leftrightarrow \nu+e^{-}$) are $\sim 50, 40, {\rm and }~30~\%$ higher for $\nu_e$, $\bar{\nu}_e$, and $\nu_x$, respectively, compared to the cases without this scattering process.
	The impact due to the electron neutrino-pair production via electron-positron annihilation ($e^{-}+e^{+} \leftrightarrow \nu_e+\bar{\nu}_e$) are minimal.
		}
	\label{fig:ns_nu_combine}
\end{figure*}

Here, we present the neutrino signals from the system. 
Figure~\ref{fig:ns_nu_combine} compares the time evolution of the average energies of neutrinos $\langle {\epsilon_\nu} \rangle$ and luminosities $L_\nu$ at 1000 km of a post-merger-like hypermassive neutron star.
All the neutrino interaction sets considered here result in qualitatively similar averaged neutrino energies.
However, this is not the case for neutrino luminosities.
Inclusion of inelastic neutrino-electron scattering ($\nu+e^{-} \leftrightarrow \nu+e^{-}$) results in higher luminosities for all species.
In addition, including the electron neutrino-pair production via electron-positron annihilation does not alter the resulting luminosities noticeably.

The increases of neutrino luminosities can be understood via the neutrino mean-free-path responses.
Specifically, the neutrino mean-free path increases significantly when neutrino energy decreases.
Since the inelastic neutrino-electron scattering enhances the energy transport from neutrinos to matter, high-energy neutrinos scatter into low-energy neutrinos, which have larger mean-free-path, can then escape relatively easier.

To better understand the role played by inelastic neutrino-electron scattering, we estimate its effective opacity.
In particular, it can be estimated as \citep{2024ApJS..272....9N}
\begin{equation}\label{eq:knes}
	\kappa_{\rm NES}\left(\varepsilon \right) = \frac{1}{\left(hc\right)^3} \int_0^{\infty} \Phi_{0, {\rm NES}}^{\rm out} \left( \varepsilon, \varepsilon_{\rm out} \right) / c \dd{V_{\varepsilon_{\rm out}}},
\end{equation}
where $\Phi_{0, {\rm NES}}^{\rm out}$ is the first Legendre coefficient of the kernel of outgoing beam scattering.
In addition, we estimate the total effective opacity by
\begin{equation}\label{eq:keff}
	\kappa_{\rm eff}^{\rm NES} = \sqrt{\left( \kappa_{\rm abs} + \kappa_{\rm NES} \right) \left( \kappa_{\rm abs} + \kappa_{\rm scat} + \kappa_{\rm NES} \right)},
\end{equation}
where $\kappa_{\rm abs}$ and $\kappa_{\rm scat}$ are the absorption and (iso-energetic) scattering opacities, respectively.
To better compare the influence due to $\kappa_{\rm NES}$, we also compute the effective opacity $\kappa_{\rm eff}$ as in equation~\eqref{eq:keff} but setting $\kappa_{\rm NES}=0$.
Here, we assume final state occupancy of the neutrino is zero.

\begin{figure*}
	\centering
	\includegraphics[width=\textwidth, angle=0]{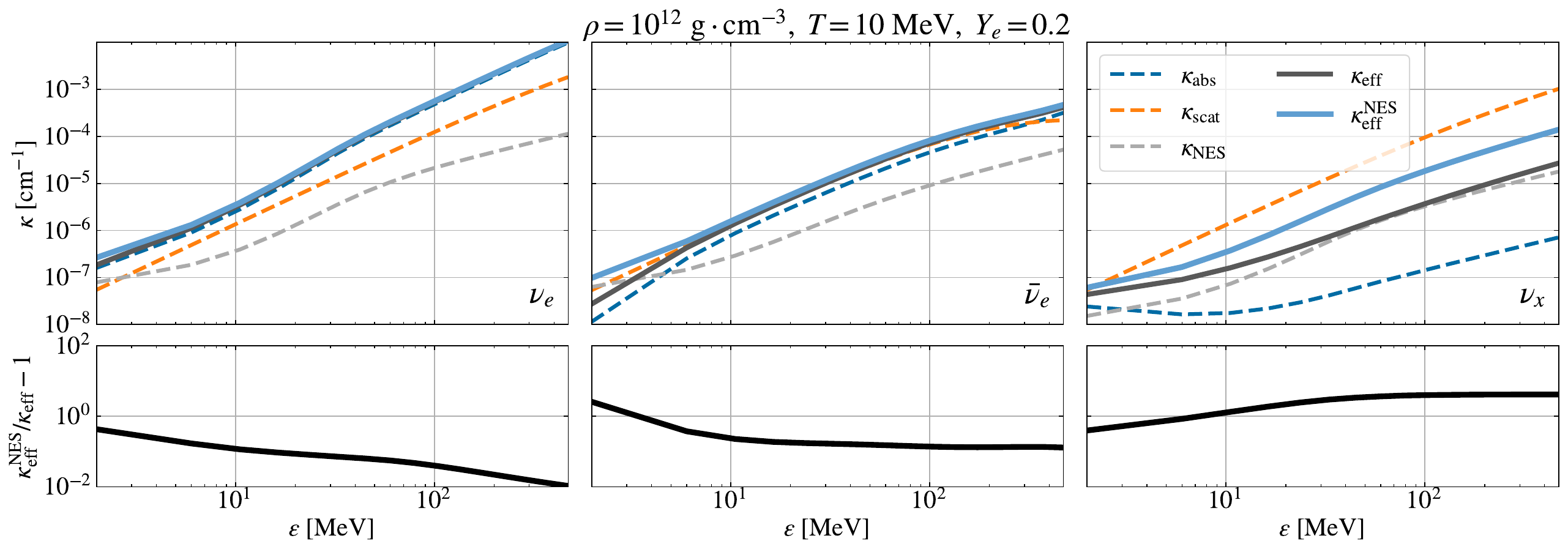}
	\caption{
		Comparison of the neutrino opacities at a hydrodynamical point $\rho=10^{12}~{\rm g \cdot cm^{-3}}, T=10~{\rm MeV}, Y_e = 0.2$ for all species.
		This hydrodynamical point is roughly the location where the cases with and without inelastic neutrino-electron scattering start to deviate (see figure~\ref{fig:ns_1d_profiles}).
		\emph{Top panel:} Dotted lines shows the absorption, (iso-energetic) scattering, and the estimated inelastic neutrino-electron scattering opacities $\kappa_{\rm abs}, \kappa_{\rm scat}, \kappa_{\rm NES}$ as functions of neutrino energy $\varepsilon$.
		Solid lines are the effective opacities with and without the contributions of $\kappa_{\rm NES}$ ($\kappa_{\rm eff}$ and $\kappa^{\rm NES}_{\rm eff}$).
		\emph{Bottom panel:} Relative error of the effective opacities with and without the contributions of $\kappa_{\rm NES}$.
		The effective opacities of electron neutrino-pair production via electron-positron annihilation is $\sim 10^{10}$ smaller, hence it is ignored here.
		At this hydrodynamical point, the impact of inelastic neutrino-electron scattering is comparable to the absorption and iso-energetic scattering opacities.
		The effective opacity $\kappa_{\rm eff}$ is at least 10\% higher for $\bar{\nu}_e$.
		Moreover, $\kappa_{\rm eff}$ is approximately enhanced by 40\% -400\% for the low to high energy $\nu_x$.
		}
	\label{fig:kappa_at_pt}
\end{figure*}

Figure~\ref{fig:kappa_at_pt} shows the neutrino opacities at a hydrodynamical point $\rho=10^{12}~{\rm g \cdot cm^{-3}}, T=10~{\rm MeV}, Y_e = 0.2$ for all species. 
This hydrodynamical point is roughly where the cases with and without inelastic neutrino-electron scattering start to deviate (see figure~\ref{fig:ns_1d_profiles}).
The strongest impact due to $\kappa_{\rm NES}$ is found in the case $\nu_x$, where $\kappa_{\rm eff}$ can be enhanced by 40\%-400\% from low to high energies.
Enhancement in the $\bar{\nu}_e$ case is also found, where the overall $\kappa_{\rm eff}$ is at least 10\% higher in the entire energy range.
Finally, this scattering has the weakest impact onto the $\nu_e$ case, which has only about 10\% to 1\% enhancement.

It is worth noting that, comparing the effective opacities $\kappa_{\rm eff}$ and $\kappa^{\rm NES}_{\rm eff}$ roughly estimates the importance of inelastic neutrino-electron scattering. 
However, the thermalisation of neutrinos, as well as the neutrino luminosities, cannot be fully estimated with this quantity. 
This is because the inelastic nature of this interaction, which is essential to neutrino-matter energy exchange and the resulting neutrino mean-free-path, cannot be captured in this estimation.
As shown in figure~\ref{fig:J_over_Jeq}, despite larger $\kappa^{\rm NES}_{\rm eff}$ of $\nu_x$ when $10^{11}~{\rm g\cdot cm^{-3} \lesssim \rho \lesssim 10^{13}~{\rm g \cdot cm^{-3}}}$, the neutrino energy ratio $J/J_{\rm eq}$ is however lower than the case without this interaction.
Similarly, as shown in figure~\ref{fig:ns_nu_combine}, the neutrino luminosities are higher in these cases.
More detailed and systematic analysis of inelasticity (e.g. \cite{2020PhRvD.102l3001F}) are left as future work.
The importance of resolving the neutrino distributions in merger simulations is further displayed via the inelastic nature of these interactions.

\begin{figure*}
	\centering
	\includegraphics[width=\textwidth, angle=0]{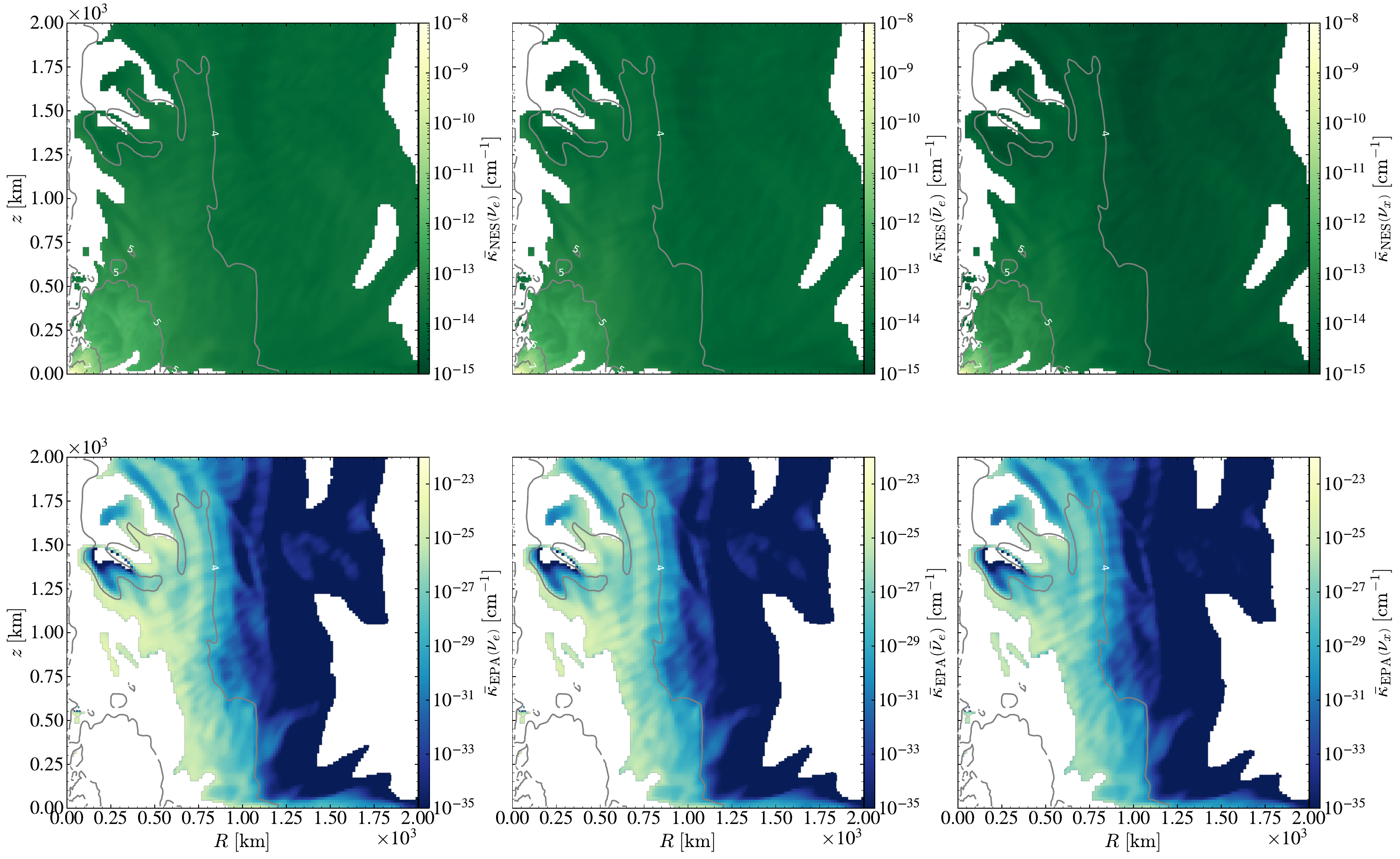}
	\caption{
		Energy-integrated effective neutrino opacities of inelastic neutrino-electron scattering $\bar{\kappa}_{\rm NES}$ and electron neutrino-pair production via electron-position annihilation $\bar{\kappa}_{\rm EPA}$ for all species at the end of the MSMG case.
		Opacities will be set to zero (white regions) when the corresponding hydrodynamical points beyond the neutrino microphysics table.
		Grey contours correspond to densities of $10^{7,6,5,4}~{\rm g \cdot cm^{-3}}$.
		$\bar{\kappa}_{\rm NES}$ and $\bar{\kappa}_{\rm EPA}$ have at least 10 orders of magnitude differences and are effective in different density ranges.
		}
	\label{fig:kappa_nes_and_epa}
\end{figure*}

All the results presented above are insensitive to the electron neutrino-pair production via electron-positron annihilation ($e^{-}+e^{+} \leftrightarrow \nu_e+\bar{\nu}_e$).
In fact, this interaction is expected to be important in the low-density and low-temperature regions.
Nevertheless, this interaction is almost negligible everywhere in our simulations.
To visualise this, we plot the energy-integrated effective neutrino opacities for inelastic neutrino-electron scattering and electron neutrino-pair production.
Specifically, we compute 
\begin{equation}
	\bar{\kappa}_{\rm NES} = \frac{\int_0^\infty \kappa_{\rm NES}\left(\varepsilon\right) \varepsilon f_{\nu} \dd{V_\varepsilon}}{\int_0^\infty \varepsilon f_{\nu} \dd{V_\varepsilon}},
\end{equation}
where $f_{\nu}$ is the neutrino distribution function extracted from simulations, and $\kappa_{\rm NES}$ is computed in equation~\eqref{eq:knes}.
We calculate the same quantity for electron neutrino-pair production ($\bar{\kappa}_{\rm EPA}$) with the same procedure.
Figure~\ref{fig:kappa_nes_and_epa} shows the profiles of $\bar{\kappa}_{\rm NES}$ and $\bar{\kappa}_{\rm EPA}$ for all species at the end of the MSMG simulation.
With the neutrino interactions provided by \texttt{NuLib} in this work, the inelastic neutrino-electron scattering not only has wider rest mass density coverage but is also at least 10 orders of magnitude stronger than the electron neutrino-pair production via electron-position annihilation.
Therefore, the influence of electron neutrino-pair production is minimal in this work.

\section{\label{sec:discussion}Discussion}
In this work, we perform general-relativistic radiation magnetohydrodynamics simulations of high angular momentum hypermassive neutron star with energy-dependent two-moment scheme, with 3 different sets of input neutrino microphysics.
We study particularly the impact due to inelastic neutrino-electron scattering and electron neutrino-pair production via electron-positron annihilation onto the disc and ejecta properties, as well as neutrino signatures.

Inelastic neutrino-electron scattering is known to be very effective in neutrino-matter energy exchange. 
With this scattering, energy is transported efficiently from neutrinos to matter in the regions where neutrinos have a higher temperature than matter, which could enhance the disc and ejecta mass.
Moreover, the neutrino mean-free path increases significantly when its energy decreases. 
Transporting energies from neutrinos to matter could result in mean-free-path increases, meaning that neutrinos can escape easier, and hence increases the neutrino luminosities.

We observe all the phenomena above-mentioned in our simulations.
In particular, in the cases that include inelastic neutrino-electron scattering, the disc and ejecta mass are about 75\% and 18\% higher, respectively.
Despite the disc being slightly more neutron-poor in these cases, the properties of the ejecta (i.e. distributions of temperature, electron fractions, entropy and asymptotic velocity) are qualitatively the same.
The enhancement of disc and ejecta mass results in stronger baryon pollution, leading to less favourable jet launching environments.
Finally, the neutrino luminosities of $\nu_e$, $\bar{\nu}_e$, and $\nu_x$ are $\sim 50, 40, {\rm and }~30~\%$ higher, respectively.

We also include electron neutrino-pair production/annihilation in this work.
In our simulations, however, no significant impact is observed by including this interaction.
By comparison, this interaction is nearly negligible compare to other neutrino interactions considered in this work.
Although the electron-positron annihilation does not have short terms effects in the simulations, it is expected to be important in the low-density and low-temperature regions, and has been proposed to power relativistic jets. 
Determining whether this interaction can power relativistic jets requires long-term simulations with advanced neutrino transport that can better resolve the polar regions, which will be addressed in the future.

Our results suggest that the neutrino-electron inelastic scattering, despite being subdominant compared to beta processes and elastic scatterings, is non-negligible in matter outflow modellings.
On the other hand, the electron-positron annihilation does not have short terms effects in the simulations.
Note that, our simulations are evolved only with a short timescale (up to 50~ms), the impact due to the completeness of neutrino microphysics can be higher at later times (e.g. when $t \gtrsim \mathcal{O}\left(1~{\rm s}\right)$).
At later times, the total ejected mass could be noticeably higher and the properties of the ejecta (e.g. composition, temperature, velocity, etc.) could be affected indirectly via more comprehensive neutrino microphysics over such long timescale.
We also note that neutrino-nucleus inelastic scattering is about one-third as effective as the neutrino-electron inelastic scattering \cite{1991ApJ...376..678B}; inclusion of this interaction may be necessary to better capture the energy changes between neutrinos and matter.

In the future, we plan to extend several aspects of this work.
Firstly, the initial profile is constructed here by solving hydrostatic equations by assuming constant entropy and neutrinoless beta-equilibrium.
Mapping from actual neutron star merger simulations with neutrino transport as more realistic initial conditions will be explored.
Secondly, the neutrino interactions considered in this work are far from complete.
We plan to input more complete neutrino interactions~\citep{2024ApJS..272....9N} for this series of studies.
Although the computational cost is expected to be very high, it is necessary to include all the possible neutrino interaction into simulations, because the impacts due to neutrino microphysics is known to be highly nonlinear~\cite{2020LRCA....6....4M}.

\begin{acknowledgments}
P.C.-K.C. gratefully acknowledges support from NSF Grant PHY-2020275 (Network for Neutrinos, Nuclear Astrophysics, and Symmetries (N3AS)).
F.F. gratefully acknowledge support from the Department of Energy, Office of Science, Office of Nuclear Physics, under contract number DE-AC02-05CH11231 and from the NSF through grant AST-2107932. 
H.H.Y.N. is supported by the ERC Advanced Grant “JETSET: Launching, propagation and emission of relativistic jets from binary mergers and across mass scales” (Grant No. 884631).
M.D. gratefully acknowledges support from the NSF through grant PHY-2110287.  
M.D. and F.F. gratefully acknowledge support from NASA through grant 80NSSC22K0719. 
This work was partially supported by grants from the Research Foundation - Flanders (G086722N and I002123N).

The simulations in this work have been performed on the UNH supercomputer Marvin, also known as Plasma, which is supported by NSF/MRI program under grant number AGS-1919310. 
This work also used Expanse cluster at San Diego Supercomputer Centre through allocation PHY240222 from the Advanced Cyberinfrastructure Coordination Ecosystem: Services \& Support (ACCESS) program~\cite{10.1145/3569951.3597559}, which is supported by National Science Foundation grants \#2138259, \#2138286, \#2138307, \#2137603, and \#2138296. 

Conformally flat, axisymmetric, differentially rotating hot neutron stars in quasi-equilibrium are constructed using the \texttt{RotNS} code~\citep{1994ApJ...422..227C, 2024PhRvD.110d3015C, 2024arXiv240305642M}.
The results of this work were produced by utilising \texttt{Gmunu}~\citep{2020CQGra..37n5015C, 2021MNRAS.508.2279C, 2022ApJS..261...22C, 2023ApJS..267...38C, 2024ApJS..272....9N}, where the tabulated neutrino interaction were provided with \texttt{NuLib}~\citep{2015ApJS..219...24O}.
The data of the simulations were post-processed and visualised with 
\texttt{yt}~\citep{2011ApJS..192....9T},
\texttt{NumPy}~\citep{harris2020array}, 
\texttt{pandas}~\citep{reback2020pandas, mckinney-proc-scipy-2010},
\texttt{SciPy}~\citep{2020SciPy-NMeth} and
\texttt{Matplotlib}~\citep{2007CSE.....9...90H, thomas_a_caswell_2023_7697899}.
\end{acknowledgments}

\appendix

\section{\label{sec:epa_approx} Validity of approximations of $e^+ + e^- \leftrightarrow \nu_{x}+\bar{\nu}_{x}$}
The neutrino-pair process for $\nu_x$ (i.e. $e^+ + e^- \leftrightarrow \nu_{x}+\bar{\nu}_{x}$, process (h) in table~\ref{tab:nu_int}) is treated approximately as effective emissivity/absorption opacity \citep{2006NuPhA.777..356B} when the SSSG implicit solver is used, while it is described as the full production/annihilation kernels \citep{1985ApJS...58..771B} when the SSMG/MSMG solver is used.
Approximating this interaction from kernels to emissivity/absorption opacity reduces the needs of energy coupled implcit solvers.
This approximation has been shown to be effective in the context of core-collapse supernovae~\citep{2015ApJS..219...24O, 2020PhRvD.102l3015B}, and generally reasonable in the context of neutron star mergers~\citep{2023LRCA....9....1F}.
We have adopted this approximation in our previous work~\cite{2024arXiv240716017C}, which is present here as the SSSG simulation.

\begin{figure*}
	\centering
	\includegraphics[width=\textwidth, angle=0]{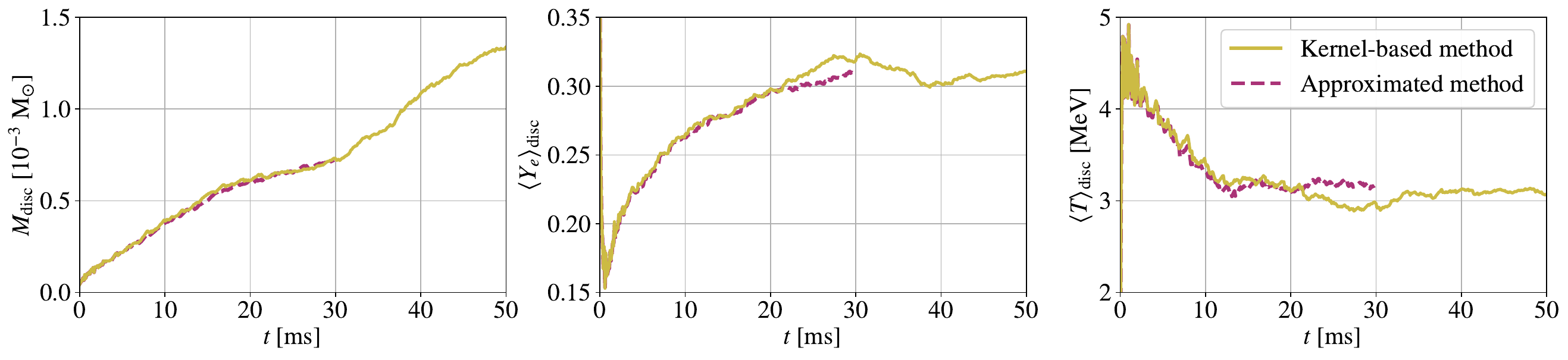}
	\caption{
		The total rest-mass, mass-averaged electron fraction, temperature of the disc as functions of time with different treatments for $e^+ + e^- \leftrightarrow \nu_{x}+\bar{\nu}_{x}$.
		}
	\label{fig:ssmg_disc_avg}
\end{figure*}

\begin{figure*}
	\centering
	\includegraphics[width=\textwidth, angle=0]{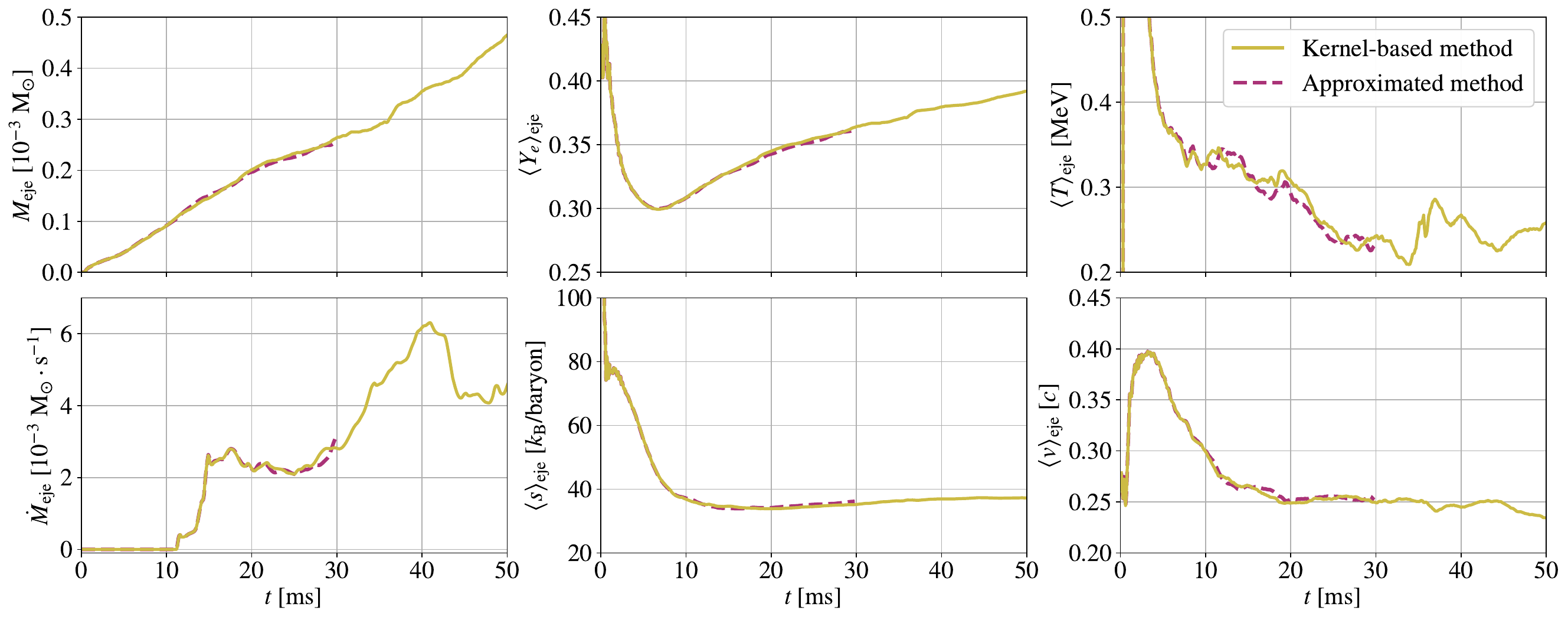}
	\caption{
		The total rest-mass and its ejection rate, mass-averaged electron fraction, temperature, entropy, and asymptotic velocity of the ejecta as functions of time with different treatments for $e^+ + e^- \leftrightarrow \nu_{x}+\bar{\nu}_{x}$.
		}
	\label{fig:ssmg_eje_avg}
\end{figure*}

\begin{figure*}
	\centering
	\includegraphics[width=\textwidth, angle=0]{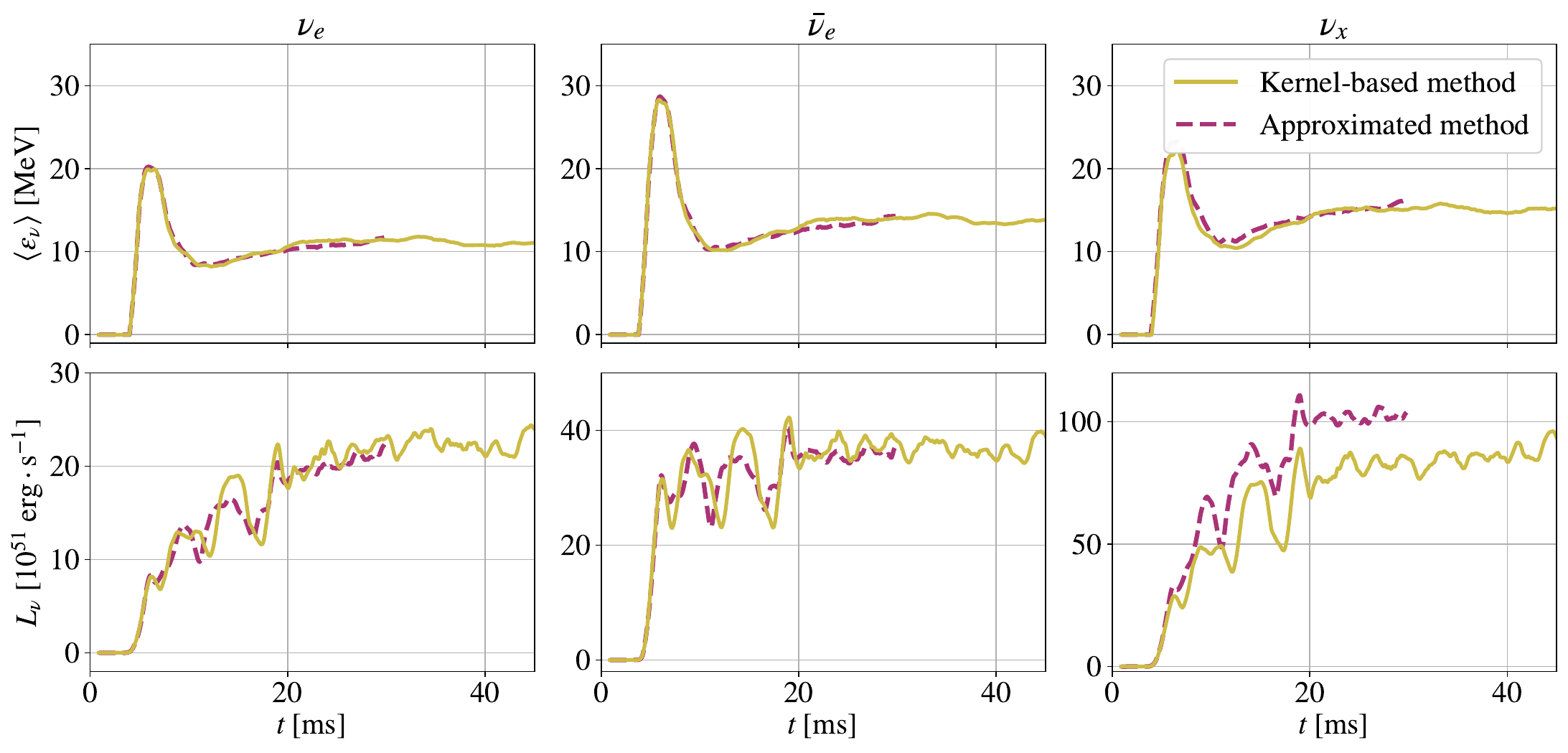}
	\caption{
		Time evolution of far-field averaged neutrino energies ($\langle {\epsilon_\nu} \rangle$, \emph{upper panels}) and luminosities ($L_\nu$, \emph{lower panels}) measured by an observer comoving with fluid at 1000 km of a post-merger-like hypermassive neutron star.
		}
	\label{fig:ssmg_nu_combine}
\end{figure*}

To validate this approximation, we perform another SSMG simulation that adopts the approximation of $e^+ + e^- \leftrightarrow \nu_{x}+\bar{\nu}_{x}$.
We do not find any significant differences in the matter evolutions, as shown in figure~\ref{fig:ssmg_disc_avg} and \ref{fig:ssmg_eje_avg}.
However, as shown in figure~\ref{fig:ssmg_nu_combine}, the approximation overestimates the neutrino luminosity for $\nu_x$ for about 25\%.
As a comparison, this approximation is shown to have about 10\% level deviations in neutrino signals in the context of core-collapse supernovae~\citep{2020PhRvD.102l3015B}. 
Nevertheless, despite larger impacts in the neutrino signal for $\nu_x$, these differences do not alter our conclusions presented in this work.


\bibliography{references}{}

\begin{thebibliography}{69}%
\makeatletter
\providecommand \@ifxundefined [1]{%
 \@ifx{#1\undefined}
}%
\providecommand \@ifnum [1]{%
 \ifnum #1\expandafter \@firstoftwo
 \else \expandafter \@secondoftwo
 \fi
}%
\providecommand \@ifx [1]{%
 \ifx #1\expandafter \@firstoftwo
 \else \expandafter \@secondoftwo
 \fi
}%
\providecommand \natexlab [1]{#1}%
\providecommand \enquote  [1]{``#1''}%
\providecommand \bibnamefont  [1]{#1}%
\providecommand \bibfnamefont [1]{#1}%
\providecommand \citenamefont [1]{#1}%
\providecommand \href@noop [0]{\@secondoftwo}%
\providecommand \href [0]{\begingroup \@sanitize@url \@href}%
\providecommand \@href[1]{\@@startlink{#1}\@@href}%
\providecommand \@@href[1]{\endgroup#1\@@endlink}%
\providecommand \@sanitize@url [0]{\catcode `\\12\catcode `\$12\catcode
  `\&12\catcode `\#12\catcode `\^12\catcode `\_12\catcode `\%12\relax}%
\providecommand \@@startlink[1]{}%
\providecommand \@@endlink[0]{}%
\providecommand \url  [0]{\begingroup\@sanitize@url \@url }%
\providecommand \@url [1]{\endgroup\@href {#1}{\urlprefix }}%
\providecommand \urlprefix  [0]{URL }%
\providecommand \Eprint [0]{\href }%
\providecommand \doibase [0]{https://doi.org/}%
\providecommand \selectlanguage [0]{\@gobble}%
\providecommand \bibinfo  [0]{\@secondoftwo}%
\providecommand \bibfield  [0]{\@secondoftwo}%
\providecommand \translation [1]{[#1]}%
\providecommand \BibitemOpen [0]{}%
\providecommand \bibitemStop [0]{}%
\providecommand \bibitemNoStop [0]{.\EOS\space}%
\providecommand \EOS [0]{\spacefactor3000\relax}%
\providecommand \BibitemShut  [1]{\csname bibitem#1\endcsname}%
\let\auto@bib@innerbib\@empty
\bibitem [{\citenamefont {{Li}}\ and\ \citenamefont
  {{Paczy{\'n}ski}}(1998)}]{1998ApJ...507L..59L}%
  \BibitemOpen
  \bibfield  {author} {\bibinfo {author} {\bibfnamefont {L.-X.}\ \bibnamefont
  {{Li}}}\ and\ \bibinfo {author} {\bibfnamefont {B.}~\bibnamefont
  {{Paczy{\'n}ski}}},\ }\bibfield  {title} {\bibinfo {title} {{Transient Events
  from Neutron Star Mergers}},\ }\href {https://doi.org/10.1086/311680}
  {\bibfield  {journal} {\bibinfo  {journal} {\apjl}\ }\textbf {\bibinfo
  {volume} {507}},\ \bibinfo {pages} {L59} (\bibinfo {year} {1998})},\ \Eprint
  {https://arxiv.org/abs/astro-ph/9807272} {arXiv:astro-ph/9807272 [astro-ph]}
  \BibitemShut {NoStop}%
\bibitem [{\citenamefont {{Metzger}}\ \emph {et~al.}(2010)\citenamefont
  {{Metzger}}, \citenamefont {{Mart{\'\i}nez-Pinedo}}, \citenamefont
  {{Darbha}}, \citenamefont {{Quataert}}, \citenamefont {{Arcones}},
  \citenamefont {{Kasen}}, \citenamefont {{Thomas}}, \citenamefont {{Nugent}},
  \citenamefont {{Panov}},\ and\ \citenamefont
  {{Zinner}}}]{2010MNRAS.406.2650M}%
  \BibitemOpen
  \bibfield  {author} {\bibinfo {author} {\bibfnamefont {B.~D.}\ \bibnamefont
  {{Metzger}}}, \bibinfo {author} {\bibfnamefont {G.}~\bibnamefont
  {{Mart{\'\i}nez-Pinedo}}}, \bibinfo {author} {\bibfnamefont {S.}~\bibnamefont
  {{Darbha}}}, \bibinfo {author} {\bibfnamefont {E.}~\bibnamefont
  {{Quataert}}}, \bibinfo {author} {\bibfnamefont {A.}~\bibnamefont
  {{Arcones}}}, \bibinfo {author} {\bibfnamefont {D.}~\bibnamefont {{Kasen}}},
  \bibinfo {author} {\bibfnamefont {R.}~\bibnamefont {{Thomas}}}, \bibinfo
  {author} {\bibfnamefont {P.}~\bibnamefont {{Nugent}}}, \bibinfo {author}
  {\bibfnamefont {I.~V.}\ \bibnamefont {{Panov}}},\ and\ \bibinfo {author}
  {\bibfnamefont {N.~T.}\ \bibnamefont {{Zinner}}},\ }\bibfield  {title}
  {\bibinfo {title} {{Electromagnetic counterparts of compact object mergers
  powered by the radioactive decay of r-process nuclei}},\ }\href
  {https://doi.org/10.1111/j.1365-2966.2010.16864.x} {\bibfield  {journal}
  {\bibinfo  {journal} {\mnras}\ }\textbf {\bibinfo {volume} {406}},\ \bibinfo
  {pages} {2650} (\bibinfo {year} {2010})},\ \Eprint
  {https://arxiv.org/abs/1001.5029} {arXiv:1001.5029 [astro-ph.HE]}
  \BibitemShut {NoStop}%
\bibitem [{\citenamefont {{Tanaka}}(2016)}]{2016AdAst2016E...8T}%
  \BibitemOpen
  \bibfield  {author} {\bibinfo {author} {\bibfnamefont {M.}~\bibnamefont
  {{Tanaka}}},\ }\bibfield  {title} {\bibinfo {title} {{Kilonova/Macronova
  Emission from Compact Binary Mergers}},\ }\href
  {https://doi.org/10.1155/2016/6341974} {\bibfield  {journal} {\bibinfo
  {journal} {Advances in Astronomy}\ }\textbf {\bibinfo {volume} {2016}},\
  \bibinfo {eid} {634197} (\bibinfo {year} {2016})},\ \Eprint
  {https://arxiv.org/abs/1605.07235} {arXiv:1605.07235 [astro-ph.HE]}
  \BibitemShut {NoStop}%
\bibitem [{\citenamefont {{Metzger}}(2019)}]{2019LRR....23....1M}%
  \BibitemOpen
  \bibfield  {author} {\bibinfo {author} {\bibfnamefont {B.~D.}\ \bibnamefont
  {{Metzger}}},\ }\bibfield  {title} {\bibinfo {title} {{Kilonovae}},\ }\href
  {https://doi.org/10.1007/s41114-019-0024-0} {\bibfield  {journal} {\bibinfo
  {journal} {Living Reviews in Relativity}\ }\textbf {\bibinfo {volume} {23}},\
  \bibinfo {eid} {1} (\bibinfo {year} {2019})},\ \Eprint
  {https://arxiv.org/abs/1910.01617} {arXiv:1910.01617 [astro-ph.HE]}
  \BibitemShut {NoStop}%
\bibitem [{\citenamefont {{Metzger}}(2017)}]{2017arXiv171005931M}%
  \BibitemOpen
  \bibfield  {author} {\bibinfo {author} {\bibfnamefont {B.~D.}\ \bibnamefont
  {{Metzger}}},\ }\bibfield  {title} {\bibinfo {title} {{Welcome to the
  Multi-Messenger Era! Lessons from a Neutron Star Merger and the Landscape
  Ahead}},\ }\href@noop {} {\bibfield  {journal} {\bibinfo  {journal} {arXiv
  e-prints}\ ,\ \bibinfo {eid} {arXiv:1710.05931}} (\bibinfo {year} {2017})},\
  \Eprint {https://arxiv.org/abs/1710.05931} {arXiv:1710.05931 [astro-ph.HE]}
  \BibitemShut {NoStop}%
\bibitem [{\citenamefont {{Rezzolla}}\ \emph {et~al.}(2018)\citenamefont
  {{Rezzolla}}, \citenamefont {{Pizzochero}}, \citenamefont {{Jones}},
  \citenamefont {{Rea}},\ and\ \citenamefont
  {{Vida{\~n}a}}}]{2018ASSL..457.....R}%
  \BibitemOpen
  \bibfield  {author} {\bibinfo {author} {\bibfnamefont {L.}~\bibnamefont
  {{Rezzolla}}}, \bibinfo {author} {\bibfnamefont {P.}~\bibnamefont
  {{Pizzochero}}}, \bibinfo {author} {\bibfnamefont {D.~I.}\ \bibnamefont
  {{Jones}}}, \bibinfo {author} {\bibfnamefont {N.}~\bibnamefont {{Rea}}},\
  and\ \bibinfo {author} {\bibfnamefont {I.}~\bibnamefont {{Vida{\~n}a}}},\
  }\bibfield  {title} {\bibinfo {title} {{The Physics and Astrophysics of
  Neutron Stars}},\ }\bibfield  {journal} {\bibinfo  {journal} {Astrophysics
  and Space Science Library}\ }\textbf {\bibinfo {volume} {457}},\ \href
  {https://doi.org/10.1007/978-3-319-97616-7} {10.1007/978-3-319-97616-7}
  (\bibinfo {year} {2018})\BibitemShut {NoStop}%
\bibitem [{\citenamefont {{Abbott}}\ \emph
  {et~al.}(2017{\natexlab{a}})\citenamefont {{Abbott}}, \citenamefont
  {{Abbott}}, \citenamefont {{Abbott}}, \citenamefont {{Acernese}},
  \citenamefont {{Ackley}}, \citenamefont {{Adams}},\ and\ \citenamefont
  {{Adams}}}]{2017PhRvL.119p1101A}%
  \BibitemOpen
  \bibfield  {author} {\bibinfo {author} {\bibfnamefont {B.~P.}\ \bibnamefont
  {{Abbott}}}, \bibinfo {author} {\bibfnamefont {R.}~\bibnamefont {{Abbott}}},
  \bibinfo {author} {\bibfnamefont {T.~D.}\ \bibnamefont {{Abbott}}}, \bibinfo
  {author} {\bibfnamefont {F.}~\bibnamefont {{Acernese}}}, \bibinfo {author}
  {\bibfnamefont {K.}~\bibnamefont {{Ackley}}}, \bibinfo {author}
  {\bibfnamefont {C.}~\bibnamefont {{Adams}}},\ and\ \bibinfo {author}
  {\bibfnamefont {T.}~\bibnamefont {{Adams}}},\ }\bibfield  {title} {\bibinfo
  {title} {{GW170817: Observation of Gravitational Waves from a Binary Neutron
  Star Inspiral}},\ }\href {https://doi.org/10.1103/PhysRevLett.119.161101}
  {\bibfield  {journal} {\bibinfo  {journal} {\prl}\ }\textbf {\bibinfo
  {volume} {119}},\ \bibinfo {eid} {161101} (\bibinfo {year}
  {2017}{\natexlab{a}})},\ \Eprint {https://arxiv.org/abs/1710.05832}
  {arXiv:1710.05832 [gr-qc]} \BibitemShut {NoStop}%
\bibitem [{\citenamefont {{Abbott}}\ \emph
  {et~al.}(2017{\natexlab{b}})\citenamefont {{Abbott}}, \citenamefont
  {{Abbott}}, \citenamefont {{Abbott}}, \citenamefont {{Acernese}},
  \citenamefont {{Ackley}}, \citenamefont {{Adams}},\ and\ \citenamefont
  {{Adams}}}]{2017ApJ...848L..13A}%
  \BibitemOpen
  \bibfield  {author} {\bibinfo {author} {\bibfnamefont {B.~P.}\ \bibnamefont
  {{Abbott}}}, \bibinfo {author} {\bibfnamefont {R.}~\bibnamefont {{Abbott}}},
  \bibinfo {author} {\bibfnamefont {T.~D.}\ \bibnamefont {{Abbott}}}, \bibinfo
  {author} {\bibfnamefont {F.}~\bibnamefont {{Acernese}}}, \bibinfo {author}
  {\bibfnamefont {K.}~\bibnamefont {{Ackley}}}, \bibinfo {author}
  {\bibfnamefont {C.}~\bibnamefont {{Adams}}},\ and\ \bibinfo {author}
  {\bibfnamefont {T.}~\bibnamefont {{Adams}}},\ }\bibfield  {title} {\bibinfo
  {title} {{Gravitational Waves and Gamma-Rays from a Binary Neutron Star
  Merger: GW170817 and GRB 170817A}},\ }\href
  {https://doi.org/10.3847/2041-8213/aa920c} {\bibfield  {journal} {\bibinfo
  {journal} {\apjl}\ }\textbf {\bibinfo {volume} {848}},\ \bibinfo {eid} {L13}
  (\bibinfo {year} {2017}{\natexlab{b}})},\ \Eprint
  {https://arxiv.org/abs/1710.05834} {arXiv:1710.05834 [astro-ph.HE]}
  \BibitemShut {NoStop}%
\bibitem [{\citenamefont {{Abbott}}\ \emph
  {et~al.}(2017{\natexlab{c}})\citenamefont {{Abbott}}, \citenamefont
  {{Abbott}}, \citenamefont {{Abbott}}, \citenamefont {{Acernese}},
  \citenamefont {{Ackley}}, \citenamefont {{Adams}},\ and\ \citenamefont
  {{Adams}}}]{2017ApJ...848L..12A}%
  \BibitemOpen
  \bibfield  {author} {\bibinfo {author} {\bibfnamefont {B.~P.}\ \bibnamefont
  {{Abbott}}}, \bibinfo {author} {\bibfnamefont {R.}~\bibnamefont {{Abbott}}},
  \bibinfo {author} {\bibfnamefont {T.~D.}\ \bibnamefont {{Abbott}}}, \bibinfo
  {author} {\bibfnamefont {F.}~\bibnamefont {{Acernese}}}, \bibinfo {author}
  {\bibfnamefont {K.}~\bibnamefont {{Ackley}}}, \bibinfo {author}
  {\bibfnamefont {C.}~\bibnamefont {{Adams}}},\ and\ \bibinfo {author}
  {\bibfnamefont {T.}~\bibnamefont {{Adams}}},\ }\bibfield  {title} {\bibinfo
  {title} {{Multi-messenger Observations of a Binary Neutron Star Merger}},\
  }\href {https://doi.org/10.3847/2041-8213/aa91c9} {\bibfield  {journal}
  {\bibinfo  {journal} {\apjl}\ }\textbf {\bibinfo {volume} {848}},\ \bibinfo
  {eid} {L12} (\bibinfo {year} {2017}{\natexlab{c}})},\ \Eprint
  {https://arxiv.org/abs/1710.05833} {arXiv:1710.05833 [astro-ph.HE]}
  \BibitemShut {NoStop}%
\bibitem [{\citenamefont {{Barnes}}\ and\ \citenamefont
  {{Kasen}}(2013)}]{2013ApJ...775...18B}%
  \BibitemOpen
  \bibfield  {author} {\bibinfo {author} {\bibfnamefont {J.}~\bibnamefont
  {{Barnes}}}\ and\ \bibinfo {author} {\bibfnamefont {D.}~\bibnamefont
  {{Kasen}}},\ }\bibfield  {title} {\bibinfo {title} {{Effect of a High Opacity
  on the Light Curves of Radioactively Powered Transients from Compact Object
  Mergers}},\ }\href {https://doi.org/10.1088/0004-637X/775/1/18} {\bibfield
  {journal} {\bibinfo  {journal} {\apj}\ }\textbf {\bibinfo {volume} {775}},\
  \bibinfo {eid} {18} (\bibinfo {year} {2013})},\ \Eprint
  {https://arxiv.org/abs/1303.5787} {arXiv:1303.5787 [astro-ph.HE]}
  \BibitemShut {NoStop}%
\bibitem [{\citenamefont {{Wanajo}}\ \emph {et~al.}(2014)\citenamefont
  {{Wanajo}}, \citenamefont {{Sekiguchi}}, \citenamefont {{Nishimura}},
  \citenamefont {{Kiuchi}}, \citenamefont {{Kyutoku}},\ and\ \citenamefont
  {{Shibata}}}]{2014ApJ...789L..39W}%
  \BibitemOpen
  \bibfield  {author} {\bibinfo {author} {\bibfnamefont {S.}~\bibnamefont
  {{Wanajo}}}, \bibinfo {author} {\bibfnamefont {Y.}~\bibnamefont
  {{Sekiguchi}}}, \bibinfo {author} {\bibfnamefont {N.}~\bibnamefont
  {{Nishimura}}}, \bibinfo {author} {\bibfnamefont {K.}~\bibnamefont
  {{Kiuchi}}}, \bibinfo {author} {\bibfnamefont {K.}~\bibnamefont
  {{Kyutoku}}},\ and\ \bibinfo {author} {\bibfnamefont {M.}~\bibnamefont
  {{Shibata}}},\ }\bibfield  {title} {\bibinfo {title} {{Production of All the
  r-process Nuclides in the Dynamical Ejecta of Neutron Star Mergers}},\ }\href
  {https://doi.org/10.1088/2041-8205/789/2/L39} {\bibfield  {journal} {\bibinfo
   {journal} {\apjl}\ }\textbf {\bibinfo {volume} {789}},\ \bibinfo {eid} {L39}
  (\bibinfo {year} {2014})},\ \Eprint {https://arxiv.org/abs/1402.7317}
  {arXiv:1402.7317 [astro-ph.SR]} \BibitemShut {NoStop}%
\bibitem [{\citenamefont {{Sekiguchi}}\ \emph {et~al.}(2015)\citenamefont
  {{Sekiguchi}}, \citenamefont {{Kiuchi}}, \citenamefont {{Kyutoku}},\ and\
  \citenamefont {{Shibata}}}]{2015PhRvD..91f4059S}%
  \BibitemOpen
  \bibfield  {author} {\bibinfo {author} {\bibfnamefont {Y.}~\bibnamefont
  {{Sekiguchi}}}, \bibinfo {author} {\bibfnamefont {K.}~\bibnamefont
  {{Kiuchi}}}, \bibinfo {author} {\bibfnamefont {K.}~\bibnamefont
  {{Kyutoku}}},\ and\ \bibinfo {author} {\bibfnamefont {M.}~\bibnamefont
  {{Shibata}}},\ }\bibfield  {title} {\bibinfo {title} {{Dynamical mass
  ejection from binary neutron star mergers: Radiation-hydrodynamics study in
  general relativity}},\ }\href {https://doi.org/10.1103/PhysRevD.91.064059}
  {\bibfield  {journal} {\bibinfo  {journal} {\prd}\ }\textbf {\bibinfo
  {volume} {91}},\ \bibinfo {eid} {064059} (\bibinfo {year} {2015})},\ \Eprint
  {https://arxiv.org/abs/1502.06660} {arXiv:1502.06660 [astro-ph.HE]}
  \BibitemShut {NoStop}%
\bibitem [{\citenamefont {{Foucart}}\ \emph {et~al.}(2015)\citenamefont
  {{Foucart}}, \citenamefont {{O'Connor}}, \citenamefont {{Roberts}},
  \citenamefont {{Duez}}, \citenamefont {{Haas}}, \citenamefont {{Kidder}},
  \citenamefont {{Ott}}, \citenamefont {{Pfeiffer}}, \citenamefont {{Scheel}},\
  and\ \citenamefont {{Szilagyi}}}]{2015PhRvD..91l4021F}%
  \BibitemOpen
  \bibfield  {author} {\bibinfo {author} {\bibfnamefont {F.}~\bibnamefont
  {{Foucart}}}, \bibinfo {author} {\bibfnamefont {E.}~\bibnamefont
  {{O'Connor}}}, \bibinfo {author} {\bibfnamefont {L.}~\bibnamefont
  {{Roberts}}}, \bibinfo {author} {\bibfnamefont {M.~D.}\ \bibnamefont
  {{Duez}}}, \bibinfo {author} {\bibfnamefont {R.}~\bibnamefont {{Haas}}},
  \bibinfo {author} {\bibfnamefont {L.~E.}\ \bibnamefont {{Kidder}}}, \bibinfo
  {author} {\bibfnamefont {C.~D.}\ \bibnamefont {{Ott}}}, \bibinfo {author}
  {\bibfnamefont {H.~P.}\ \bibnamefont {{Pfeiffer}}}, \bibinfo {author}
  {\bibfnamefont {M.~A.}\ \bibnamefont {{Scheel}}},\ and\ \bibinfo {author}
  {\bibfnamefont {B.}~\bibnamefont {{Szilagyi}}},\ }\bibfield  {title}
  {\bibinfo {title} {{Post-merger evolution of a neutron star-black hole binary
  with neutrino transport}},\ }\href
  {https://doi.org/10.1103/PhysRevD.91.124021} {\bibfield  {journal} {\bibinfo
  {journal} {\prd}\ }\textbf {\bibinfo {volume} {91}},\ \bibinfo {eid} {124021}
  (\bibinfo {year} {2015})},\ \Eprint {https://arxiv.org/abs/1502.04146}
  {arXiv:1502.04146 [astro-ph.HE]} \BibitemShut {NoStop}%
\bibitem [{\citenamefont {{Foucart}}\ \emph {et~al.}(2016)\citenamefont
  {{Foucart}}, \citenamefont {{O'Connor}}, \citenamefont {{Roberts}},
  \citenamefont {{Kidder}}, \citenamefont {{Pfeiffer}},\ and\ \citenamefont
  {{Scheel}}}]{2016PhRvD..94l3016F}%
  \BibitemOpen
  \bibfield  {author} {\bibinfo {author} {\bibfnamefont {F.}~\bibnamefont
  {{Foucart}}}, \bibinfo {author} {\bibfnamefont {E.}~\bibnamefont
  {{O'Connor}}}, \bibinfo {author} {\bibfnamefont {L.}~\bibnamefont
  {{Roberts}}}, \bibinfo {author} {\bibfnamefont {L.~E.}\ \bibnamefont
  {{Kidder}}}, \bibinfo {author} {\bibfnamefont {H.~P.}\ \bibnamefont
  {{Pfeiffer}}},\ and\ \bibinfo {author} {\bibfnamefont {M.~A.}\ \bibnamefont
  {{Scheel}}},\ }\bibfield  {title} {\bibinfo {title} {{Impact of an improved
  neutrino energy estimate on outflows in neutron star merger simulations}},\
  }\href {https://doi.org/10.1103/PhysRevD.94.123016} {\bibfield  {journal}
  {\bibinfo  {journal} {\prd}\ }\textbf {\bibinfo {volume} {94}},\ \bibinfo
  {eid} {123016} (\bibinfo {year} {2016})},\ \Eprint
  {https://arxiv.org/abs/1607.07450} {arXiv:1607.07450 [astro-ph.HE]}
  \BibitemShut {NoStop}%
\bibitem [{\citenamefont {{Foucart}}(2018)}]{2018MNRAS.475.4186F}%
  \BibitemOpen
  \bibfield  {author} {\bibinfo {author} {\bibfnamefont {F.}~\bibnamefont
  {{Foucart}}},\ }\bibfield  {title} {\bibinfo {title} {{Monte Carlo closure
  for moment-based transport schemes in general relativistic radiation
  hydrodynamic simulations}},\ }\href {https://doi.org/10.1093/mnras/sty108}
  {\bibfield  {journal} {\bibinfo  {journal} {\mnras}\ }\textbf {\bibinfo
  {volume} {475}},\ \bibinfo {pages} {4186} (\bibinfo {year} {2018})},\ \Eprint
  {https://arxiv.org/abs/1708.08452} {arXiv:1708.08452 [astro-ph.HE]}
  \BibitemShut {NoStop}%
\bibitem [{\citenamefont {{Radice}}\ \emph {et~al.}(2022)\citenamefont
  {{Radice}}, \citenamefont {{Bernuzzi}}, \citenamefont {{Perego}},\ and\
  \citenamefont {{Haas}}}]{2022MNRAS.512.1499R}%
  \BibitemOpen
  \bibfield  {author} {\bibinfo {author} {\bibfnamefont {D.}~\bibnamefont
  {{Radice}}}, \bibinfo {author} {\bibfnamefont {S.}~\bibnamefont
  {{Bernuzzi}}}, \bibinfo {author} {\bibfnamefont {A.}~\bibnamefont
  {{Perego}}},\ and\ \bibinfo {author} {\bibfnamefont {R.}~\bibnamefont
  {{Haas}}},\ }\bibfield  {title} {\bibinfo {title} {{A new moment-based
  general-relativistic neutrino-radiation transport code: Methods and first
  applications to neutron star mergers}},\ }\href
  {https://doi.org/10.1093/mnras/stac589} {\bibfield  {journal} {\bibinfo
  {journal} {\mnras}\ }\textbf {\bibinfo {volume} {512}},\ \bibinfo {pages}
  {1499} (\bibinfo {year} {2022})},\ \Eprint {https://arxiv.org/abs/2111.14858}
  {arXiv:2111.14858 [astro-ph.HE]} \BibitemShut {NoStop}%
\bibitem [{\citenamefont {{Sun}}\ \emph {et~al.}(2022)\citenamefont {{Sun}},
  \citenamefont {{Ruiz}}, \citenamefont {{Shapiro}},\ and\ \citenamefont
  {{Tsokaros}}}]{2022PhRvD.105j4028S}%
  \BibitemOpen
  \bibfield  {author} {\bibinfo {author} {\bibfnamefont {L.}~\bibnamefont
  {{Sun}}}, \bibinfo {author} {\bibfnamefont {M.}~\bibnamefont {{Ruiz}}},
  \bibinfo {author} {\bibfnamefont {S.~L.}\ \bibnamefont {{Shapiro}}},\ and\
  \bibinfo {author} {\bibfnamefont {A.}~\bibnamefont {{Tsokaros}}},\ }\bibfield
   {title} {\bibinfo {title} {{Jet launching from binary neutron star mergers:
  Incorporating neutrino transport and magnetic fields}},\ }\href
  {https://doi.org/10.1103/PhysRevD.105.104028} {\bibfield  {journal} {\bibinfo
   {journal} {\prd}\ }\textbf {\bibinfo {volume} {105}},\ \bibinfo {eid}
  {104028} (\bibinfo {year} {2022})},\ \Eprint
  {https://arxiv.org/abs/2202.12901} {arXiv:2202.12901 [astro-ph.HE]}
  \BibitemShut {NoStop}%
\bibitem [{\citenamefont {{Schianchi}}\ \emph {et~al.}(2024)\citenamefont
  {{Schianchi}}, \citenamefont {{Gieg}}, \citenamefont {{Nedora}},
  \citenamefont {{Neuweiler}}, \citenamefont {{Ujevic}}, \citenamefont
  {{Bulla}},\ and\ \citenamefont {{Dietrich}}}]{2024PhRvD.109d4012S}%
  \BibitemOpen
  \bibfield  {author} {\bibinfo {author} {\bibfnamefont {F.}~\bibnamefont
  {{Schianchi}}}, \bibinfo {author} {\bibfnamefont {H.}~\bibnamefont {{Gieg}}},
  \bibinfo {author} {\bibfnamefont {V.}~\bibnamefont {{Nedora}}}, \bibinfo
  {author} {\bibfnamefont {A.}~\bibnamefont {{Neuweiler}}}, \bibinfo {author}
  {\bibfnamefont {M.}~\bibnamefont {{Ujevic}}}, \bibinfo {author}
  {\bibfnamefont {M.}~\bibnamefont {{Bulla}}},\ and\ \bibinfo {author}
  {\bibfnamefont {T.}~\bibnamefont {{Dietrich}}},\ }\bibfield  {title}
  {\bibinfo {title} {{M 1 neutrino transport within the numerical-relativistic
  code BAM with application to low mass binary neutron star mergers}},\ }\href
  {https://doi.org/10.1103/PhysRevD.109.044012} {\bibfield  {journal} {\bibinfo
   {journal} {\prd}\ }\textbf {\bibinfo {volume} {109}},\ \bibinfo {eid}
  {044012} (\bibinfo {year} {2024})},\ \Eprint
  {https://arxiv.org/abs/2307.04572} {arXiv:2307.04572 [gr-qc]} \BibitemShut
  {NoStop}%
\bibitem [{\citenamefont {{Izquierdo}}\ \emph {et~al.}(2024)\citenamefont
  {{Izquierdo}}, \citenamefont {{Abalos}},\ and\ \citenamefont
  {{Palenzuela}}}]{2024PhRvD.109d3044I}%
  \BibitemOpen
  \bibfield  {author} {\bibinfo {author} {\bibfnamefont {M.~R.}\ \bibnamefont
  {{Izquierdo}}}, \bibinfo {author} {\bibfnamefont {F.}~\bibnamefont
  {{Abalos}}},\ and\ \bibinfo {author} {\bibfnamefont {C.}~\bibnamefont
  {{Palenzuela}}},\ }\bibfield  {title} {\bibinfo {title} {{Guided moments
  formalism: A new efficient full-neutrino treatment for astrophysical
  simulations}},\ }\href {https://doi.org/10.1103/PhysRevD.109.043044}
  {\bibfield  {journal} {\bibinfo  {journal} {\prd}\ }\textbf {\bibinfo
  {volume} {109}},\ \bibinfo {eid} {043044} (\bibinfo {year} {2024})},\ \Eprint
  {https://arxiv.org/abs/2312.09275} {arXiv:2312.09275 [astro-ph.HE]}
  \BibitemShut {NoStop}%
\bibitem [{\citenamefont {{Musolino}}\ and\ \citenamefont
  {{Rezzolla}}(2024)}]{2024MNRAS.528.5952M}%
  \BibitemOpen
  \bibfield  {author} {\bibinfo {author} {\bibfnamefont {C.}~\bibnamefont
  {{Musolino}}}\ and\ \bibinfo {author} {\bibfnamefont {L.}~\bibnamefont
  {{Rezzolla}}},\ }\bibfield  {title} {\bibinfo {title} {{A practical guide to
  a moment approach for neutrino transport in numerical relativity}},\ }\href
  {https://doi.org/10.1093/mnras/stae224} {\bibfield  {journal} {\bibinfo
  {journal} {\mnras}\ }\textbf {\bibinfo {volume} {528}},\ \bibinfo {pages}
  {5952} (\bibinfo {year} {2024})},\ \Eprint {https://arxiv.org/abs/2304.09168}
  {arXiv:2304.09168 [gr-qc]} \BibitemShut {NoStop}%
\bibitem [{\citenamefont {{Foucart}}\ \emph {et~al.}(2020)\citenamefont
  {{Foucart}}, \citenamefont {{Duez}}, \citenamefont {{Hebert}}, \citenamefont
  {{Kidder}}, \citenamefont {{Pfeiffer}},\ and\ \citenamefont
  {{Scheel}}}]{2020ApJ...902L..27F}%
  \BibitemOpen
  \bibfield  {author} {\bibinfo {author} {\bibfnamefont {F.}~\bibnamefont
  {{Foucart}}}, \bibinfo {author} {\bibfnamefont {M.~D.}\ \bibnamefont
  {{Duez}}}, \bibinfo {author} {\bibfnamefont {F.}~\bibnamefont {{Hebert}}},
  \bibinfo {author} {\bibfnamefont {L.~E.}\ \bibnamefont {{Kidder}}}, \bibinfo
  {author} {\bibfnamefont {H.~P.}\ \bibnamefont {{Pfeiffer}}},\ and\ \bibinfo
  {author} {\bibfnamefont {M.~A.}\ \bibnamefont {{Scheel}}},\ }\bibfield
  {title} {\bibinfo {title} {{Monte-Carlo Neutrino Transport in Neutron Star
  Merger Simulations}},\ }\href {https://doi.org/10.3847/2041-8213/abbb87}
  {\bibfield  {journal} {\bibinfo  {journal} {\apjl}\ }\textbf {\bibinfo
  {volume} {902}},\ \bibinfo {eid} {L27} (\bibinfo {year} {2020})},\ \Eprint
  {https://arxiv.org/abs/2008.08089} {arXiv:2008.08089 [astro-ph.HE]}
  \BibitemShut {NoStop}%
\bibitem [{\citenamefont {{Foucart}}\ \emph {et~al.}(2024)\citenamefont
  {{Foucart}}, \citenamefont {{Cheong}}, \citenamefont {{Duez}}, \citenamefont
  {{Kidder}}, \citenamefont {{Pfeiffer}},\ and\ \citenamefont
  {{Scheel}}}]{2024arXiv240715989F}%
  \BibitemOpen
  \bibfield  {author} {\bibinfo {author} {\bibfnamefont {F.}~\bibnamefont
  {{Foucart}}}, \bibinfo {author} {\bibfnamefont {C.-K.~P.}\ \bibnamefont
  {{Cheong}}}, \bibinfo {author} {\bibfnamefont {M.~D.}\ \bibnamefont
  {{Duez}}}, \bibinfo {author} {\bibfnamefont {L.~E.}\ \bibnamefont
  {{Kidder}}}, \bibinfo {author} {\bibfnamefont {H.~P.}\ \bibnamefont
  {{Pfeiffer}}},\ and\ \bibinfo {author} {\bibfnamefont {M.~A.}\ \bibnamefont
  {{Scheel}}},\ }\bibfield  {title} {\bibinfo {title} {{Robustness of neutron
  star merger simulations to changes in neutrino transport and neutrino-matter
  interactions}},\ }\href@noop {} {\bibfield  {journal} {\bibinfo  {journal}
  {arXiv e-prints}\ ,\ \bibinfo {eid} {arXiv:2407.15989}} (\bibinfo {year}
  {2024})},\ \Eprint {https://arxiv.org/abs/2407.15989} {arXiv:2407.15989
  [astro-ph.HE]} \BibitemShut {NoStop}%
\bibitem [{\citenamefont {{Cheong}}\ \emph
  {et~al.}(2024{\natexlab{a}})\citenamefont {{Cheong}}, \citenamefont
  {{Foucart}}, \citenamefont {{Duez}}, \citenamefont {{Offermans}},
  \citenamefont {{Muhammed}},\ and\ \citenamefont
  {{Chawhan}}}]{2024arXiv240716017C}%
  \BibitemOpen
  \bibfield  {author} {\bibinfo {author} {\bibfnamefont {C.-K.~P.}\
  \bibnamefont {{Cheong}}}, \bibinfo {author} {\bibfnamefont {F.}~\bibnamefont
  {{Foucart}}}, \bibinfo {author} {\bibfnamefont {M.~D.}\ \bibnamefont
  {{Duez}}}, \bibinfo {author} {\bibfnamefont {A.}~\bibnamefont {{Offermans}}},
  \bibinfo {author} {\bibfnamefont {N.}~\bibnamefont {{Muhammed}}},\ and\
  \bibinfo {author} {\bibfnamefont {P.}~\bibnamefont {{Chawhan}}},\ }\bibfield
  {title} {\bibinfo {title} {{Energy-dependent and energy-integrated two-moment
  general-relativistic neutrino transport simulations of hypermassive neutron
  star}},\ }\href {https://doi.org/10.48550/arXiv.2407.16017} {\bibfield
  {journal} {\bibinfo  {journal} {arXiv e-prints}\ ,\ \bibinfo {eid}
  {arXiv:2407.16017}} (\bibinfo {year} {2024}{\natexlab{a}})},\ \Eprint
  {https://arxiv.org/abs/2407.16017} {arXiv:2407.16017 [astro-ph.HE]}
  \BibitemShut {NoStop}%
\bibitem [{\citenamefont {{Richers}}\ \emph {et~al.}(2015)\citenamefont
  {{Richers}}, \citenamefont {{Kasen}}, \citenamefont {{O'Connor}},
  \citenamefont {{Fern{\'a}ndez}},\ and\ \citenamefont
  {{Ott}}}]{2015ApJ...813...38R}%
  \BibitemOpen
  \bibfield  {author} {\bibinfo {author} {\bibfnamefont {S.}~\bibnamefont
  {{Richers}}}, \bibinfo {author} {\bibfnamefont {D.}~\bibnamefont {{Kasen}}},
  \bibinfo {author} {\bibfnamefont {E.}~\bibnamefont {{O'Connor}}}, \bibinfo
  {author} {\bibfnamefont {R.}~\bibnamefont {{Fern{\'a}ndez}}},\ and\ \bibinfo
  {author} {\bibfnamefont {C.~D.}\ \bibnamefont {{Ott}}},\ }\bibfield  {title}
  {\bibinfo {title} {{Monte Carlo Neutrino Transport through Remnant Disks from
  Neutron Star Mergers}},\ }\href {https://doi.org/10.1088/0004-637X/813/1/38}
  {\bibfield  {journal} {\bibinfo  {journal} {\apj}\ }\textbf {\bibinfo
  {volume} {813}},\ \bibinfo {eid} {38} (\bibinfo {year} {2015})},\ \Eprint
  {https://arxiv.org/abs/1507.03606} {arXiv:1507.03606 [astro-ph.HE]}
  \BibitemShut {NoStop}%
\bibitem [{\citenamefont {{Perego}}\ \emph {et~al.}(2017)\citenamefont
  {{Perego}}, \citenamefont {{Yasin}},\ and\ \citenamefont
  {{Arcones}}}]{2017JPhG...44h4007P}%
  \BibitemOpen
  \bibfield  {author} {\bibinfo {author} {\bibfnamefont {A.}~\bibnamefont
  {{Perego}}}, \bibinfo {author} {\bibfnamefont {H.}~\bibnamefont {{Yasin}}},\
  and\ \bibinfo {author} {\bibfnamefont {A.}~\bibnamefont {{Arcones}}},\
  }\bibfield  {title} {\bibinfo {title} {{Neutrino pair annihilation above
  merger remnants: implications of a long-lived massive neutron star}},\ }\href
  {https://doi.org/10.1088/1361-6471/aa7bdc} {\bibfield  {journal} {\bibinfo
  {journal} {Journal of Physics G Nuclear Physics}\ }\textbf {\bibinfo {volume}
  {44}},\ \bibinfo {pages} {084007} (\bibinfo {year} {2017})},\ \Eprint
  {https://arxiv.org/abs/1701.02017} {arXiv:1701.02017 [astro-ph.HE]}
  \BibitemShut {NoStop}%
\bibitem [{\citenamefont {{Fujibayashi}}\ \emph {et~al.}(2017)\citenamefont
  {{Fujibayashi}}, \citenamefont {{Sekiguchi}}, \citenamefont {{Kiuchi}},\ and\
  \citenamefont {{Shibata}}}]{2017ApJ...846..114F}%
  \BibitemOpen
  \bibfield  {author} {\bibinfo {author} {\bibfnamefont {S.}~\bibnamefont
  {{Fujibayashi}}}, \bibinfo {author} {\bibfnamefont {Y.}~\bibnamefont
  {{Sekiguchi}}}, \bibinfo {author} {\bibfnamefont {K.}~\bibnamefont
  {{Kiuchi}}},\ and\ \bibinfo {author} {\bibfnamefont {M.}~\bibnamefont
  {{Shibata}}},\ }\bibfield  {title} {\bibinfo {title} {{Properties of
  Neutrino-driven Ejecta from the Remnant of a Binary Neutron Star Merger: Pure
  Radiation Hydrodynamics Case}},\ }\href
  {https://doi.org/10.3847/1538-4357/aa8039} {\bibfield  {journal} {\bibinfo
  {journal} {\apj}\ }\textbf {\bibinfo {volume} {846}},\ \bibinfo {eid} {114}
  (\bibinfo {year} {2017})},\ \Eprint {https://arxiv.org/abs/1703.10191}
  {arXiv:1703.10191 [astro-ph.HE]} \BibitemShut {NoStop}%
\bibitem [{\citenamefont {{Just}}\ \emph {et~al.}(2016)\citenamefont {{Just}},
  \citenamefont {{Obergaulinger}}, \citenamefont {{Janka}}, \citenamefont
  {{Bauswein}},\ and\ \citenamefont {{Schwarz}}}]{2016ApJ...816L..30J}%
  \BibitemOpen
  \bibfield  {author} {\bibinfo {author} {\bibfnamefont {O.}~\bibnamefont
  {{Just}}}, \bibinfo {author} {\bibfnamefont {M.}~\bibnamefont
  {{Obergaulinger}}}, \bibinfo {author} {\bibfnamefont {H.~T.}\ \bibnamefont
  {{Janka}}}, \bibinfo {author} {\bibfnamefont {A.}~\bibnamefont
  {{Bauswein}}},\ and\ \bibinfo {author} {\bibfnamefont {N.}~\bibnamefont
  {{Schwarz}}},\ }\bibfield  {title} {\bibinfo {title} {{Neutron-star Merger
  Ejecta as Obstacles to Neutrino-powered Jets of Gamma-Ray Bursts}},\ }\href
  {https://doi.org/10.3847/2041-8205/816/2/L30} {\bibfield  {journal} {\bibinfo
   {journal} {\apjl}\ }\textbf {\bibinfo {volume} {816}},\ \bibinfo {eid} {L30}
  (\bibinfo {year} {2016})},\ \Eprint {https://arxiv.org/abs/1510.04288}
  {arXiv:1510.04288 [astro-ph.HE]} \BibitemShut {NoStop}%
\bibitem [{\citenamefont {{Just}}\ \emph {et~al.}(2023)\citenamefont {{Just}},
  \citenamefont {{Vijayan}}, \citenamefont {{Xiong}}, \citenamefont
  {{Goriely}}, \citenamefont {{Soultanis}}, \citenamefont {{Bauswein}},
  \citenamefont {{Guilet}}, \citenamefont {{Janka}},\ and\ \citenamefont
  {{Mart{\'\i}nez-Pinedo}}}]{2023ApJ...951L..12J}%
  \BibitemOpen
  \bibfield  {author} {\bibinfo {author} {\bibfnamefont {O.}~\bibnamefont
  {{Just}}}, \bibinfo {author} {\bibfnamefont {V.}~\bibnamefont {{Vijayan}}},
  \bibinfo {author} {\bibfnamefont {Z.}~\bibnamefont {{Xiong}}}, \bibinfo
  {author} {\bibfnamefont {S.}~\bibnamefont {{Goriely}}}, \bibinfo {author}
  {\bibfnamefont {T.}~\bibnamefont {{Soultanis}}}, \bibinfo {author}
  {\bibfnamefont {A.}~\bibnamefont {{Bauswein}}}, \bibinfo {author}
  {\bibfnamefont {J.}~\bibnamefont {{Guilet}}}, \bibinfo {author}
  {\bibfnamefont {H.~T.}\ \bibnamefont {{Janka}}},\ and\ \bibinfo {author}
  {\bibfnamefont {G.}~\bibnamefont {{Mart{\'\i}nez-Pinedo}}},\ }\bibfield
  {title} {\bibinfo {title} {{End-to-end Kilonova Models of Neutron Star
  Mergers with Delayed Black Hole Formation}},\ }\href
  {https://doi.org/10.3847/2041-8213/acdad2} {\bibfield  {journal} {\bibinfo
  {journal} {\apjl}\ }\textbf {\bibinfo {volume} {951}},\ \bibinfo {eid} {L12}
  (\bibinfo {year} {2023})},\ \Eprint {https://arxiv.org/abs/2302.10928}
  {arXiv:2302.10928 [astro-ph.HE]} \BibitemShut {NoStop}%
\bibitem [{\citenamefont {{Miller}}\ \emph {et~al.}(2020)\citenamefont
  {{Miller}}, \citenamefont {{Sprouse}}, \citenamefont {{Fryer}}, \citenamefont
  {{Ryan}}, \citenamefont {{Dolence}}, \citenamefont {{Mumpower}},\ and\
  \citenamefont {{Surman}}}]{2020ApJ...902...66M}%
  \BibitemOpen
  \bibfield  {author} {\bibinfo {author} {\bibfnamefont {J.~M.}\ \bibnamefont
  {{Miller}}}, \bibinfo {author} {\bibfnamefont {T.~M.}\ \bibnamefont
  {{Sprouse}}}, \bibinfo {author} {\bibfnamefont {C.~L.}\ \bibnamefont
  {{Fryer}}}, \bibinfo {author} {\bibfnamefont {B.~R.}\ \bibnamefont {{Ryan}}},
  \bibinfo {author} {\bibfnamefont {J.~C.}\ \bibnamefont {{Dolence}}}, \bibinfo
  {author} {\bibfnamefont {M.~R.}\ \bibnamefont {{Mumpower}}},\ and\ \bibinfo
  {author} {\bibfnamefont {R.}~\bibnamefont {{Surman}}},\ }\bibfield  {title}
  {\bibinfo {title} {{Full Transport General Relativistic Radiation
  Magnetohydrodynamics for Nucleosynthesis in Collapsars}},\ }\href
  {https://doi.org/10.3847/1538-4357/abb4e3} {\bibfield  {journal} {\bibinfo
  {journal} {\apj}\ }\textbf {\bibinfo {volume} {902}},\ \bibinfo {eid} {66}
  (\bibinfo {year} {2020})},\ \Eprint {https://arxiv.org/abs/1912.03378}
  {arXiv:1912.03378 [astro-ph.HE]} \BibitemShut {NoStop}%
\bibitem [{\citenamefont {{Foucart}}(2023)}]{2023LRCA....9....1F}%
  \BibitemOpen
  \bibfield  {author} {\bibinfo {author} {\bibfnamefont {F.}~\bibnamefont
  {{Foucart}}},\ }\bibfield  {title} {\bibinfo {title} {{Neutrino transport in
  general relativistic neutron star merger simulations}},\ }\href
  {https://doi.org/10.1007/s41115-023-00016-y} {\bibfield  {journal} {\bibinfo
  {journal} {Living Reviews in Computational Astrophysics}\ }\textbf {\bibinfo
  {volume} {9}},\ \bibinfo {eid} {1} (\bibinfo {year} {2023})},\ \Eprint
  {https://arxiv.org/abs/2209.02538} {arXiv:2209.02538 [astro-ph.HE]}
  \BibitemShut {NoStop}%
\bibitem [{\citenamefont {{Ng}}\ \emph {et~al.}(2024)\citenamefont {{Ng}},
  \citenamefont {{Cheong}}, \citenamefont {{Lam}},\ and\ \citenamefont
  {{Li}}}]{2024ApJS..272....9N}%
  \BibitemOpen
  \bibfield  {author} {\bibinfo {author} {\bibfnamefont {H.~H.-Y.}\
  \bibnamefont {{Ng}}}, \bibinfo {author} {\bibfnamefont {P.~C.-K.}\
  \bibnamefont {{Cheong}}}, \bibinfo {author} {\bibfnamefont {A.~T.-L.}\
  \bibnamefont {{Lam}}},\ and\ \bibinfo {author} {\bibfnamefont {T.~G.~F.}\
  \bibnamefont {{Li}}},\ }\bibfield  {title} {\bibinfo {title}
  {{General-relativistic Radiation Transport Scheme in Gmunu. II.
  Implementation of Novel Microphysical Library for Neutrino
  Radiation{\textemdash}Weakhub}},\ }\href
  {https://doi.org/10.3847/1538-4365/ad2fbd} {\bibfield  {journal} {\bibinfo
  {journal} {\apjs}\ }\textbf {\bibinfo {volume} {272}},\ \bibinfo {eid} {9}
  (\bibinfo {year} {2024})},\ \Eprint {https://arxiv.org/abs/2309.03526}
  {arXiv:2309.03526 [astro-ph.HE]} \BibitemShut {NoStop}%
\bibitem [{\citenamefont {{Mezzacappa}}\ and\ \citenamefont
  {{Bruenn}}(1993)}]{1993ApJ...410..740M}%
  \BibitemOpen
  \bibfield  {author} {\bibinfo {author} {\bibfnamefont {A.}~\bibnamefont
  {{Mezzacappa}}}\ and\ \bibinfo {author} {\bibfnamefont {S.~W.}\ \bibnamefont
  {{Bruenn}}},\ }\bibfield  {title} {\bibinfo {title} {{Stellar Core Collapse:
  A Boltzmann Treatment of Neutrino-Electron Scattering}},\ }\href
  {https://doi.org/10.1086/172791} {\bibfield  {journal} {\bibinfo  {journal}
  {\apj}\ }\textbf {\bibinfo {volume} {410}},\ \bibinfo {pages} {740} (\bibinfo
  {year} {1993})}\BibitemShut {NoStop}%
\bibitem [{\citenamefont {{Kawaguchi}}\ \emph {et~al.}(2024)\citenamefont
  {{Kawaguchi}}, \citenamefont {{Fujibayashi}},\ and\ \citenamefont
  {{Shibata}}}]{2024arXiv241002380K}%
  \BibitemOpen
  \bibfield  {author} {\bibinfo {author} {\bibfnamefont {K.}~\bibnamefont
  {{Kawaguchi}}}, \bibinfo {author} {\bibfnamefont {S.}~\bibnamefont
  {{Fujibayashi}}},\ and\ \bibinfo {author} {\bibfnamefont {M.}~\bibnamefont
  {{Shibata}}},\ }\bibfield  {title} {\bibinfo {title} {{Long-term Monte Carlo
  based neutrino-radiation viscous-hydrodynamics simulations for a merger
  remnant black hole-torus system}},\ }\href@noop {} {\bibfield  {journal}
  {\bibinfo  {journal} {arXiv e-prints}\ ,\ \bibinfo {eid} {arXiv:2410.02380}}
  (\bibinfo {year} {2024})},\ \Eprint {https://arxiv.org/abs/2410.02380}
  {arXiv:2410.02380 [astro-ph.HE]} \BibitemShut {NoStop}%
\bibitem [{\citenamefont {{Cheong}}\ \emph
  {et~al.}(2024{\natexlab{b}})\citenamefont {{Cheong}}, \citenamefont
  {{Muhammed}}, \citenamefont {{Chawhan}}, \citenamefont {{Duez}},
  \citenamefont {{Foucart}}, \citenamefont {{Kidder}}, \citenamefont
  {{Pfeiffer}},\ and\ \citenamefont {{Scheel}}}]{2024PhRvD.110d3015C}%
  \BibitemOpen
  \bibfield  {author} {\bibinfo {author} {\bibfnamefont {P.~C.-K.}\
  \bibnamefont {{Cheong}}}, \bibinfo {author} {\bibfnamefont {N.}~\bibnamefont
  {{Muhammed}}}, \bibinfo {author} {\bibfnamefont {P.}~\bibnamefont
  {{Chawhan}}}, \bibinfo {author} {\bibfnamefont {M.~D.}\ \bibnamefont
  {{Duez}}}, \bibinfo {author} {\bibfnamefont {F.}~\bibnamefont {{Foucart}}},
  \bibinfo {author} {\bibfnamefont {L.~E.}\ \bibnamefont {{Kidder}}}, \bibinfo
  {author} {\bibfnamefont {H.~P.}\ \bibnamefont {{Pfeiffer}}},\ and\ \bibinfo
  {author} {\bibfnamefont {M.~A.}\ \bibnamefont {{Scheel}}},\ }\bibfield
  {title} {\bibinfo {title} {{High angular momentum hot differentially rotating
  equilibrium star evolutions in conformally flat spacetime}},\ }\href
  {https://doi.org/10.1103/PhysRevD.110.043015} {\bibfield  {journal} {\bibinfo
   {journal} {\prd}\ }\textbf {\bibinfo {volume} {110}},\ \bibinfo {eid}
  {043015} (\bibinfo {year} {2024}{\natexlab{b}})},\ \Eprint
  {https://arxiv.org/abs/2402.18529} {arXiv:2402.18529 [astro-ph.HE]}
  \BibitemShut {NoStop}%
\bibitem [{\citenamefont {{Hempel}}\ and\ \citenamefont
  {{Schaffner-Bielich}}(2010)}]{2010NuPhA.837..210H}%
  \BibitemOpen
  \bibfield  {author} {\bibinfo {author} {\bibfnamefont {M.}~\bibnamefont
  {{Hempel}}}\ and\ \bibinfo {author} {\bibfnamefont {J.}~\bibnamefont
  {{Schaffner-Bielich}}},\ }\bibfield  {title} {\bibinfo {title} {{A
  statistical model for a complete supernova equation of state}},\ }\href
  {https://doi.org/10.1016/j.nuclphysa.2010.02.010} {\bibfield  {journal}
  {\bibinfo  {journal} {\nphysa}\ }\textbf {\bibinfo {volume} {837}},\ \bibinfo
  {pages} {210} (\bibinfo {year} {2010})},\ \Eprint
  {https://arxiv.org/abs/0911.4073} {arXiv:0911.4073 [nucl-th]} \BibitemShut
  {NoStop}%
\bibitem [{\citenamefont {{Ury{\={u}}}}\ \emph {et~al.}(2019)\citenamefont
  {{Ury{\={u}}}}, \citenamefont {{Yoshida}}, \citenamefont {{Gourgoulhon}},
  \citenamefont {{Markakis}}, \citenamefont {{Fujisawa}}, \citenamefont
  {{Tsokaros}}, \citenamefont {{Taniguchi}},\ and\ \citenamefont
  {{Eriguchi}}}]{2019PhRvD.100l3019U}%
  \BibitemOpen
  \bibfield  {author} {\bibinfo {author} {\bibfnamefont {K.}~\bibnamefont
  {{Ury{\={u}}}}}, \bibinfo {author} {\bibfnamefont {S.}~\bibnamefont
  {{Yoshida}}}, \bibinfo {author} {\bibfnamefont {E.}~\bibnamefont
  {{Gourgoulhon}}}, \bibinfo {author} {\bibfnamefont {C.}~\bibnamefont
  {{Markakis}}}, \bibinfo {author} {\bibfnamefont {K.}~\bibnamefont
  {{Fujisawa}}}, \bibinfo {author} {\bibfnamefont {A.}~\bibnamefont
  {{Tsokaros}}}, \bibinfo {author} {\bibfnamefont {K.}~\bibnamefont
  {{Taniguchi}}},\ and\ \bibinfo {author} {\bibfnamefont {Y.}~\bibnamefont
  {{Eriguchi}}},\ }\bibfield  {title} {\bibinfo {title} {{New code for
  equilibriums and quasiequilibrium initial data of compact objects. IV.
  Rotating relativistic stars with mixed poloidal and toroidal magnetic
  fields}},\ }\href {https://doi.org/10.1103/PhysRevD.100.123019} {\bibfield
  {journal} {\bibinfo  {journal} {\prd}\ }\textbf {\bibinfo {volume} {100}},\
  \bibinfo {eid} {123019} (\bibinfo {year} {2019})},\ \Eprint
  {https://arxiv.org/abs/1906.10393} {arXiv:1906.10393 [gr-qc]} \BibitemShut
  {NoStop}%
\bibitem [{\citenamefont {{Cheong}}\ \emph {et~al.}(2020)\citenamefont
  {{Cheong}}, \citenamefont {{Lin}},\ and\ \citenamefont
  {{Li}}}]{2020CQGra..37n5015C}%
  \BibitemOpen
  \bibfield  {author} {\bibinfo {author} {\bibfnamefont {P.~C.-K.}\
  \bibnamefont {{Cheong}}}, \bibinfo {author} {\bibfnamefont {L.-M.}\
  \bibnamefont {{Lin}}},\ and\ \bibinfo {author} {\bibfnamefont {T.~G.~F.}\
  \bibnamefont {{Li}}},\ }\bibfield  {title} {\bibinfo {title} {{Gmunu: toward
  multigrid based Einstein field equations solver for general-relativistic
  hydrodynamics simulations}},\ }\href
  {https://doi.org/10.1088/1361-6382/ab8e9c} {\bibfield  {journal} {\bibinfo
  {journal} {Classical and Quantum Gravity}\ }\textbf {\bibinfo {volume}
  {37}},\ \bibinfo {eid} {145015} (\bibinfo {year} {2020})},\ \Eprint
  {https://arxiv.org/abs/2001.05723} {arXiv:2001.05723 [gr-qc]} \BibitemShut
  {NoStop}%
\bibitem [{\citenamefont {{Cheong}}\ \emph {et~al.}(2021)\citenamefont
  {{Cheong}}, \citenamefont {{Lam}}, \citenamefont {{Ng}},\ and\ \citenamefont
  {{Li}}}]{2021MNRAS.508.2279C}%
  \BibitemOpen
  \bibfield  {author} {\bibinfo {author} {\bibfnamefont {P.~C.-K.}\
  \bibnamefont {{Cheong}}}, \bibinfo {author} {\bibfnamefont {A.~T.-L.}\
  \bibnamefont {{Lam}}}, \bibinfo {author} {\bibfnamefont {H.~H.-Y.}\
  \bibnamefont {{Ng}}},\ and\ \bibinfo {author} {\bibfnamefont {T.~G.~F.}\
  \bibnamefont {{Li}}},\ }\bibfield  {title} {\bibinfo {title} {{Gmunu:
  paralleled, grid-adaptive, general-relativistic magnetohydrodynamics in
  curvilinear geometries in dynamical space-times}},\ }\href
  {https://doi.org/10.1093/mnras/stab2606} {\bibfield  {journal} {\bibinfo
  {journal} {\mnras}\ }\textbf {\bibinfo {volume} {508}},\ \bibinfo {pages}
  {2279} (\bibinfo {year} {2021})},\ \Eprint {https://arxiv.org/abs/2012.07322}
  {arXiv:2012.07322 [astro-ph.IM]} \BibitemShut {NoStop}%
\bibitem [{\citenamefont {{Cheong}}\ \emph {et~al.}(2022)\citenamefont
  {{Cheong}}, \citenamefont {{Pong}}, \citenamefont {{Yip}},\ and\
  \citenamefont {{Li}}}]{2022ApJS..261...22C}%
  \BibitemOpen
  \bibfield  {author} {\bibinfo {author} {\bibfnamefont {P.~C.-K.}\
  \bibnamefont {{Cheong}}}, \bibinfo {author} {\bibfnamefont {D.~Y.~T.}\
  \bibnamefont {{Pong}}}, \bibinfo {author} {\bibfnamefont {A.~K.~L.}\
  \bibnamefont {{Yip}}},\ and\ \bibinfo {author} {\bibfnamefont {T.~G.~F.}\
  \bibnamefont {{Li}}},\ }\bibfield  {title} {\bibinfo {title} {{An Extension
  of Gmunu: General-relativistic Resistive Magnetohydrodynamics Based on
  Staggered-meshed Constrained Transport with Elliptic Cleaning}},\ }\href
  {https://doi.org/10.3847/1538-4365/ac6cec} {\bibfield  {journal} {\bibinfo
  {journal} {\apjs}\ }\textbf {\bibinfo {volume} {261}},\ \bibinfo {eid} {22}
  (\bibinfo {year} {2022})},\ \Eprint {https://arxiv.org/abs/2110.03732}
  {arXiv:2110.03732 [astro-ph.IM]} \BibitemShut {NoStop}%
\bibitem [{\citenamefont {{Cheong}}\ \emph {et~al.}(2023)\citenamefont
  {{Cheong}}, \citenamefont {{Ng}}, \citenamefont {{Lam}},\ and\ \citenamefont
  {{Li}}}]{2023ApJS..267...38C}%
  \BibitemOpen
  \bibfield  {author} {\bibinfo {author} {\bibfnamefont {P.~C.-K.}\
  \bibnamefont {{Cheong}}}, \bibinfo {author} {\bibfnamefont {H.~H.-Y.}\
  \bibnamefont {{Ng}}}, \bibinfo {author} {\bibfnamefont {A.~T.-L.}\
  \bibnamefont {{Lam}}},\ and\ \bibinfo {author} {\bibfnamefont {T.~G.~F.}\
  \bibnamefont {{Li}}},\ }\bibfield  {title} {\bibinfo {title}
  {{General-relativistic Radiation Transport Scheme in Gmunu. I. Implementation
  of Two-moment-based Multifrequency Radiative Transfer and Code Tests}},\
  }\href {https://doi.org/10.3847/1538-4365/acd931} {\bibfield  {journal}
  {\bibinfo  {journal} {\apjs}\ }\textbf {\bibinfo {volume} {267}},\ \bibinfo
  {eid} {38} (\bibinfo {year} {2023})},\ \Eprint
  {https://arxiv.org/abs/2303.03261} {arXiv:2303.03261 [astro-ph.IM]}
  \BibitemShut {NoStop}%
\bibitem [{\citenamefont {{Evans}}\ and\ \citenamefont
  {{Hawley}}(1988)}]{1988ApJ...332..659E}%
  \BibitemOpen
  \bibfield  {author} {\bibinfo {author} {\bibfnamefont {C.~R.}\ \bibnamefont
  {{Evans}}}\ and\ \bibinfo {author} {\bibfnamefont {J.~F.}\ \bibnamefont
  {{Hawley}}},\ }\bibfield  {title} {\bibinfo {title} {{Simulation of
  Magnetohydrodynamic Flows: A Constrained Transport Model}},\ }\href
  {https://doi.org/10.1086/166684} {\bibfield  {journal} {\bibinfo  {journal}
  {\apj}\ }\textbf {\bibinfo {volume} {332}},\ \bibinfo {pages} {659} (\bibinfo
  {year} {1988})}\BibitemShut {NoStop}%
\bibitem [{\citenamefont {Harten}\ \emph {et~al.}(1983)\citenamefont {Harten},
  \citenamefont {Lax},\ and\ \citenamefont {Leer}}]{harten1983upstream}%
  \BibitemOpen
  \bibfield  {author} {\bibinfo {author} {\bibfnamefont {A.}~\bibnamefont
  {Harten}}, \bibinfo {author} {\bibfnamefont {P.}~\bibnamefont {Lax}},\ and\
  \bibinfo {author} {\bibfnamefont {B.}~\bibnamefont {Leer}},\ }\bibfield
  {title} {\bibinfo {title} {On upstream differencing and godunov-type schemes
  for hyperbolic conservation laws},\ }\href {https://doi.org/10.1137/1025002}
  {\bibfield  {journal} {\bibinfo  {journal} {SIAM Review}\ }\textbf {\bibinfo
  {volume} {25}},\ \bibinfo {pages} {35} (\bibinfo {year} {1983})},\ \Eprint
  {https://arxiv.org/abs/https://doi.org/10.1137/1025002}
  {https://doi.org/10.1137/1025002} \BibitemShut {NoStop}%
\bibitem [{\citenamefont {{Colella}}\ and\ \citenamefont
  {{Woodward}}(1984)}]{1984JCoPh..54..174C}%
  \BibitemOpen
  \bibfield  {author} {\bibinfo {author} {\bibfnamefont {P.}~\bibnamefont
  {{Colella}}}\ and\ \bibinfo {author} {\bibfnamefont {P.~R.}\ \bibnamefont
  {{Woodward}}},\ }\bibfield  {title} {\bibinfo {title} {{The Piecewise
  Parabolic Method (PPM) for Gas-Dynamical Simulations}},\ }\href
  {https://doi.org/10.1016/0021-9991(84)90143-8} {\bibfield  {journal}
  {\bibinfo  {journal} {Journal of Computational Physics}\ }\textbf {\bibinfo
  {volume} {54}},\ \bibinfo {pages} {174} (\bibinfo {year} {1984})}\BibitemShut
  {NoStop}%
\bibitem [{\citenamefont {{Cavaglieri}}\ and\ \citenamefont
  {{Bewley}}(2015)}]{2015JCoPh.286..172C}%
  \BibitemOpen
  \bibfield  {author} {\bibinfo {author} {\bibfnamefont {D.}~\bibnamefont
  {{Cavaglieri}}}\ and\ \bibinfo {author} {\bibfnamefont {T.}~\bibnamefont
  {{Bewley}}},\ }\bibfield  {title} {\bibinfo {title} {{Low-storage
  implicit/explicit Runge-Kutta schemes for the simulation of stiff
  high-dimensional ODE systems}},\ }\href
  {https://doi.org/10.1016/j.jcp.2015.01.031} {\bibfield  {journal} {\bibinfo
  {journal} {Journal of Computational Physics}\ }\textbf {\bibinfo {volume}
  {286}},\ \bibinfo {pages} {172} (\bibinfo {year} {2015})}\BibitemShut
  {NoStop}%
\bibitem [{\citenamefont {{Minerbo}}(1978)}]{1978JQSRT..20..541M}%
  \BibitemOpen
  \bibfield  {author} {\bibinfo {author} {\bibfnamefont {G.~N.}\ \bibnamefont
  {{Minerbo}}},\ }\bibfield  {title} {\bibinfo {title} {{Maximum entropy
  Eddington factors.}},\ }\href {https://doi.org/10.1016/0022-4073(78)90024-9}
  {\bibfield  {journal} {\bibinfo  {journal} {\jqsrt}\ }\textbf {\bibinfo
  {volume} {20}},\ \bibinfo {pages} {541} (\bibinfo {year} {1978})}\BibitemShut
  {NoStop}%
\bibitem [{\citenamefont {{O'Connor}}(2015)}]{2015ApJS..219...24O}%
  \BibitemOpen
  \bibfield  {author} {\bibinfo {author} {\bibfnamefont {E.}~\bibnamefont
  {{O'Connor}}},\ }\bibfield  {title} {\bibinfo {title} {{An Open-source
  Neutrino Radiation Hydrodynamics Code for Core-collapse Supernovae}},\ }\href
  {https://doi.org/10.1088/0067-0049/219/2/24} {\bibfield  {journal} {\bibinfo
  {journal} {\apjs}\ }\textbf {\bibinfo {volume} {219}},\ \bibinfo {eid} {24}
  (\bibinfo {year} {2015})},\ \Eprint {https://arxiv.org/abs/1411.7058}
  {arXiv:1411.7058 [astro-ph.HE]} \BibitemShut {NoStop}%
\bibitem [{\citenamefont {{Horowitz}}(2002)}]{2002PhRvD..65d3001H}%
  \BibitemOpen
  \bibfield  {author} {\bibinfo {author} {\bibfnamefont {C.~J.}\ \bibnamefont
  {{Horowitz}}},\ }\bibfield  {title} {\bibinfo {title} {{Weak magnetism for
  antineutrinos in supernovae}},\ }\href
  {https://doi.org/10.1103/PhysRevD.65.043001} {\bibfield  {journal} {\bibinfo
  {journal} {\prd}\ }\textbf {\bibinfo {volume} {65}},\ \bibinfo {eid} {043001}
  (\bibinfo {year} {2002})},\ \Eprint {https://arxiv.org/abs/astro-ph/0109209}
  {arXiv:astro-ph/0109209 [astro-ph]} \BibitemShut {NoStop}%
\bibitem [{\citenamefont {{Burrows}}\ \emph {et~al.}(2006)\citenamefont
  {{Burrows}}, \citenamefont {{Reddy}},\ and\ \citenamefont
  {{Thompson}}}]{2006NuPhA.777..356B}%
  \BibitemOpen
  \bibfield  {author} {\bibinfo {author} {\bibfnamefont {A.}~\bibnamefont
  {{Burrows}}}, \bibinfo {author} {\bibfnamefont {S.}~\bibnamefont {{Reddy}}},\
  and\ \bibinfo {author} {\bibfnamefont {T.~A.}\ \bibnamefont {{Thompson}}},\
  }\bibfield  {title} {\bibinfo {title} {{Neutrino opacities in nuclear
  matter}},\ }\href {https://doi.org/10.1016/j.nuclphysa.2004.06.012}
  {\bibfield  {journal} {\bibinfo  {journal} {\nphysa}\ }\textbf {\bibinfo
  {volume} {777}},\ \bibinfo {pages} {356} (\bibinfo {year} {2006})},\ \Eprint
  {https://arxiv.org/abs/astro-ph/0404432} {arXiv:astro-ph/0404432 [astro-ph]}
  \BibitemShut {NoStop}%
\bibitem [{\citenamefont {{Bruenn}}(1985)}]{1985ApJS...58..771B}%
  \BibitemOpen
  \bibfield  {author} {\bibinfo {author} {\bibfnamefont {S.~W.}\ \bibnamefont
  {{Bruenn}}},\ }\bibfield  {title} {\bibinfo {title} {{Stellar core collapse -
  Numerical model and infall epoch}},\ }\href {https://doi.org/10.1086/191056}
  {\bibfield  {journal} {\bibinfo  {journal} {\apjs}\ }\textbf {\bibinfo
  {volume} {58}},\ \bibinfo {pages} {771} (\bibinfo {year} {1985})}\BibitemShut
  {NoStop}%
\bibitem [{\citenamefont {{Horowitz}}(1997)}]{1997PhRvD..55.4577H}%
  \BibitemOpen
  \bibfield  {author} {\bibinfo {author} {\bibfnamefont {C.~J.}\ \bibnamefont
  {{Horowitz}}},\ }\bibfield  {title} {\bibinfo {title} {{Neutrino trapping in
  a supernova and the screening of weak neutral currents}},\ }\href
  {https://doi.org/10.1103/PhysRevD.55.4577} {\bibfield  {journal} {\bibinfo
  {journal} {\prd}\ }\textbf {\bibinfo {volume} {55}},\ \bibinfo {pages} {4577}
  (\bibinfo {year} {1997})},\ \Eprint {https://arxiv.org/abs/astro-ph/9603138}
  {arXiv:astro-ph/9603138 [astro-ph]} \BibitemShut {NoStop}%
\bibitem [{\citenamefont {{Hammond}}\ \emph {et~al.}(2021)\citenamefont
  {{Hammond}}, \citenamefont {{Hawke}},\ and\ \citenamefont
  {{Andersson}}}]{2021PhRvD.104j3006H}%
  \BibitemOpen
  \bibfield  {author} {\bibinfo {author} {\bibfnamefont {P.}~\bibnamefont
  {{Hammond}}}, \bibinfo {author} {\bibfnamefont {I.}~\bibnamefont {{Hawke}}},\
  and\ \bibinfo {author} {\bibfnamefont {N.}~\bibnamefont {{Andersson}}},\
  }\bibfield  {title} {\bibinfo {title} {{Thermal aspects of neutron star
  mergers}},\ }\href {https://doi.org/10.1103/PhysRevD.104.103006} {\bibfield
  {journal} {\bibinfo  {journal} {\prd}\ }\textbf {\bibinfo {volume} {104}},\
  \bibinfo {eid} {103006} (\bibinfo {year} {2021})},\ \Eprint
  {https://arxiv.org/abs/2108.08649} {arXiv:2108.08649 [astro-ph.HE]}
  \BibitemShut {NoStop}%
\bibitem [{\citenamefont {{Longo Micchi}}\ \emph {et~al.}(2023)\citenamefont
  {{Longo Micchi}}, \citenamefont {{Radice}},\ and\ \citenamefont
  {{Chirenti}}}]{2023MNRAS.525.6359L}%
  \BibitemOpen
  \bibfield  {author} {\bibinfo {author} {\bibfnamefont {L.~F.}\ \bibnamefont
  {{Longo Micchi}}}, \bibinfo {author} {\bibfnamefont {D.}~\bibnamefont
  {{Radice}}},\ and\ \bibinfo {author} {\bibfnamefont {C.}~\bibnamefont
  {{Chirenti}}},\ }\bibfield  {title} {\bibinfo {title} {{Multimessenger
  emission from the accretion-induced collapse of white dwarfs}},\ }\href
  {https://doi.org/10.1093/mnras/stad2420} {\bibfield  {journal} {\bibinfo
  {journal} {\mnras}\ }\textbf {\bibinfo {volume} {525}},\ \bibinfo {pages}
  {6359} (\bibinfo {year} {2023})},\ \Eprint {https://arxiv.org/abs/2306.04711}
  {arXiv:2306.04711 [astro-ph.HE]} \BibitemShut {NoStop}%
\bibitem [{\citenamefont {{Espino}}\ \emph {et~al.}(2024)\citenamefont
  {{Espino}}, \citenamefont {{Hammond}}, \citenamefont {{Radice}},
  \citenamefont {{Bernuzzi}}, \citenamefont {{Gamba}}, \citenamefont {{Zappa}},
  \citenamefont {{Micchi}},\ and\ \citenamefont
  {{Perego}}}]{2024PhRvL.132u1001E}%
  \BibitemOpen
  \bibfield  {author} {\bibinfo {author} {\bibfnamefont {P.~L.}\ \bibnamefont
  {{Espino}}}, \bibinfo {author} {\bibfnamefont {P.}~\bibnamefont {{Hammond}}},
  \bibinfo {author} {\bibfnamefont {D.}~\bibnamefont {{Radice}}}, \bibinfo
  {author} {\bibfnamefont {S.}~\bibnamefont {{Bernuzzi}}}, \bibinfo {author}
  {\bibfnamefont {R.}~\bibnamefont {{Gamba}}}, \bibinfo {author} {\bibfnamefont
  {F.}~\bibnamefont {{Zappa}}}, \bibinfo {author} {\bibfnamefont {L.~F.~L.}\
  \bibnamefont {{Micchi}}},\ and\ \bibinfo {author} {\bibfnamefont
  {A.}~\bibnamefont {{Perego}}},\ }\bibfield  {title} {\bibinfo {title}
  {{Neutrino Trapping and Out-of-Equilibrium Effects in Binary Neutron-Star
  Merger Remnants}},\ }\href {https://doi.org/10.1103/PhysRevLett.132.211001}
  {\bibfield  {journal} {\bibinfo  {journal} {\prl}\ }\textbf {\bibinfo
  {volume} {132}},\ \bibinfo {eid} {211001} (\bibinfo {year} {2024})},\ \Eprint
  {https://arxiv.org/abs/2311.00031} {arXiv:2311.00031 [astro-ph.HE]}
  \BibitemShut {NoStop}%
\bibitem [{\citenamefont {{Perego}}\ \emph {et~al.}(2019)\citenamefont
  {{Perego}}, \citenamefont {{Bernuzzi}},\ and\ \citenamefont
  {{Radice}}}]{2019EPJA...55..124P}%
  \BibitemOpen
  \bibfield  {author} {\bibinfo {author} {\bibfnamefont {A.}~\bibnamefont
  {{Perego}}}, \bibinfo {author} {\bibfnamefont {S.}~\bibnamefont
  {{Bernuzzi}}},\ and\ \bibinfo {author} {\bibfnamefont {D.}~\bibnamefont
  {{Radice}}},\ }\bibfield  {title} {\bibinfo {title} {{Thermodynamics
  conditions of matter in neutron star mergers}},\ }\href
  {https://doi.org/10.1140/epja/i2019-12810-7} {\bibfield  {journal} {\bibinfo
  {journal} {European Physical Journal A}\ }\textbf {\bibinfo {volume} {55}},\
  \bibinfo {eid} {124} (\bibinfo {year} {2019})},\ \Eprint
  {https://arxiv.org/abs/1903.07898} {arXiv:1903.07898 [gr-qc]} \BibitemShut
  {NoStop}%
\bibitem [{\citenamefont {{Kuroda}}\ \emph {et~al.}(2016)\citenamefont
  {{Kuroda}}, \citenamefont {{Takiwaki}},\ and\ \citenamefont
  {{Kotake}}}]{2016ApJS..222...20K}%
  \BibitemOpen
  \bibfield  {author} {\bibinfo {author} {\bibfnamefont {T.}~\bibnamefont
  {{Kuroda}}}, \bibinfo {author} {\bibfnamefont {T.}~\bibnamefont
  {{Takiwaki}}},\ and\ \bibinfo {author} {\bibfnamefont {K.}~\bibnamefont
  {{Kotake}}},\ }\bibfield  {title} {\bibinfo {title} {{A New Multi-energy
  Neutrino Radiation-Hydrodynamics Code in Full General Relativity and Its
  Application to the Gravitational Collapse of Massive Stars}},\ }\href
  {https://doi.org/10.3847/0067-0049/222/2/20} {\bibfield  {journal} {\bibinfo
  {journal} {\apjs}\ }\textbf {\bibinfo {volume} {222}},\ \bibinfo {eid} {20}
  (\bibinfo {year} {2016})},\ \Eprint {https://arxiv.org/abs/1501.06330}
  {arXiv:1501.06330 [astro-ph.HE]} \BibitemShut {NoStop}%
\bibitem [{\citenamefont {{Fischer}}\ \emph {et~al.}(2020)\citenamefont
  {{Fischer}}, \citenamefont {{Guo}}, \citenamefont {{Mart{\'\i}nez-Pinedo}},
  \citenamefont {{Liebend{\"o}rfer}},\ and\ \citenamefont
  {{Mezzacappa}}}]{2020PhRvD.102l3001F}%
  \BibitemOpen
  \bibfield  {author} {\bibinfo {author} {\bibfnamefont {T.}~\bibnamefont
  {{Fischer}}}, \bibinfo {author} {\bibfnamefont {G.}~\bibnamefont {{Guo}}},
  \bibinfo {author} {\bibfnamefont {G.}~\bibnamefont {{Mart{\'\i}nez-Pinedo}}},
  \bibinfo {author} {\bibfnamefont {M.}~\bibnamefont {{Liebend{\"o}rfer}}},\
  and\ \bibinfo {author} {\bibfnamefont {A.}~\bibnamefont {{Mezzacappa}}},\
  }\bibfield  {title} {\bibinfo {title} {{Muonization of supernova matter}},\
  }\href {https://doi.org/10.1103/PhysRevD.102.123001} {\bibfield  {journal}
  {\bibinfo  {journal} {\prd}\ }\textbf {\bibinfo {volume} {102}},\ \bibinfo
  {eid} {123001} (\bibinfo {year} {2020})},\ \Eprint
  {https://arxiv.org/abs/2008.13628} {arXiv:2008.13628 [astro-ph.HE]}
  \BibitemShut {NoStop}%
\bibitem [{\citenamefont {{Bruenn}}\ and\ \citenamefont
  {{Haxton}}(1991)}]{1991ApJ...376..678B}%
  \BibitemOpen
  \bibfield  {author} {\bibinfo {author} {\bibfnamefont {S.~W.}\ \bibnamefont
  {{Bruenn}}}\ and\ \bibinfo {author} {\bibfnamefont {W.~C.}\ \bibnamefont
  {{Haxton}}},\ }\bibfield  {title} {\bibinfo {title} {{Neutrino-Nucleus
  Interactions in Core-Collapse Supernovae}},\ }\href
  {https://doi.org/10.1086/170316} {\bibfield  {journal} {\bibinfo  {journal}
  {\apj}\ }\textbf {\bibinfo {volume} {376}},\ \bibinfo {pages} {678} (\bibinfo
  {year} {1991})}\BibitemShut {NoStop}%
\bibitem [{\citenamefont {{Mezzacappa}}\ \emph {et~al.}(2020)\citenamefont
  {{Mezzacappa}}, \citenamefont {{Endeve}}, \citenamefont {{Messer}},\ and\
  \citenamefont {{Bruenn}}}]{2020LRCA....6....4M}%
  \BibitemOpen
  \bibfield  {author} {\bibinfo {author} {\bibfnamefont {A.}~\bibnamefont
  {{Mezzacappa}}}, \bibinfo {author} {\bibfnamefont {E.}~\bibnamefont
  {{Endeve}}}, \bibinfo {author} {\bibfnamefont {O.~E.~B.}\ \bibnamefont
  {{Messer}}},\ and\ \bibinfo {author} {\bibfnamefont {S.~W.}\ \bibnamefont
  {{Bruenn}}},\ }\bibfield  {title} {\bibinfo {title} {{Physical, numerical,
  and computational challenges of modeling neutrino transport in core-collapse
  supernovae}},\ }\href {https://doi.org/10.1007/s41115-020-00010-8} {\bibfield
   {journal} {\bibinfo  {journal} {Living Reviews in Computational
  Astrophysics}\ }\textbf {\bibinfo {volume} {6}},\ \bibinfo {eid} {4}
  (\bibinfo {year} {2020})},\ \Eprint {https://arxiv.org/abs/2010.09013}
  {arXiv:2010.09013 [astro-ph.HE]} \BibitemShut {NoStop}%
\bibitem [{\citenamefont {Boerner}\ \emph {et~al.}(2023)\citenamefont
  {Boerner}, \citenamefont {Deems}, \citenamefont {Furlani}, \citenamefont
  {Knuth},\ and\ \citenamefont {Towns}}]{10.1145/3569951.3597559}%
  \BibitemOpen
  \bibfield  {author} {\bibinfo {author} {\bibfnamefont {T.~J.}\ \bibnamefont
  {Boerner}}, \bibinfo {author} {\bibfnamefont {S.}~\bibnamefont {Deems}},
  \bibinfo {author} {\bibfnamefont {T.~R.}\ \bibnamefont {Furlani}}, \bibinfo
  {author} {\bibfnamefont {S.~L.}\ \bibnamefont {Knuth}},\ and\ \bibinfo
  {author} {\bibfnamefont {J.}~\bibnamefont {Towns}},\ }\bibfield  {title}
  {\bibinfo {title} {Access: Advancing innovation: Nsf’s advanced
  cyberinfrastructure coordination ecosystem: Services \& support},\ }in\ \href
  {https://doi.org/10.1145/3569951.3597559} {\emph {\bibinfo {booktitle}
  {Practice and Experience in Advanced Research Computing}}},\ \bibinfo {series
  and number} {PEARC '23}\ (\bibinfo  {publisher} {Association for Computing
  Machinery},\ \bibinfo {address} {New York, NY, USA},\ \bibinfo {year}
  {2023})\ p.\ \bibinfo {pages} {173–176}\BibitemShut {NoStop}%
\bibitem [{\citenamefont {{Cook}}\ \emph {et~al.}(1994)\citenamefont {{Cook}},
  \citenamefont {{Shapiro}},\ and\ \citenamefont
  {{Teukolsky}}}]{1994ApJ...422..227C}%
  \BibitemOpen
  \bibfield  {author} {\bibinfo {author} {\bibfnamefont {G.~B.}\ \bibnamefont
  {{Cook}}}, \bibinfo {author} {\bibfnamefont {S.~L.}\ \bibnamefont
  {{Shapiro}}},\ and\ \bibinfo {author} {\bibfnamefont {S.~A.}\ \bibnamefont
  {{Teukolsky}}},\ }\bibfield  {title} {\bibinfo {title} {{Rapidly Rotating
  Polytropes in General Relativity}},\ }\href {https://doi.org/10.1086/173721}
  {\bibfield  {journal} {\bibinfo  {journal} {\apj}\ }\textbf {\bibinfo
  {volume} {422}},\ \bibinfo {pages} {227} (\bibinfo {year}
  {1994})}\BibitemShut {NoStop}%
\bibitem [{\citenamefont {{Muhammed}}\ \emph {et~al.}(2024)\citenamefont
  {{Muhammed}}, \citenamefont {{Duez}}, \citenamefont {{Chawhan}},
  \citenamefont {{Ghadiri}}, \citenamefont {{Buchman}}, \citenamefont
  {{Foucart}}, \citenamefont {{Chi-Kit Cheong}}, \citenamefont {{Kidder}},
  \citenamefont {{Pfeiffer}},\ and\ \citenamefont
  {{Scheel}}}]{2024arXiv240305642M}%
  \BibitemOpen
  \bibfield  {author} {\bibinfo {author} {\bibfnamefont {N.}~\bibnamefont
  {{Muhammed}}}, \bibinfo {author} {\bibfnamefont {M.~D.}\ \bibnamefont
  {{Duez}}}, \bibinfo {author} {\bibfnamefont {P.}~\bibnamefont {{Chawhan}}},
  \bibinfo {author} {\bibfnamefont {N.}~\bibnamefont {{Ghadiri}}}, \bibinfo
  {author} {\bibfnamefont {L.~T.}\ \bibnamefont {{Buchman}}}, \bibinfo {author}
  {\bibfnamefont {F.}~\bibnamefont {{Foucart}}}, \bibinfo {author}
  {\bibfnamefont {P.}~\bibnamefont {{Chi-Kit Cheong}}}, \bibinfo {author}
  {\bibfnamefont {L.~E.}\ \bibnamefont {{Kidder}}}, \bibinfo {author}
  {\bibfnamefont {H.~P.}\ \bibnamefont {{Pfeiffer}}},\ and\ \bibinfo {author}
  {\bibfnamefont {M.~A.}\ \bibnamefont {{Scheel}}},\ }\bibfield  {title}
  {\bibinfo {title} {{Stability of hypermassive neutron stars with realistic
  rotation and entropy profiles}},\ }\href
  {https://doi.org/10.48550/arXiv.2403.05642} {\bibfield  {journal} {\bibinfo
  {journal} {arXiv e-prints}\ ,\ \bibinfo {eid} {arXiv:2403.05642}} (\bibinfo
  {year} {2024})},\ \Eprint {https://arxiv.org/abs/2403.05642}
  {arXiv:2403.05642 [gr-qc]} \BibitemShut {NoStop}%
\bibitem [{\citenamefont {{Turk}}\ \emph {et~al.}(2011)\citenamefont {{Turk}},
  \citenamefont {{Smith}}, \citenamefont {{Oishi}}, \citenamefont {{Skory}},
  \citenamefont {{Skillman}}, \citenamefont {{Abel}},\ and\ \citenamefont
  {{Norman}}}]{2011ApJS..192....9T}%
  \BibitemOpen
  \bibfield  {author} {\bibinfo {author} {\bibfnamefont {M.~J.}\ \bibnamefont
  {{Turk}}}, \bibinfo {author} {\bibfnamefont {B.~D.}\ \bibnamefont {{Smith}}},
  \bibinfo {author} {\bibfnamefont {J.~S.}\ \bibnamefont {{Oishi}}}, \bibinfo
  {author} {\bibfnamefont {S.}~\bibnamefont {{Skory}}}, \bibinfo {author}
  {\bibfnamefont {S.~W.}\ \bibnamefont {{Skillman}}}, \bibinfo {author}
  {\bibfnamefont {T.}~\bibnamefont {{Abel}}},\ and\ \bibinfo {author}
  {\bibfnamefont {M.~L.}\ \bibnamefont {{Norman}}},\ }\bibfield  {title}
  {\bibinfo {title} {{yt: A Multi-code Analysis Toolkit for Astrophysical
  Simulation Data}},\ }\href {https://doi.org/10.1088/0067-0049/192/1/9}
  {\bibfield  {journal} {\bibinfo  {journal} {The Astrophysical Journal
  Supplement Series}\ }\textbf {\bibinfo {volume} {192}},\ \bibinfo {eid} {9}
  (\bibinfo {year} {2011})},\ \Eprint {https://arxiv.org/abs/1011.3514}
  {arXiv:1011.3514 [astro-ph.IM]} \BibitemShut {NoStop}%
\bibitem [{\citenamefont {Harris}\ \emph {et~al.}(2020)\citenamefont {Harris},
  \citenamefont {Millman}, \citenamefont {van~der Walt}, \citenamefont
  {Gommers}, \citenamefont {Virtanen}, \citenamefont {Cournapeau},
  \citenamefont {Wieser}, \citenamefont {Taylor}, \citenamefont {Berg},
  \citenamefont {Smith}, \citenamefont {Kern}, \citenamefont {Picus},
  \citenamefont {Hoyer}, \citenamefont {van Kerkwijk}, \citenamefont {Brett},
  \citenamefont {Haldane}, \citenamefont {del R{\'{i}}o}, \citenamefont
  {Wiebe}, \citenamefont {Peterson}, \citenamefont {G{\'{e}}rard-Marchant},
  \citenamefont {Sheppard}, \citenamefont {Reddy}, \citenamefont {Weckesser},
  \citenamefont {Abbasi}, \citenamefont {Gohlke},\ and\ \citenamefont
  {Oliphant}}]{harris2020array}%
  \BibitemOpen
  \bibfield  {author} {\bibinfo {author} {\bibfnamefont {C.~R.}\ \bibnamefont
  {Harris}}, \bibinfo {author} {\bibfnamefont {K.~J.}\ \bibnamefont {Millman}},
  \bibinfo {author} {\bibfnamefont {S.~J.}\ \bibnamefont {van~der Walt}},
  \bibinfo {author} {\bibfnamefont {R.}~\bibnamefont {Gommers}}, \bibinfo
  {author} {\bibfnamefont {P.}~\bibnamefont {Virtanen}}, \bibinfo {author}
  {\bibfnamefont {D.}~\bibnamefont {Cournapeau}}, \bibinfo {author}
  {\bibfnamefont {E.}~\bibnamefont {Wieser}}, \bibinfo {author} {\bibfnamefont
  {J.}~\bibnamefont {Taylor}}, \bibinfo {author} {\bibfnamefont
  {S.}~\bibnamefont {Berg}}, \bibinfo {author} {\bibfnamefont {N.~J.}\
  \bibnamefont {Smith}}, \bibinfo {author} {\bibfnamefont {R.}~\bibnamefont
  {Kern}}, \bibinfo {author} {\bibfnamefont {M.}~\bibnamefont {Picus}},
  \bibinfo {author} {\bibfnamefont {S.}~\bibnamefont {Hoyer}}, \bibinfo
  {author} {\bibfnamefont {M.~H.}\ \bibnamefont {van Kerkwijk}}, \bibinfo
  {author} {\bibfnamefont {M.}~\bibnamefont {Brett}}, \bibinfo {author}
  {\bibfnamefont {A.}~\bibnamefont {Haldane}}, \bibinfo {author} {\bibfnamefont
  {J.~F.}\ \bibnamefont {del R{\'{i}}o}}, \bibinfo {author} {\bibfnamefont
  {M.}~\bibnamefont {Wiebe}}, \bibinfo {author} {\bibfnamefont
  {P.}~\bibnamefont {Peterson}}, \bibinfo {author} {\bibfnamefont
  {P.}~\bibnamefont {G{\'{e}}rard-Marchant}}, \bibinfo {author} {\bibfnamefont
  {K.}~\bibnamefont {Sheppard}}, \bibinfo {author} {\bibfnamefont
  {T.}~\bibnamefont {Reddy}}, \bibinfo {author} {\bibfnamefont
  {W.}~\bibnamefont {Weckesser}}, \bibinfo {author} {\bibfnamefont
  {H.}~\bibnamefont {Abbasi}}, \bibinfo {author} {\bibfnamefont
  {C.}~\bibnamefont {Gohlke}},\ and\ \bibinfo {author} {\bibfnamefont {T.~E.}\
  \bibnamefont {Oliphant}},\ }\bibfield  {title} {\bibinfo {title} {Array
  programming with {NumPy}},\ }\href
  {https://doi.org/10.1038/s41586-020-2649-2} {\bibfield  {journal} {\bibinfo
  {journal} {Nature}\ }\textbf {\bibinfo {volume} {585}},\ \bibinfo {pages}
  {357} (\bibinfo {year} {2020})}\BibitemShut {NoStop}%
\bibitem [{\citenamefont {pandas~development team}(2020)}]{reback2020pandas}%
  \BibitemOpen
  \bibfield  {author} {\bibinfo {author} {\bibfnamefont {T.}~\bibnamefont
  {pandas~development team}},\ }\href {https://doi.org/10.5281/zenodo.3509134}
  {\bibinfo {title} {pandas-dev/pandas: Pandas}} (\bibinfo {year}
  {2020})\BibitemShut {NoStop}%
\bibitem [{\citenamefont {{W}es
  {M}c{K}inney}(2010)}]{mckinney-proc-scipy-2010}%
  \BibitemOpen
  \bibfield  {author} {\bibinfo {author} {\bibnamefont {{W}es {M}c{K}inney}},\
  }\bibfield  {title} {\bibinfo {title} {{D}ata {S}tructures for {S}tatistical
  {C}omputing in {P}ython},\ }in\ \href
  {https://doi.org/10.25080/Majora-92bf1922-00a} {\emph {\bibinfo {booktitle}
  {{P}roceedings of the 9th {P}ython in {S}cience {C}onference}}},\ \bibinfo
  {editor} {edited by\ \bibinfo {editor} {\bibnamefont {{S}t\'efan van~der
  {W}alt}}\ and\ \bibinfo {editor} {\bibnamefont {{J}arrod {M}illman}}}\
  (\bibinfo {year} {2010})\ pp.\ \bibinfo {pages} {56 -- 61}\BibitemShut
  {NoStop}%
\bibitem [{\citenamefont {Virtanen}\ \emph {et~al.}(2020)\citenamefont
  {Virtanen}, \citenamefont {Gommers}, \citenamefont {Oliphant}, \citenamefont
  {Haberland}, \citenamefont {Reddy}, \citenamefont {Cournapeau}, \citenamefont
  {Burovski}, \citenamefont {Peterson}, \citenamefont {Weckesser},
  \citenamefont {Bright}, \citenamefont {{van der Walt}}, \citenamefont
  {Brett}, \citenamefont {Wilson}, \citenamefont {Millman}, \citenamefont
  {Mayorov}, \citenamefont {Nelson}, \citenamefont {Jones}, \citenamefont
  {Kern}, \citenamefont {Larson}, \citenamefont {Carey}, \citenamefont {Polat},
  \citenamefont {Feng}, \citenamefont {Moore}, \citenamefont {{VanderPlas}},
  \citenamefont {Laxalde}, \citenamefont {Perktold}, \citenamefont {Cimrman},
  \citenamefont {Henriksen}, \citenamefont {Quintero}, \citenamefont {Harris},
  \citenamefont {Archibald}, \citenamefont {Ribeiro}, \citenamefont
  {Pedregosa}, \citenamefont {{van Mulbregt}},\ and\ \citenamefont {{SciPy 1.0
  Contributors}}}]{2020SciPy-NMeth}%
  \BibitemOpen
  \bibfield  {author} {\bibinfo {author} {\bibfnamefont {P.}~\bibnamefont
  {Virtanen}}, \bibinfo {author} {\bibfnamefont {R.}~\bibnamefont {Gommers}},
  \bibinfo {author} {\bibfnamefont {T.~E.}\ \bibnamefont {Oliphant}}, \bibinfo
  {author} {\bibfnamefont {M.}~\bibnamefont {Haberland}}, \bibinfo {author}
  {\bibfnamefont {T.}~\bibnamefont {Reddy}}, \bibinfo {author} {\bibfnamefont
  {D.}~\bibnamefont {Cournapeau}}, \bibinfo {author} {\bibfnamefont
  {E.}~\bibnamefont {Burovski}}, \bibinfo {author} {\bibfnamefont
  {P.}~\bibnamefont {Peterson}}, \bibinfo {author} {\bibfnamefont
  {W.}~\bibnamefont {Weckesser}}, \bibinfo {author} {\bibfnamefont
  {J.}~\bibnamefont {Bright}}, \bibinfo {author} {\bibfnamefont {S.~J.}\
  \bibnamefont {{van der Walt}}}, \bibinfo {author} {\bibfnamefont
  {M.}~\bibnamefont {Brett}}, \bibinfo {author} {\bibfnamefont
  {J.}~\bibnamefont {Wilson}}, \bibinfo {author} {\bibfnamefont {K.~J.}\
  \bibnamefont {Millman}}, \bibinfo {author} {\bibfnamefont {N.}~\bibnamefont
  {Mayorov}}, \bibinfo {author} {\bibfnamefont {A.~R.~J.}\ \bibnamefont
  {Nelson}}, \bibinfo {author} {\bibfnamefont {E.}~\bibnamefont {Jones}},
  \bibinfo {author} {\bibfnamefont {R.}~\bibnamefont {Kern}}, \bibinfo {author}
  {\bibfnamefont {E.}~\bibnamefont {Larson}}, \bibinfo {author} {\bibfnamefont
  {C.~J.}\ \bibnamefont {Carey}}, \bibinfo {author} {\bibfnamefont
  {{\.I}.}~\bibnamefont {Polat}}, \bibinfo {author} {\bibfnamefont
  {Y.}~\bibnamefont {Feng}}, \bibinfo {author} {\bibfnamefont {E.~W.}\
  \bibnamefont {Moore}}, \bibinfo {author} {\bibfnamefont {J.}~\bibnamefont
  {{VanderPlas}}}, \bibinfo {author} {\bibfnamefont {D.}~\bibnamefont
  {Laxalde}}, \bibinfo {author} {\bibfnamefont {J.}~\bibnamefont {Perktold}},
  \bibinfo {author} {\bibfnamefont {R.}~\bibnamefont {Cimrman}}, \bibinfo
  {author} {\bibfnamefont {I.}~\bibnamefont {Henriksen}}, \bibinfo {author}
  {\bibfnamefont {E.~A.}\ \bibnamefont {Quintero}}, \bibinfo {author}
  {\bibfnamefont {C.~R.}\ \bibnamefont {Harris}}, \bibinfo {author}
  {\bibfnamefont {A.~M.}\ \bibnamefont {Archibald}}, \bibinfo {author}
  {\bibfnamefont {A.~H.}\ \bibnamefont {Ribeiro}}, \bibinfo {author}
  {\bibfnamefont {F.}~\bibnamefont {Pedregosa}}, \bibinfo {author}
  {\bibfnamefont {P.}~\bibnamefont {{van Mulbregt}}},\ and\ \bibinfo {author}
  {\bibnamefont {{SciPy 1.0 Contributors}}},\ }\bibfield  {title} {\bibinfo
  {title} {{{SciPy} 1.0: Fundamental Algorithms for Scientific Computing in
  Python}},\ }\href {https://doi.org/10.1038/s41592-019-0686-2} {\bibfield
  {journal} {\bibinfo  {journal} {Nature Methods}\ }\textbf {\bibinfo {volume}
  {17}},\ \bibinfo {pages} {261} (\bibinfo {year} {2020})}\BibitemShut
  {NoStop}%
\bibitem [{\citenamefont {{Hunter}}(2007)}]{2007CSE.....9...90H}%
  \BibitemOpen
  \bibfield  {author} {\bibinfo {author} {\bibfnamefont {J.~D.}\ \bibnamefont
  {{Hunter}}},\ }\bibfield  {title} {\bibinfo {title} {{Matplotlib: A 2D
  Graphics Environment}},\ }\href {https://doi.org/10.1109/MCSE.2007.55}
  {\bibfield  {journal} {\bibinfo  {journal} {Computing in Science and
  Engineering}\ }\textbf {\bibinfo {volume} {9}},\ \bibinfo {pages} {90}
  (\bibinfo {year} {2007})}\BibitemShut {NoStop}%
\bibitem [{\citenamefont {Caswell}\ \emph {et~al.}(2023)\citenamefont
  {Caswell}, \citenamefont {Lee}, \citenamefont {de~Andrade}, \citenamefont
  {Droettboom}, \citenamefont {Hoffmann}, \citenamefont {Klymak}, \citenamefont
  {Hunter}, \citenamefont {Firing}, \citenamefont {Stansby}, \citenamefont
  {Varoquaux}, \citenamefont {Nielsen}, \citenamefont {Root}, \citenamefont
  {May}, \citenamefont {Gustafsson}, \citenamefont {Elson}, \citenamefont
  {Seppänen}, \citenamefont {Lee}, \citenamefont {Dale}, \citenamefont
  {hannah}, \citenamefont {McDougall}, \citenamefont {Straw}, \citenamefont
  {Hobson}, \citenamefont {Sunden}, \citenamefont {Lucas}, \citenamefont
  {Gohlke}, \citenamefont {Vincent}, \citenamefont {Yu}, \citenamefont {Ma},
  \citenamefont {Silvester},\ and\ \citenamefont
  {Moad}}]{thomas_a_caswell_2023_7697899}%
  \BibitemOpen
  \bibfield  {author} {\bibinfo {author} {\bibfnamefont {T.~A.}\ \bibnamefont
  {Caswell}}, \bibinfo {author} {\bibfnamefont {A.}~\bibnamefont {Lee}},
  \bibinfo {author} {\bibfnamefont {E.~S.}\ \bibnamefont {de~Andrade}},
  \bibinfo {author} {\bibfnamefont {M.}~\bibnamefont {Droettboom}}, \bibinfo
  {author} {\bibfnamefont {T.}~\bibnamefont {Hoffmann}}, \bibinfo {author}
  {\bibfnamefont {J.}~\bibnamefont {Klymak}}, \bibinfo {author} {\bibfnamefont
  {J.}~\bibnamefont {Hunter}}, \bibinfo {author} {\bibfnamefont
  {E.}~\bibnamefont {Firing}}, \bibinfo {author} {\bibfnamefont
  {D.}~\bibnamefont {Stansby}}, \bibinfo {author} {\bibfnamefont
  {N.}~\bibnamefont {Varoquaux}}, \bibinfo {author} {\bibfnamefont {J.~H.}\
  \bibnamefont {Nielsen}}, \bibinfo {author} {\bibfnamefont {B.}~\bibnamefont
  {Root}}, \bibinfo {author} {\bibfnamefont {R.}~\bibnamefont {May}}, \bibinfo
  {author} {\bibfnamefont {O.}~\bibnamefont {Gustafsson}}, \bibinfo {author}
  {\bibfnamefont {P.}~\bibnamefont {Elson}}, \bibinfo {author} {\bibfnamefont
  {J.~K.}\ \bibnamefont {Seppänen}}, \bibinfo {author} {\bibfnamefont {J.-J.}\
  \bibnamefont {Lee}}, \bibinfo {author} {\bibfnamefont {D.}~\bibnamefont
  {Dale}}, \bibinfo {author} {\bibnamefont {hannah}}, \bibinfo {author}
  {\bibfnamefont {D.}~\bibnamefont {McDougall}}, \bibinfo {author}
  {\bibfnamefont {A.}~\bibnamefont {Straw}}, \bibinfo {author} {\bibfnamefont
  {P.}~\bibnamefont {Hobson}}, \bibinfo {author} {\bibfnamefont
  {K.}~\bibnamefont {Sunden}}, \bibinfo {author} {\bibfnamefont
  {G.}~\bibnamefont {Lucas}}, \bibinfo {author} {\bibfnamefont
  {C.}~\bibnamefont {Gohlke}}, \bibinfo {author} {\bibfnamefont {A.~F.}\
  \bibnamefont {Vincent}}, \bibinfo {author} {\bibfnamefont {T.~S.}\
  \bibnamefont {Yu}}, \bibinfo {author} {\bibfnamefont {E.}~\bibnamefont {Ma}},
  \bibinfo {author} {\bibfnamefont {S.}~\bibnamefont {Silvester}},\ and\
  \bibinfo {author} {\bibfnamefont {C.}~\bibnamefont {Moad}},\ }\href
  {https://doi.org/10.5281/zenodo.7697899} {\bibinfo {title}
  {matplotlib/matplotlib: Rel: v3.7.1}} (\bibinfo {year} {2023})\BibitemShut
  {NoStop}%
\bibitem [{\citenamefont {{Betranhandy}}\ and\ \citenamefont
  {{O'Connor}}(2020)}]{2020PhRvD.102l3015B}%
  \BibitemOpen
  \bibfield  {author} {\bibinfo {author} {\bibfnamefont {A.}~\bibnamefont
  {{Betranhandy}}}\ and\ \bibinfo {author} {\bibfnamefont {E.}~\bibnamefont
  {{O'Connor}}},\ }\bibfield  {title} {\bibinfo {title} {{Impact of neutrino
  pair-production rates in core-collapse supernovae}},\ }\href
  {https://doi.org/10.1103/PhysRevD.102.123015} {\bibfield  {journal} {\bibinfo
   {journal} {\prd}\ }\textbf {\bibinfo {volume} {102}},\ \bibinfo {eid}
  {123015} (\bibinfo {year} {2020})},\ \Eprint
  {https://arxiv.org/abs/2010.02261} {arXiv:2010.02261 [astro-ph.HE]}
  \BibitemShut {NoStop}%
\end{thebibliography}%

\end{document}